\documentclass[pra,eqsecnum,twocolumn,amsmath,amssymb]{revtex4}

\usepackage{graphicx}
\usepackage{bm}
\usepackage[usenames,dvipsnames]{color}
\definecolor{altncolor}{rgb}{0,0,0.8}
\usepackage[colorlinks=true, linkcolor=green, anchorcolor=altncolor,
citecolor=altncolor, filecolor=altncolor, menucolor=altncolor,
urlcolor=altncolor]{hyperref}

\begin{document}

\title{Quantum-enhanced algorithms for classical target detection in complex environments}

\author{Peter B. Weichman}

\affiliation{BAE Systems, FAST Labs, 600 District Avenue, Burlington, MA 01803}




\begin{abstract}

Quantum computational approaches to some classic target identification and localization algorithms, especially for radar images, are investigated, and are found to raise a number of quantum statistics and quantum measurement issues with much broader applicability. Such algorithms are computationally intensive, involving coherent processing of large sensor data sets in order to extract a small number of low profile targets from a cluttered background. Target enhancement is accomplished through accurate statistical characterization of the environment, followed by optimal identification of statistical outliers. The key result of the work is that the environmental covariance matrix estimation and manipulation at the heart of the statistical analysis actually enables a highly efficient quantum implementation. The algorithm is inspired by recent approaches to quantum machine learning, but requires significant extensions, including previously overlooked `quantum analog--digital' conversion steps (which are found to substantially increase the required number of qubits), `quantum statistical' generalization of the classic phase estimation and Grover search algorithms, and careful consideration of projected measurement operations. Application regimes where quantum efficiencies could enable significant overall algorithm speedup are identified. Key possible bottlenecks, such as data loading and conversion, are identified as well.

\end{abstract}

\maketitle

\section{Introduction}
\label{sec:intro}

Although numerous technological uncertainties remain, quantum computers have been growing rapidly in size over the past few years \cite{qcsize,qsupreme}. Continuing on the current trajectory, platforms capable of implementing `intermediate-scale' quantum-enhanced algorithms with important practical applications could emerge in the 3--5 year time frame. This paper is aimed at expanding the space of such possible applications by examining the problem of moving target detection in a noisy, cluttered background, most commonly encountered in long-range radar imaging. Though motivated by this problem, the work highlights a number of more general quantum computational features that should find broader applicability.

The space-time adaptive processing (STAP) technique is a method for coherently combining and statistically processing signals from multiple receivers in order to suppress clutter and noise, and highlight moving targets of interest \cite{Klemm2004,Melvin2004}. Given the amount of input data and the number of mathematical operations, the technique is often computationally limited. On the other hand, the required output, namely the position and speed of a handful of above-threshold targets in what may be a very large image space, is remarkably low dimensional. Given the extreme sensitivity of multi-qubit entangled states to external measurement, such limited output requirements are a rather general feature of quantum algorithms \cite{NC2000}, and it is therefore natural to explore the possibility of a quantum STAP (QSTAP) implementation.

In the following we examine the various elements of the STAP algorithm and show that efficient quantum implementations indeed exist, with speed-ups varying from exponential to polynomial. The main bottleneck is data loading, which occurs repeatedly since each datum is in general accessed multiple times. For now the assumption is made that this operation may be run in parallel so that the data is continuously available from quantum memory as needed. This will clearly need to be tested on future platforms.

The results of the investigation may be summarized as follows. In Sec.\ \ref{sec:radarspbgnd} the classical STAP algorithm is summarized, including the clutter covariance matrix estimation and the `detection statistic' derived from it. Computation of the latter requires as input the inverse of the covariance matrix, providing the immediate motivation for adapting quantum linear algebra algorithms \cite{HHL2009} to the problem. However, although the covariance matrix is of low rank, it is far from sparse, and so direct application of these algorithms not possible.

An alternative formulation, known as density matrix exponentiation, has been proposed for certain problems in the quantum machine learning literature \cite{QML2017}, especially quantum principal component analysis \cite{QPCA2013}. The approach is inherently quantum statistical: rather than evolving a fixed quantum state, the computation proceeds by entangling the `working qubit space' of the desired state with a very high dimensional data space---namely with each of the very large number of copies of the radar data set used at each stage of the computation. On the face of it, this would seem to require a prohibitively large number of qubits, but the key insight is that the algorithm output requires measurements \emph{only} on the working space, and the results of these measurements do not depend at all on the subsequent evolution of a particular data vector copy following its brief interaction with the working space. In particular, coherence need not be maintained within the data space, allowing the corresponding qubits to be re-initialized and recycled (with previous state dissipated into the broader environment). Measurements on the working qubits, now heavily entangled with the environment, correspond to averages with respect to the reduced (working space) density matrix, itself a quantum average over all environmental degrees of freedom.

The evolution of the density matrix, although straightforward to define, nevertheless requires a careful reformulation of a number of other associated quantum algorithms normally encoded in isolated systems of qubits. Most importantly, its evolution is a form of quantum simulation, but standard simulation algorithms rely on `digital' representations of the wavefunction, in which values are stored bitwise in quantum registers rather than in the standard `analog' representation as complex amplitudes. Unfortunately, the rules of quantum statistics underlying the definition of the reduced density matrix, in particular the Born rule associating probabilities with wavefunction inner products, \emph{require} the quantum analog representation.

A new quantum digital-to-analog (qD/A) conversion, apparently previously overlooked \cite{QML2017,QPCA2013,QSVM2014}, must therefore be applied to the input data, and the reverse quantum analog-to-digital (qA/D) transformation must be applied to the output working space state before the final target identification algorithm may be applied (an extension of Grover search \cite{NC2000}). Since the analog state lives in a much lower dimensional space (a single complex number replaces a multi-qubit register), the qD/A transformation is extremely efficient, essentially a form of subspace projection. However, the reverse transformation requires full reconstruction of all the higher dimensional subspaces, and this requires a significant number of copies (exponential in the number of register bits) of the analog state. These may all be produced in parallel, so do not necessarily slow the computation, but it does significantly increase the number of qubits.

In addition to the digital--analog conversion steps, the density matrix formulation also requires a careful reformulation of a number of underlying standard algorithms that operate on the output working qubit subspace, accounting for the new environmental entanglement. These include phase estimation \cite{NC2000} (implementing diagonalization of the covariance matrix), the HHL matrix inversion algorithm \cite{HHL2009}, and the Grover search and quantum counting algorithms \cite{NC2000}. For example, phase estimation requires evaluation of the evolving state at a sequence of times, and subsequent quantum Fourier transform. A certain multi-time factorization property for the density matrix required to make this work turns out to place additional constraints on the organization of the data qubit space. More generally, extensive use is made of the fact that linear operations on the working qubit space factor out of the environmental average, and hence may be applied directly to the reduced density matrix. Equally important, unitary operations acting only on the data and/or environmental space leave the density matrix invariant.

The existence of the environmental entanglement, and of various effective subspace projection operations, raises a number of interesting quantum measurement issues that are carefully disentangled. For example, the re-initialization and recycling of the data qubits might themselves be viewed as a form of quantum measurement, whether deliberate or not, and one might worry that this influences ones knowledge of the state of the working space qubits. It is shown, however, that this is not the case---the former are all different forms of unitary operation on the environment and therefore have no effect on the reduced density matrix.

The outline of remainder of this paper is as follows. Following the classical STAP algorithm summary in Sec.\ \ref{sec:radarspbgnd}, the STAP algorithm reformulation as a quantum computation is summarized in Sec.\ \ref{sec:qimpsummary} and the key quantum speed-ups are identified. Details of each are divided among later sections: Drawing from the quantum machine learning literature \cite{QML2017}, and extending it in various important ways, the algorithm combines quantum simulation (Sec.\ \ref{sec:qsim}), quantum environmental interaction (Sec.\ \ref{sec:qdiss}), quantum phase estimation (Sec.\ \ref{sec:qphase}) and associated quantum linear algebra (Sec.\ \ref{sec:qlinalg}), and quantum search (Secs.\ \ref{sec:qtgtdetect} and \ref{sec:groveroraclegen}) in a very interesting way. The paper is concluded in Sec.\ \ref{sec:conclude}). Appendix \ref{app:qad} provides technical details of the qD/A and qA/D algorithms.

\section{Radar signal processing background}
\label{sec:radarspbgnd}

\begin{figure}
\includegraphics[width=3.3in,viewport=150 80 720 520,clip]{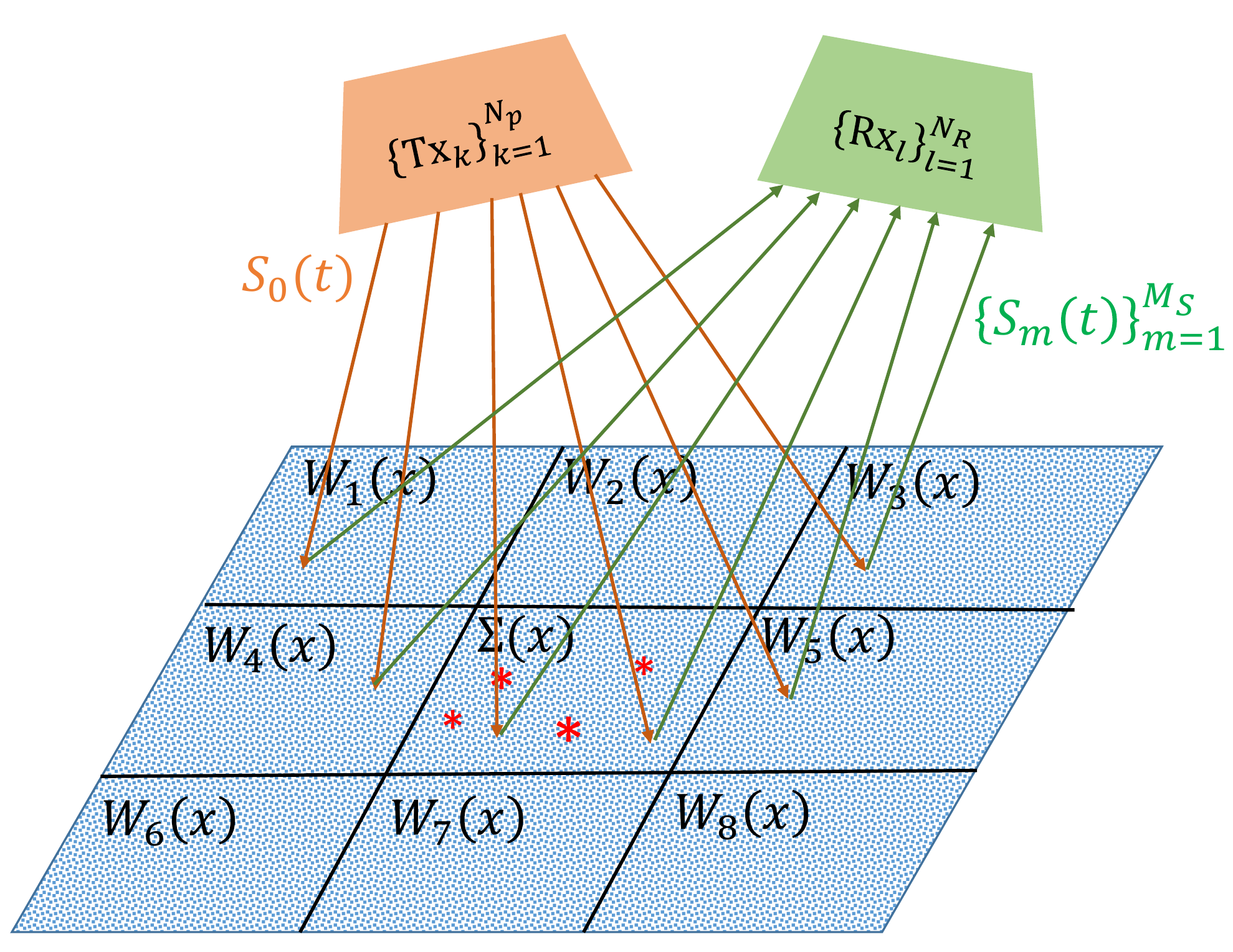}

\caption{\textbf{Radar data collection geometry:} Scene reflections from $N_p$ transmitted pulses $S_0(t)$ are detected by $N_R$ receivers, generating $M_S = N_p N_R$ received signals $S_m(t)$. Each pulse consists of an outgoing spherical wave, visualized here as a set of rays, so that $S_m(t)$ is a superposition of returns from every point in the scene---see (\ref{2.3}) and (\ref{2.6}). The scene consists of a target containing region (red stars, in general moving) surrounded by $K$ ($=8$ here) target-free clutter regions used to characterize the background statistics. The pulse compression operation (\ref{2.8}) converts the signal time traces into image functions $\Sigma(x)$ and $\{W_k(x) \}_{k=1}^K$, with $x = ({\bf x},{\bf v})$ some combination of spatial and velocity degrees of freedom depending on the measurement geometry. The $W_k$ are used to estimate the environmental covariance matrix $\hat G$ via (\ref{2.18}), and its inverse is used to optimally filter the target region data---see (\ref{2.20}). As described in Sec.\ \ref{sec:dataload}, the quantum implementation of this filtering operation is based on importing these data into appropriately formatted quantum states $|\psi_\Sigma \rangle$ and $|\psi_W \rangle$.}

\label{fig:radar}
\end{figure}

As illustrated in Fig.\ \ref{fig:radar}, measured radar data generally consist of scalar time traces $S_m(t)$, $m=1,2,3,\ldots,M_S$, corresponding to environmentally scattered returns due to a set of $M_S$ carefully designed transmitted broadband (usually `chirp' \cite{Chirp}) pulses. One may typically decompose $M_S = N_R N_p$ where $N_R$ is the number of platform receiver elements and $N_p$ is the number of transmitted pulses launched periodically along the trajectory of the moving platform. Clearly, in a real application the time $t$ will be sampled at $N_t$ discrete points as well \cite{foot:centerfreq}, for a total of $N_D = M_S N_t$ data samples. With very liberal estimates, one might have $M_S \sim N_t = O(10^5)$ (enabling imaging, say of a $10 \times 10$ km$^2$ area with 10 cm resolution), and assign $\sigma = 24$ bits to each recorded signal level, accumulating a roughly 30 GB dataset.

As discussed below, the STAP processing scheme entails some initial (classical) preprocessing of the data stream, and divides the imaged area into some number $K$ of subareas. The latter are used to statistically characterize the environment in order to highlight `outlying' target signals. As discussed in Sec.\ \ref{sec:qimpsummary}, the preprocessed data will be stored in a quantum state $|\Psi_D \rangle$ from which the data for each subarea is assumed separately accessible \cite{QRAM2008}.

\subsection{Signal model}
\label{sec:sigmod}

Let $s_0(t)$ be the transmitted pulse waveform. A typical signal model for each time trace $m$ is of the form
\begin{equation}
S_m(t) = S^m_\mathrm{tgt}(t) + S^m_\mathrm{bg}(t) + S^m_\mathrm{no}(t)
\label{2.1}
\end{equation}
consisting, respectively, of discrete target, background, and noise contributions. The target contribution takes the form
\begin{equation}
S^m_\mathrm{tgt}(t) = \sum_{j=1}^{N_\mathrm{tgt}} f_j
\frac{s_0(\eta_{mj}[t-R_m({\bf x}_j)/c])}
{[2\pi R_m({\bf x}_j)]^2},
\label{2.2}
\end{equation}
where $f_j$ is the scattering amplitude of target $j$, and
\begin{equation}
R_m({\bf x}) = |{\bf x}_T^m - {\bf x}| + |{\bf x}_R^m - {\bf x}|
\label{2.3}
\end{equation}
is the round trip distance from the launch point ${\bf x}_T^m$ of return $m$, to the point ${\bf x}$, and back to the corresponding receiver point ${\bf x}_R^m$. The Doppler factor, under the assumed conditions that all speeds are very small to the speed of light $c$, is given by
\begin{equation}
\eta_{mj} = \eta_m({\bf x}_j,{\bf v}_j - {\bf v}_0),
\label{2.4}
\end{equation}
with two-way Doppler function defined by
\begin{equation}
\eta_m({\bf x},{\bf v}) \equiv 1 - 2{\bf \hat n}_m({\bf x})
\cdot {\bf v}/c,
\label{2.5}
\end{equation}
where ${\bf \hat n}_m({\bf x}) = \frac{{\bf x} - {\bf x}_T^m}{|{\bf x} - {\bf x}_T^m|}$ is the look direction, and ${\bf v}_0,{\bf v}_j$ are the platform and target velocities, respectively. We neglect in (\ref{2.5}) and in the geometrical spreading denominator of (\ref{2.2}) the higher order difference $|{\bf x}_R^m - {\bf x}_T^m| \ll |{\bf x}-{\bf x}_R^m|, |{\bf x}-{\bf x}_T^m|$.

The background scene contribution takes the form
\begin{equation}
S^m_\mathrm{bg}(t) = \int_A d^2r f({\bf r})
\frac{s_0(\eta_m({\bf r}) [t-R_m({\bf r})/c])}
{[2\pi R_m({\bf r})]^2},
\label{2.6}
\end{equation}
where $A$ is the ground area of interest, and
\begin{equation}
\eta_m({\bf r}) \equiv \eta_m({\bf r},-{\bf v}_0)
= 1 + 2{\bf \hat n}_m({\bf r}) \cdot {\bf v}_0/c.
\label{2.7}
\end{equation}
The areal reflectivity $f({\bf r})$ is typically treated as a random field with certain prescribed statistics over the region of interest.

Finally, the noise signal $S_\mathrm{no}(t)$ contains all other residual contributions, including measurement uncertainty, background noise (including in-band radio stations, deliberate jamming signals, sky noise), and instrument noise, and is often modeled as Gaussian white noise over the frequency band of interest.

\subsection{Pulse and cross-range compression}
\label{sec:compress}

Given the chirp-like nature \cite{Chirp} of the typical transmitted waveform $s_0(t)$, even the return signal (\ref{2.2}) from discrete targets will be extended in time. At a given time, $S_m(t)$ also contains returns from a large cross-range swathe. As a first step, therefore, one performs a simultaneous pulse and cross-range compression transformation \cite{foot:qcompress}. This is accomplished through the inner product
\begin{equation}
\Sigma({\bf x},{\bf v}) = \sum_{m=1}^{M_S} \int dt S_m(t)
\sigma_m({\bf x},{\bf v};t)^*
\label{2.8}
\end{equation}
of the data with the expected signal from a target if it were present at a given position ${\bf x}$ with velocity ${\bf v}$:
\begin{equation}
\sigma_m({\bf x},{\bf v};t) \equiv
s_0(\eta_m({\bf x},{\bf v}) [t - R_m({\bf x})/c]).
\label{2.9}
\end{equation}
Equation (\ref{2.8}) reorganizes the raw received time traces into a set parameterized by position-Doppler indices, and forms the basis for all that follows. In applications, a suitable $N_x \times N_v$ gridding of this space would be used to preserve the correct number of degrees of freedom $N_D = N_x N_v = M_S N_t$.

If the platform track length is much smaller than the range, there will be Doppler sensitivity only to the radial component of the velocity $v = {\bf v} \cdot {\bf \hat n}({\bf x})$, where ${\bf \hat n}({\bf x})$ represents the mean look direction (e.g., from the track center). Given a sufficiently diverse dataset, properly designed pulse waveform, and strong targets, $\Sigma$ will have strong narrow peaks at the true target values ${\bf x}_j, {\bf \hat n}({\bf x}_j) \cdot ({\bf v}_j - {\bf v}_0)$ \cite{foot:motionmodel}. However, for low profile, slow moving targets in a cluttered background [e.g., with scene reflectivity $f({\bf r})$ also containing point-like features], robustly distinguishing target from clutter becomes more difficult. It is here that a quantitative characterization of the environment statistics becomes important, and forms the basis for the STAP algorithm.

The degree of cross-range localization relies on the diversity of transmitter-receiver locations ${\bf x}_T^m,{\bf x}_R^m$, while range--Doppler localization relies on the spectral properties of the pulse waveform. Thus, if we define
\begin{eqnarray}
F(\tau,\eta) &=& \int dt s_0(t) s_0[\eta(t + \tau)]^*
\nonumber \\
&=& \int \frac{d\omega}{2\pi \eta} \hat s_0(\omega)
\hat s_0(\omega/\eta)^* e^{i\omega \tau},
\label{2.10}
\end{eqnarray}
where the pulse spectrum $\hat s_0(\omega)$ is the Fourier transform of $s_0(t)$, then $F(\tau,\eta)$ will be strongly peaked about $\tau = 0$, $\eta = 1$, with peak width in $\tau$ governed by the pulse bandwidth $\Delta f$ via $\Delta \tau = 1/2\Delta f$, while the peak width in $\eta$ is governed by the center frequency $f_0$ and temporal pulse length $\Delta t$ via $\Delta \eta = \Delta v/c = 1/2f_0 \Delta t = 1/2N_0$ where $N_0$ is the number of wave periods in the pulse. In both cases the factor of two is due to two-way propagation. Well designed range--Doppler pulses are therefore broad in both frequency and time. For example, $\Delta f = 150$ MHz provides 1 m range resolution, while $N_0 = 10^7$ (e.g., $f_0 = 10$ GHz, $\Delta t = 1$ ms) provides 15 m/s range-rate resolution.

Inserting the target signal model (\ref{2.2}) into (\ref{2.8}) one obtains
\begin{equation}
\Sigma_\mathrm{tgt}({\bf x},{\bf v}) = \sum_{j=1}^{N_\mathrm{tgt}} f_j
{\cal S}({\bf x},{\bf v};{\bf x}_j,{\bf v}_j-{\bf v}_0)
\label{2.11}
\end{equation}
in which the `point spread function' is given by
\begin{eqnarray}
&&{\cal S}({\bf x},{\bf v};{\bf x}',{\bf v}')
\label{2.12} \\
&&\ \ \ \ =\ \sum_{m=1}^{M_S} \frac{F\left[\eta_m({\bf x}',{\bf v}')
\frac{R_m({\bf x}') - R_m({\bf x})}{c},
\frac{\eta_m({\bf x},{\bf v})}{\eta_m({\bf x}',{\bf v}')} \right]}
{\eta_m({\bf x}',{\bf v}') [2\pi R_m({\bf x}')]^2}.
\nonumber
\end{eqnarray}
Although each individual term in the sum is peaked only in range-Doppler [near $R_m({\bf x}) = R_m({\bf x}_j)$ and $\eta({\bf x},{\bf v}) = \eta({\bf x}_j,{\bf v}_j - {\bf v}_0)$], the effect of the sum is to use the real-plus-synthetic antenna aperture $L_\mathrm{ap}$ to introduce a peak in cross-range as well. The angular width will be governed by the ratio $\lambda/L_\mathrm{ap}$ where $\lambda = f_0/c$ is the center wavelength. If the physical antenna is two-dimensional, with height $H_\mathrm{ap}$, the beam will have a vertical focus as well, governed by the ratio $\lambda/H_\mathrm{ap}$. If one restricts ${\bf x}$ to a ground plane area of interest, a plot of $\Sigma$ produces an area--Doppler image cube [which degenerates to a range--Doppler image in the case of a narrow aperture, $\lambda/L_\mathrm{ap} = O(1)$].

Similar to (\ref{2.9}), the background contribution to the compressed signal is given by
\begin{equation}
\Sigma_\mathrm{bg}({\bf x},{\bf v}) = \int_A d^2r
f({\bf r}) {\cal S}({\bf x},{\bf v};{\bf r},-{\bf v}_0)
\label{2.13}
\end{equation}
which represents a weighted average of the background scattering function over the support of the point spread function.

\subsection{Clutter covariance matrix and clutter cancelation}
\label{sec:cluttercov}

In the presence of the non-target terms in (\ref{2.1}), the compression operation (\ref{2.8}) will generate additional strong features in $\Sigma({\bf x},{\bf v})$. The stationary background $S_\mathrm{bg}$ will generate a `clutter ridge' at all ${\bf x}$, but localized in speed near the platform Doppler $v({\bf x}) = -{\bf \hat n}({\bf x}) \cdot {\bf v}_0$. The noise term will generally produce a background spread over all ${\bf x},{\bf v}$, though a jamming signal originating from a position ${\bf x}_J$ will produce a strong peak over all Doppler and all ranges, but localized near cross-range ${\bf \hat n}({\bf x}_J)$.

We assume the focus to be on a region of essentially fixed topography so that the background statistics are near-stationary. A jamming source signal $s_J(t)$ would similarly be assumed to be temporally random with stationary statistics. The STAP algorithm, to be described now, is then able to cancel out much of their effect and highlight the moving targets. In effect, one is able to narrow the background clutter ridge, and substantially null out a jamming signal. The cancelation obviously relies on first obtaining accurate estimates for the background statistics. In applications, this means prior analysis of radar data from nearby regions with the same statistics for $f({\bf r})$, but known \emph{not} to contain targets (see below). These statistical estimates are then applied to (assumed) target-containing regions, highlighting statistical outliers.

As usual, with an assumed Gaussian statistical model, the key quantity is the clutter covariance matrix generated by the compressed data,
\begin{equation}
G({\bf x},{\bf v};{\bf x}',{\bf v}')
= \langle \Sigma_\mathrm{bg}({\bf x},{\bf v})
\Sigma_\mathrm{bg}({\bf x}',{\bf v}')^* \rangle.
\label{2.14}
\end{equation}
Given $\langle f({\bf r}) \rangle = 0$, one defines the clutter covariance
\begin{equation}
g({\bf r},{\bf r}') = \langle f({\bf r}) f({\bf r}')^* \rangle,
\label{2.15}
\end{equation}
which will be assumed to vanish rapidly for $|{\bf r}-{\bf r}'| > \xi$, where $\xi$ is a characteristic correlation length. Within the signal model (\ref{2.7}), one obtains
\begin{eqnarray}
G({\bf x},{\bf v};{\bf x}',{\bf v}')
&=& \int d^2r \int d^2r' g({\bf r},{\bf r}')
\label{2.16} \\
&&\times\ {\cal S}({\bf x},{\bf v};{\bf r},-{\bf v}_0)
{\cal S}({\bf x}',{\bf v}';{\bf r}',-{\bf v}_0)^*,
\nonumber
\end{eqnarray}
in which, for simplicity, we specialize now to a clutter-only model, assuming negligible noise and no active jamming. This form is essentially a convolution of the point spread functions about ${\bf r}$ and ${\bf r}'$ with the clutter covariance. As such, if one restricts ${\bf x},{\bf x}'$ to the area $A$, and restricts attention as well to the radial components $v,v'$ of the velocities, $G$ should be near-diagonal and well conditioned.

\subsubsection{Covariance matrix estimation}
\label{subsec:covest}

If the clutter covariance $g$ is not known \emph{a priori}, the covariance matrix is estimated from the data itself. Specifically, for each ${\bf x},{\bf x}'$ in an area $A$ of interest (and all ${\bf v},{\bf v}'$), one computes
\begin{equation}
G_\mathrm{est}({\bf x},{\bf v};{\bf x}',{\bf v}')
= \sum_{k=1}^K w_k
\Sigma({\bf x}+{\bf X}_k,{\bf v})
\Sigma({\bf x}'+{\bf X}_k,{\bf v}')^*
\label{2.17}
\end{equation}
in which ${\bf X}_k$ is a sequence of nonoverlapping (target free) translations of the area $A$, and $w_k$ is a weight factor, which can be used, for example, to compensate for the different $1/R^2$ spherical spreading factors in different areas. One may also write this in the form
\begin{equation}
\hat G_\mathrm{est} = \hat W \hat W^\dagger
\label{2.18}
\end{equation}
in which $\hat W$ is the $N_D \times K$ matrix formed from the shifted area data:
\begin{equation}
W_{{\bf x},{\bf v};k} = \sqrt{w_k}
\Sigma({\bf x}+{\bf X}_k,{\bf v}).
\label{2.19}
\end{equation}
Note that $\hat W$ will not be sparse (since one expects returns from all areas of the scene). The near diagonal property of $\hat G_\mathrm{est}$ must follow from the destructive phase interference induced by the sum over $k$ in (\ref{2.17}).

\subsubsection{Constant false alarm rate (CFAR) detection}
\label{subsec:cfar}

Given that $G$ is known or estimated, one can now use it to enhance the target signal relative to the background. A common criterion is to maximize the target signal to clutter ratio. The optimal cancelation is then achieved by filtering the target region compressed signal $\Sigma({\bf x},{\bf v})$ through the inverse of the matrix $G$ \cite{Klemm2004,Melvin2004},
\begin{equation}
\Sigma_G({\bf x},v) = \int_A d^2x' \int dv'
[G^{-1}]({\bf x},v;{\bf x}',v') \Sigma({\bf x}',v').
\label{2.20}
\end{equation}
From a decision theory point of view, one now forms the detection statistic
\begin{equation}
h_G({\bf x},v)
= \frac{|\Sigma_G({\bf x},v)|^2}{\Sigma_{0,G}({\bf x},v)}
\label{2.21}
\end{equation}
in which the normalization is given by
\begin{equation}
\Sigma_{0,G}({\bf x},v) = [G^{-1}]({\bf x},v; {\bf x},v).
\label{2.22}
\end{equation}
Under the Gaussian assumption, for a given choice of threshold $h_0$, fixed by the permitted false alarm rate (probability that a presumed target is actually a background statistical outlier), deciding that a target is present in pixel $({\bf x},v)$ whenever $h_G({\bf x},v) > h_0$ provides the highest probability of detection \cite{Klemm2004,Melvin2004}. The scaling by $\Sigma_{0,G}$ accounts for possible slow variation in the background statistics, and ensures that the false alarm rate remains constant over the position--velocity region of interest (CFAR detection). For small $h_0$, targets will be almost certainly be detected, but many bright background features will be misidentified as targets as well. For large $h_0$ false alarms are suppressed at the expense of detecting only the brightest targets. The choice of $h_0$ depends on the mission (e.g., passive monitoring vs.\ active targeting).

Note that the independent treatment of each pixel entailed by (\ref{2.21}), including false alarm rate, is a critical consequence of the compression operation (\ref{2.8}). Target signatures that are not localized to a single pixel would require a more involved decision process.

Note also that the estimate (\ref{2.17}) generates a rank-deficient matrix which is therefore technically not invertible. In this case the correct generalization is to limit the inverse to the $K$-dimensional subspace ${\cal K}$ spanned by the data vectors (\ref{2.19}). Interestingly, it turns out that there is a very natural quantum procedure for this. The operation (\ref{2.20}) then projects $\Sigma$ into this subspace and produces a filtered signal $\Sigma_G$ lying in this subspace as well. The restricted inverse convention also serves to fully define the detection procedure.

\section{Quantum implementation summary}
\label{sec:qimpsummary}

\begin{figure*}
\includegraphics[width=6.0in,viewport=35 60 925 470,clip]{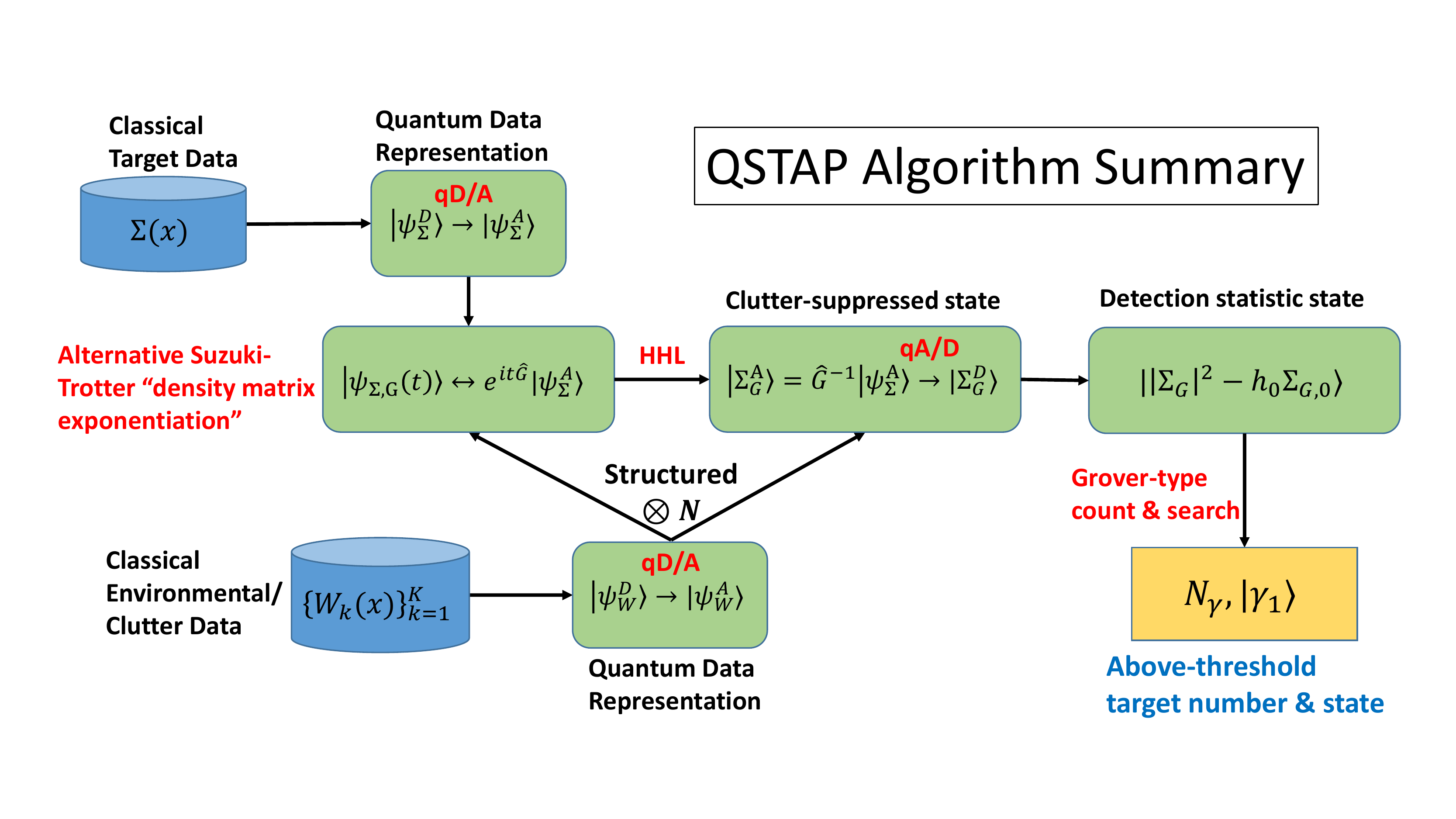}

\caption{\textbf{QSTAP machine learning algorithm flow chart:} The classical target and background compressed datasets $\Sigma(x)$ and $\{W_k(x)\}_{k=1}^K$, respectively, are loaded into quantum memory in standard `quantum digital' register format---see (\ref{3.1}) and (\ref{3.6}). Each must then be converted to quantum analog form (qD/A operation
described in App.\ \ref{app:qd2a}) in order to implement the unitary evolution (\ref{3.12}) via the density matrix exponentiation approach (Sec.\ \ref{sec:qsim}). The latter underlies the quantum phase estimation (Sec.\ \ref{sec:qphase}) and linear algebra (Sec.\ \ref{sec:qlinalg}) algorithms composing the HHL algorithm, and implementing the clutter suppression operation (\ref{2.20}). The `Structured $\otimes N$' label signifies that many copies of the background dataset are required [essentially one for each micro-time step in the evolution (Sec.\ \ref{sec:altST})], and that an additional binary hierarchical structure must be imposed on the space of these copies to maintain efficient density matrix phase estimation (see Sec.\ \ref{subsec:hloracle}). The analog form HHL output state $|\Sigma^{\cal A}_G \rangle \rangle$ must first be converted back to digital form (qA/D operation described in Sec.\ \ref{app:qa2d}) in order to construct the detection statistic (\ref{2.21}) in the form of a state $||\Sigma_G|^2 - h_0 \Sigma_{0,G} \rangle$. An extension of the Grover algorithm is then used to estmate the number of targets $N_\gamma$ and the quantum state $|\gamma_1 \rangle$ describing their locations (Secs.\ \ref{sec:qtgtdetect} and \ref{sec:groveroraclegen}).}

\label{fig:qstap}
\end{figure*}

This section provides an overview of the proposed quantum implementation of the STAP algorithm, summarized in Fig.\ \ref{fig:qstap}, with details relegated to later sections. For generality and clarity of presentation we now combine $x = ({\bf x},{\bf v})$ into a single index. With the formulation (\ref{2.20})--(\ref{2.22}) of the detection problem, the character of the underlying physical spaces play no further role. This notation conveniently also includes cases, alluded to above, in which the sensor geometry permits focus only within reduced dimensions, e.g., $x = {\bf x}$ for non-Doppler resolving pulses, and $x = (\hat {\bf n} \cdot {\bf x},\hat {\bf n} \cdot {\bf v})$ in absence of cross-range resolution.

\subsection{Data loading}
\label{sec:dataload}

The compressed data $\Sigma(x)$ from the putative target-containing region is initially stored in quantum memory in the standard `quantum digital' format
\begin{equation}
|\psi_\Sigma^{\cal D} \rangle = \frac{1}{\sqrt{N_D}}
\sum_x |\Sigma(x) \rangle |x \rangle,
\label{3.1}
\end{equation}
and only needs to be uploaded once. Here, using binary expansion $x = x_{p-1} x_{p-2} \ldots x_0$, the index register $|x \rangle = \otimes_{l = 1}^p |x_l \rangle$ is represented by $p = \lceil \log_2(N_D) \rceil$ qubits. Similarly, each corresponding compressed signal, suitably scaled here to integer levels, is represented by a binary expansion $\Sigma = \Sigma_{\sigma-1} \Sigma_{\sigma-2} \ldots \Sigma_0$ and the data register $|\Sigma \rangle = \otimes_{l=1}^\sigma |\Sigma_l \rangle$ is represented by $\sigma$ qubits \cite{foot:cmplx}. As alluded to earlier, preserving the number of degrees of freedom, one generally chooses $N_D \approx M N_t$, with pixel physical dimensions chosen to correspond roughly to the expected resolution based on the compression operation (\ref{2.8}). Based on the estimates at the beginning of Sec.\ \ref{sec:radarspbgnd}, one might have $N_D = O(10^{10})$, hence $p \simeq 34$ qubits. Using again $\sigma = 24$, one obtains for this example a total memory register of $\sim 58$ qubits.

Essentially all common quantum algorithms \cite{NC2000} are based on the format (\ref{3.1}), including quantum simulation and phase estimation (see below). It will transpire, however, that the reformulation of the phase estimation algorithm for low rank, nonsparse matrices requires as input instead the `quantum analog' format
\begin{eqnarray}
|\psi_\Sigma^{\cal A} \rangle &=& \frac{1}{{\cal N}_\Sigma}
\sum_x \Sigma(x) |x \rangle
\nonumber \\
{\cal N}_\Sigma^2 &=& \sum_x |\Sigma(x)|^2,
\label{3.2}
\end{eqnarray}
in which $\Sigma(x)$ now appears as a conventional wavefunction amplitude at coordinate $x$. The qD/A transformation
\begin{equation}
|\psi_\Sigma^{\cal D} \rangle \to |\psi_\Sigma^{\cal A} \rangle,
\label{3.3}
\end{equation}
that outputs (\ref{3.2}) from (\ref{3.1}) represents a kind of projection in data space (see App.\ \ref{app:qd2a}), compressing the $\sigma$-qubit register $|\Sigma(x) \rangle$ into the single complex amplitude $\Sigma(x)$. All information in the original state is preservable because only a very small subspace of the full register space---namely, the $2^\sigma$ possible discrete values associated with each simultaneous, normalized eigenstate of the qubit vertical spin operators $\{\hat Z_j \}_{j=1}^\sigma$. This operation must be considered as additional overhead on the data loading step. The inverse qA/D transformation will be encountered later. As described in App.\ \ref{app:qa2d}, the latter is significantly more involved since it requires reconstituting the larger state space from the complex number.

The requirement for the analog form is that wavefunction inner products
\begin{eqnarray}
&&\langle \psi_{\Sigma'}^{\cal A}|\psi_\Sigma^{\cal A} \rangle \
=\ \sum_x \psi_{\Sigma'}(x)^* \psi_\Sigma(x)
\label{3.4} \\
&&\ \ \ \ \ \ =\ \frac{1}{{\cal N}_\Sigma {\cal N}_{\Sigma'}}
\sum_x \left(\sum_{j=0}^{\sigma-1} 2^j \Sigma'_j(x)^* \right)
\left(\sum_{j=0}^{\sigma-1} 2^j \Sigma_j(x) \right)
\nonumber
\end{eqnarray}
follow directly from (\ref{3.2}) in the usual way (in a slight abuse of notation for complex registers \cite{foot:cmplx}). In contrast,
\begin{equation}
\langle \psi_{\Sigma'}^{\cal D}|\psi_\Sigma^{\cal D} \rangle
= \frac{L_{\Sigma \Sigma'}}{N_D}
\label{3.5}
\end{equation}
instead counts the number $L_{\Sigma \Sigma'}$ of pixels for which $\Sigma(x) = \Sigma'(x)$. Standard quantum algorithms are designed to account for this difference without the explicit conversion step (\ref{3.3}). In the present case, this turns out not to be possible. The disadvantage is that (\ref{3.2}) much more densely encodes the wavefunction in the scalar amplitude, rather than spreading it across a multi-qubit register. This likely impacts error analysis, but such considerations lie beyond the scope of the present work.

The STAP algorithm will be based on the estimated covariance (\ref{2.17}). As alluded to in Sec.\ \ref{sec:intro} the $K$ target-free, compressed data vectors (\ref{2.19}) need to be accessed by the algorithm multiple times. We assume therefore that the quantum memory provides access in the form of an oracle operation
\begin{eqnarray}
\hat U_W |k \rangle |0 \rangle
&=& |k \rangle |\psi_{W,k}^{\cal D} \rangle
\nonumber \\
|\psi_{W,k}^{\cal D} \rangle &=& \frac{1}{\sqrt{N_D}}
\sum_x |W_{x;k} \rangle |x \rangle
\nonumber \\
&\to& |\psi_{W,k}^{\cal A} \rangle.
\label{3.6}
\end{eqnarray}
in which the second register is identical in form to (\ref{3.1}), and the last line is the result of a subsequent qD/A conversion step. Given this data storage format, a call to $\hat U_W$ with input state
\begin{equation}
\hat H^{\otimes \kappa} |0 \rangle = \frac{1}{\sqrt{K}} \sum_{k=0}^{K-1} |k \rangle,
\label{3.7}
\end{equation}
where $\hat H^{\otimes \kappa} = \otimes_{j=1}^\kappa \hat H_j$ is a Hadamard gate product \cite{NC2000,foot:hadamard}, allows one to construct the $p+\kappa$ ($\kappa = \lceil \log_2(K) \rceil$) qubit state
\begin{equation}
|\psi_W \rangle = \sum_{k=1}^K {\cal N}_{W,k}
|k \rangle |\psi_{W,k}^{\cal A} \rangle.
\label{3.8}
\end{equation}
Here, for consistency, we assume that the additional overall normalization
\begin{equation}
\sum_{k=1}^K {\cal N}_{W,k}^2 = 1.
\label{3.9}
\end{equation}
has been applied to the data, amounting to simple rescaling of the detection statistic (\ref{2.21}) and threshold parameter $h_0$. The key property of this state is that it is equivalent to the matrix ${\bf W}$ in (\ref{2.18}), and the associated density matrix, obtained by averaging over the first register
\begin{eqnarray}
\hat \rho_G &=& \mathrm{tr}_1[|\psi_W \rangle \langle\psi_W|]
\nonumber \\
&=& \sum_{k=1}^K {\cal N}_{W,k}^2 |\psi_{W,k}^{\cal A} \rangle
\langle \psi_{W,k}^{\cal A}|
\label{3.10}
\end{eqnarray}
is precisely the quantum representation of $\hat G_\mathrm{est}$. Consistently, $\mathrm{tr}[\hat \rho_G] = 1$ follows from (\ref{3.9}). The corresponding density matrix formed from the digital states \cite{QPCA2013,QSVM2014} will be seen to generate incorrect measurement statistics.

\subsection{Matrix inversion via quantum simulation and phase estimation}
\label{sec:matinvqsim}

In order to compute the clutter-suppressed signal (\ref{2.20}) one is effectively solving the linear equation
\begin{equation}
\hat G_\mathrm{est} {\bm \Sigma}_G = \hat P_{\cal K} {\bm \Sigma}.
\label{3.11}
\end{equation}
in which $\hat P_{\cal K}$ is the orthogonal projection onto the subspace ${\cal K}$ spanned by the data vectors (\ref{2.19}). The HHL algorithm \cite{HHL2009} is designed to solve precisely this problem, including the subspace projection operation, so long as the unitary evolution
\begin{equation}
|\psi_{\Sigma,G}(t) \rangle = e^{i t \hat G_\mathrm{est}} |\psi_\Sigma \rangle
\label{3.12}
\end{equation}
can be efficiently simulated for a sufficiently large range of times $t$. Exponential speed-up estimates were originally based on sparse forms of $\hat G_\mathrm{est}$ \cite{HHL2009}, but the simulation algorithm was then extended \cite{QPCA2013,QSVM2014} to include non-sparse but low rank matrices of precisely the form (\ref{2.18}).

Sparse matrices (number of nonzero entries in any given row or column $\ll N_D$) are handled by first decomposing them into a sum of 1-sparse matrices (exactly one nonzero entry in each row and column), and then applying the Suzuki--Trotter decomposition \cite{HS2005} to the sum [in a way that implements the intrinsically analog form (\ref{3.12}) using the quantum digital format] \cite{A2004,BACS2006}.

In contrast, direct exponentiation of the form (\ref{2.18}) for non-sparse matrices is avoided via an alternative version of the Suzuki--Trotter decomposition whose efficiency relies instead on small $K/N_D \ll 1$, but requires the analog form of all the data states. Unusually, this formulation makes use of a second set of `environmental' qubits that are loaded via (\ref{3.6}), dynamically entangled with the set comprising the state (\ref{3.2}), and then continuously reinitialized and recycled during the course of the evolution (which, in effect, dissipates their state into the broader environment). This is done in such a way as to preserve quantum statistics (wavefunction inner products and operator averages with respect to the system density matrix), hence ensures correct measurement output from the quantum computer. But the method does raise some interesting quantum measurement questions \cite{W2019} that are discussed in Sec.\ \ref{sec:qdiss}.

Fourier analysis of the time series (\ref{3.12}), however derived, enables the phase estimation algorithm \cite{NC2000} which resolves $|\psi_\Sigma \rangle$ into a superposition of the eigenstates of $\hat G_\mathrm{est}$, with eigenvalue information supplied as well. From there one may construct a quantum state close to the one representing the vector ${\bm \Sigma}_G = \hat G_\mathrm{est}^{-1} \hat P_{\cal K} {\bm \Sigma}$ \cite{HHL2009}.

Computation of the detection statistic denominator (\ref{2.22}) proceeds similarly, but with initial states given by the set of individual register index values $|x \rangle$ (see Sec.\ \ref{sec:qtgtdetect}).

\subsection{Quantum search and quantum counting}
\label{sec:qsearchqcount}

Given the output state $|\Sigma_G \rangle$ one seeks, according to (\ref{2.21}), to identify pixels satisfying
\begin{equation}
|\Sigma_G(x)|^2 > h_0 \Sigma_{0,G}(x)
\label{3.13}
\end{equation}
for some threshold choice $h_0$. Defining the corresponding logical function
\begin{equation}
\gamma(x) = \left\{
\begin{array}{ll}
1, & |\Sigma_G(x)|^2 - h_0 \Sigma_{0,G}(x) > 0 \\
0, & \mbox{otherwise},
\end{array} \right.
\label{3.14}
\end{equation}
the core of the Grover search algorithm \cite{NC2000} is a black box (oracle) which applies the unitary transformation
\begin{equation}
\hat U_\gamma |x \rangle |q \rangle = |x \rangle |q \oplus \gamma(x) \rangle,
\label{3.15}
\end{equation}
thus flipping the single qubit $|q \rangle$ if (and only if) $\gamma = 1$ (a conditional $\hat X$ gate) \cite{foot:checkoracle}. In particular, choosing $|q_- \rangle = \hat H|1 \rangle = \frac{1}{\sqrt{2}}(|0 \rangle - |1 \rangle)$, one obtains $\hat U_\gamma |x\rangle |q_- \rangle = (-1)^{\gamma(x)} |x \rangle |q_- \rangle$. Since $|q_- \rangle$ is unchanged, we can drop it from the notation and adopt the shorthand convention
\begin{equation}
\hat U_\gamma |x \rangle = (-1)^{\gamma(x)} |x \rangle.
\label{3.16}
\end{equation}
This oracle obviously must take as input the state encoding the vector $|{\bm \Sigma}_G|^2 - h_0 {\bm \Sigma}_{0,G}$ (and multiple calls to $\hat U_\gamma$ require re-computation of this vector). Details of its construction are presented in Secs.\ \ref{sec:qtgtdetect} and \ref{sec:groveroraclegen}.

The key property is the identity
\begin{equation}
\hat U_\gamma \sum_x a(x) |x \rangle
= \sum_x a(x) \delta_{\gamma(x),1} |x \rangle
- \sum_x a(x) \delta_{\gamma(x),0} |x \rangle,
\label{3.17}
\end{equation}
for any amplitude $a(x)$, implying that the two subspaces $\gamma(x) = 0,1$ are invariant: $\hat U_\gamma$ preserves any superposition of solution pixels (eigenvalue $+1$), and reverses the sign of any superposition of non-solution pixels (eigenvalue $-1$). If one knows in advance the number of pixels $N_\gamma = N_\gamma(h_0)$ satisfying (\ref{3.13}), then this property allows one, with high probability, to construct accurate approximations to the uniform superpositions
\begin{eqnarray}
|\gamma_0 \rangle &=& \frac{1}{\sqrt{N_D - N_\gamma}}
\sum_x \delta_{\gamma(x),0} |x \rangle
\nonumber \\
|\gamma_1 \rangle &=& \frac{1}{\sqrt{N_\gamma}}
\sum_x \delta_{\gamma(x),1} |x \rangle,
\label{3.18}
\end{eqnarray}
also being opposite sign eigenvectors of $\hat U_\gamma$ \cite{NC2000}. Measurements on $|\gamma_1 \rangle$ allow one to extract information about the solution pixels (though repeated measurements require repeated calls to $\hat U_\gamma$).

If $N_\gamma$ is not known in advance, as will generally be the case for the radar target problem, then one must first estimate it to sufficient accuracy. This is accomplished by applying the phase estimation algorithm to the Grover operator
\begin{eqnarray}
\hat U_\mathrm{gr} &=& (2|\xi_0 \rangle \langle \xi_0| - \hat I_\Sigma) \hat U_\gamma,
\nonumber \\
|\xi_0 \rangle &\equiv& \hat H^{\otimes p} |0 \rangle
= \frac{1}{\sqrt{N_D}} \sum_x |x \rangle
\nonumber \\
&=& \sqrt{1 - \frac{N_\gamma}{N_D}} |\gamma_0 \rangle
+ \sqrt{\frac{N_\gamma}{N_D}} |\gamma_1 \rangle,
\label{3.19}
\end{eqnarray}
derived from $\hat U_\gamma$. Here $\hat I_\Sigma$ is the identity operator on the $|x \rangle$ register, and $|\xi_0 \rangle \langle \xi_0|$ projects onto the uniform superposition state \cite{foot:hadamard}. The corresponding difference operator $2|\xi_0 \rangle \langle \xi_0| - \hat I_\Sigma$ (which may be constructed from a series of controlled operations on each qubit plus an ancilla \cite{NC2000}) acts as the identity on $|\xi_0 \rangle$ and reverses the sign of any orthogonal state. One obtains
\begin{equation}
\hat U_\mathrm{gr} |\gamma_\pm \rangle
= e^{\pm i \theta_\gamma} |\gamma_\pm \rangle
\label{3.20}
\end{equation}
with eigenstates
\begin{equation}
|\gamma_\pm \rangle = \frac{1}{\sqrt{2}}
(|\gamma_0 \rangle \mp i |\gamma_1 \rangle),
\label{3.21}
\end{equation}
and eigenvalue $\theta_\gamma$ defined by
\begin{equation}
\cos(\theta_\gamma/2) = \sqrt{1 - \frac{N_\gamma}{N_D}},\ \
\sin(\theta_\gamma/2) = \sqrt{\frac{N_\gamma}{N_D}}.
\label{3.22}
\end{equation}
Phase estimation (which also requires repeated calls to $\hat U_\gamma$) produces a high accuracy approximation to $\theta_\gamma$, hence to $N_\gamma$. Using this value, one may derive the desired states (\ref{3.18}).

Further details of the these algorithms are presented in Secs.\ \ref{sec:qtgtdetect} and \ref{sec:groveroraclegen}.

\section{Quantum simulation of low rank matrices}
\label{sec:qsim}

Efficient simulation of the unitary evolution (\ref{3.12}) is not possible by standard methods involving Trotter decomposition of sparse matrices \cite{A2004,BACS2006}. The following trick, however, effectively replaces sparseness of $\hat G_\mathrm{est}$ with low rank $K$.

\subsection{Alternative Suzuki--Trotter decomposition}
\label{sec:altST}

To begin, let $\hat S$ be the (self adjoint) swap operator acting on the product ${\cal H} \otimes {\cal H}$ of two copies of some Hilbert space ${\cal H}$:
\begin{equation}
\hat S |\psi_1 \rangle |\psi_2 \rangle
= |\psi_2 \rangle |\psi_1 \rangle
\label{4.1}
\end{equation}
for any pair of states $|\psi_1\rangle, |\psi_2 \rangle \in {\cal H}$. For the present application ${\cal H}$ will represent the space of possible compressed data vectors (\ref{3.2}). Similarly, it follows that
\begin{equation}
\hat S \hat F \otimes \hat G \hat S = \hat G \otimes \hat F.
\label{4.2}
\end{equation}
for any pair of self-adjoint operators $\hat F$, $\hat G$ on ${\cal H}$.

Next consider the unitary evolution generated by $\hat S$:
\begin{eqnarray}
|\psi_S(t) \rangle &=& e^{it \hat S}
|\psi_1 \rangle |\psi_2 \rangle
\nonumber \\
&=& \cos(t) |\psi_1 \rangle |\psi_2 \rangle
+\ i \sin(t) |\psi_2 \rangle |\psi_1 \rangle,
\label{4.3}
\end{eqnarray}
and similarly for operators
\begin{eqnarray}
e^{i \hat S t} \hat G \otimes \hat F e^{-i \hat S t}
&=& \cos^2(t) \hat G \otimes \hat F + \sin^2(t) \hat F \otimes \hat G
\nonumber \\
&&+ \frac{i}{2} \sin(2t) \hat S
(\hat G \otimes \hat F - \hat F \otimes \hat G).
\nonumber \\
\label{4.4}
\end{eqnarray}
The key identity now emerges by averaging, via a trace operation, over the first subspace degrees of freedom:
\begin{eqnarray}
\hat F_S(t) &\equiv & {\cal L}_{t,G}[\hat F]
\nonumber \\
&\equiv& \mathrm{tr}_1\left[e^{it \hat S} \hat G
\otimes \hat F e^{-it \hat S} \right]
\nonumber \\
&=& \cos^2(t) \mathrm{tr}[\hat G] \hat F
+ \sin^2(t) \mathrm{tr}[\hat F] \hat G
\nonumber \\
&&- \frac{i}{2} \sin(2t) [\hat F,\hat G]
\nonumber \\
&=& \mathrm{tr}[\hat G] \hat F - it [\hat F,\hat G] + O(t^2),
\label{4.5}
\end{eqnarray}
which may be compared to the unitary evolution generated by $\hat G$:
\begin{eqnarray}
\hat F(t) &=& e^{it\hat G} \hat F e^{-it\hat G}
\nonumber \\
&=& \hat F - it [\hat F,\hat G] + O(t^2).
\label{4.6}
\end{eqnarray}
One sees that if one normalizes
\begin{equation}
\mathrm{tr}[\hat G] =  1,
\label{4.7}
\end{equation}
then (\ref{4.5}) and (\ref{4.6}) coincide. The two disagree beyond linear order, but (\ref{4.6}) and (\ref{4.7}) suffice to construct \cite{QPCA2013,QSVM2014}
\begin{equation}
\hat F(t) = \lim_{N \to \infty} ({\cal L}_{t/N,G})^N[\hat F],
\label{4.8}
\end{equation}
which may be viewed as an alternative Suzuki--Trotter formula \cite{HS2005}. At the expense of sequentially adjoining an extra copy of ${\cal H}$, and then averaging over it, one has reduced the evolution generated by $\hat G$ to that generated by the 1-sparse matrix $\hat S$.

\subsubsection{Two-time operators}
\label{subsec:2top}

For future reference, we consider as well a generalization to two-time operators
\begin{equation}
\hat F(t_1,t_2) = e^{it_1 \hat G} \hat F e^{-it_2 \hat G},
\label{4.9}
\end{equation}
from which one obtains
\begin{eqnarray}
\hat F(t_1+t,t_2) &=& (\hat I + it\hat G)\hat F(t_1,t_2) + O(t^2)
\nonumber \\
\hat F(t_1,t_2+t) &=& \hat F(t_1,t_2)(\hat I - it\hat G) + O(t^2).\ \ \ \ \ \
\label{4.10}
\end{eqnarray}
In comparison,
\begin{eqnarray}
{\cal L}^{(1)}_{t,G}[\hat F(t_1,t_2)] &\equiv &
\mathrm{tr}_1[e^{i\hat S t} \hat G \otimes \hat F(t_1,t_2)]
\nonumber \\
&=& \mathrm{tr}[\hat G] \hat F(t_1,t_2) + it \hat G \hat F(t_1,t_2)
+ O(t^2)
\nonumber \\
{\cal L}^{(2)}_{t,G}[\hat F(t_1,t_2)] &\equiv &
\mathrm{tr}_1[\hat G \otimes \hat F(t_1,t_2) e^{-i\hat S t}]
\label{4.11} \\
&=& \mathrm{tr}[\hat G] \hat F(t_1,t_2)
- it \hat F(t_1,t_2) \hat G + O(t^2),
\nonumber
\end{eqnarray}
which, to linear order in $t$, is identical to (\ref{4.10}) under the trace condition (\ref{4.7}). One may now iterate (\ref{4.11}) to obtain
\begin{equation}
\hat F(t_1,t_2) = \lim_{N_1,N_2 \to \infty}
({\cal L}_{t_1/N_1,G}^{(1)})^{N_1} ({\cal L}_{t_2/N_2,G}^{(2)})^{N_2}[\hat F].
\label{4.12}
\end{equation}
which is also equivalent to (\ref{4.8}) when $t_1 = t_2$. For large $N_1,N_2$ the order of operations all commute here.

In the application to follow (see especially Sec.\ \ref{subsec:factor}) we will encounter cases with a mixture of simultaneous and separate time evolutions. Thus, with the convention ${\cal L}^{(0)}_{t,G} \equiv {\cal L}_{t,G}$, by breaking up the time intervals $t_1,t_2$ into steps $\{\tau_j \}$ one may generalize both (\ref{4.8}) and (\ref{4.12}) in the form
\begin{equation}
\hat F(t_1,t_2) = \lim_{{\bf N} \to \infty}
\prod_j ({\cal L}^{(\nu_j)}_{\tau_j/N_j,G})^{N_j}[\hat F]
\label{4.13}
\end{equation}
in which $\nu_j \in \{0,1,2\}$, the limit notation indicates that all $N_j \to \infty$, and the only constraints on the segments $\tau_j$ are
\begin{equation}
t_\alpha = \sum_j (\delta_{\nu_j 0} + \delta_{\nu_j \alpha}) \tau_j,\ \
\alpha = 1,2.
\label{4.14}
\end{equation}

\subsubsection{Density matrix time reversal}
\label{subsec:treverse}

There is in fact no constraint on the signs of the $\tau_j$ (or of $t_1,t_2$) since (\ref{4.5}) and (\ref{4.11}) remain perfectly valid for $t < 0$. Since the same trace operation over the data space is applied in all cases, although this reversal of time operates as desired on the reduced density matrix it does \emph{not} correspond to true time reversal in the full Hilbert space. The latter would require effectively undoing the trace operations by sequentially de-computing the data registers, restoring them to their original $|\psi_W \rangle$ states. Such a prescription certainly becomes problematical for large $N_j$ (requiring a huge number of perfectly maintained error-free qubits), and fails for negative times for which there are no remaining data registers to de-compute.

The need for time reversal will be encountered as part of the HHL algorithm generalization (Sec.\ \ref{sec:qlinalg}) which requires reversal of the phase estimation algorithm. It will be shown that the `density matrix time reversal' operation indeed accomplishes the corresponding task here, maintaining the error-free qubit requirement at a reasonable level.

\subsection{Adaptation to low rank matrices}
\label{sec:lowrankadapt}

For general $\hat G$, equation (\ref{4.8}) does not necessarily lead to an algorithmic advantage since one still needs to generate the $N_D^2$ entries of $\hat G$. Moreover, in the present application $\hat F = |\psi_\Sigma \rangle \langle \psi_\Sigma|$ is the density matrix generated by the target region data, which is also not in general sparse. However, for low rank matrices of the form (\ref{2.18}), it will be seen that only the $N_D K$ entries of ${\bf W}$, along with the $N_D$ dimensional state $|\psi_\Sigma \rangle$, are needed, and that matrix evolution (\ref{4.6}) may be represented by an alternative `quantum statistical' version of the desired evolution (\ref{3.12}) which is still guaranteed to generate identical qubit measurement outcomes. The latter leads to the quantum analog format requirement. Henceforth, unless otherwise stated, this format is assumed and the explicit ${\cal A}$ label is dropped for notational simplicity.

Paralleling (\ref{4.6}) consider first the state evolution
\begin{eqnarray}
|\psi_{W,\Sigma}(t) \rangle
&=& e^{it\hat S} |\psi_W \rangle |\psi_\Sigma \rangle
\nonumber \\
&=& \sum_{k=1}^K {\cal N}_{W,k} |k\rangle
[\cos(t)|\psi_{W,k} \rangle |\psi_\Sigma \rangle
\nonumber \\
&&\ \ \ \ \ \
+\ i \sin(t) |\psi_\Sigma \rangle |\psi_{W,k} \rangle]
\label{4.15}
\end{eqnarray}
in which $\hat S$ swaps the last two registers (acting as the identity on the $|k\rangle$ register). Using (\ref{3.10}), the corresponding reduced density matrix is
\begin{eqnarray}
\hat F_{S,\Sigma}(t) &\equiv& \mathrm{tr}_{1,2}
[|\psi_{W,\Sigma}(t) \rangle \langle \psi_{W,\Sigma}(t)|]
\nonumber \\
&=& \mathrm{tr}_2\left[e^{it\hat S} \hat \rho_G
|\psi_\Sigma \rangle \langle \psi_\Sigma| e^{-i t\hat S} \right]
\nonumber \\
&=& |\psi_\Sigma \rangle \langle \psi_\Sigma|
\nonumber \\
&&-\ it \sum_{k=1}^K {\cal N}_{W,k}^2
[\langle \psi_\Sigma |\psi_{W,k} \rangle
|\psi_\Sigma \rangle \langle \psi_{W,k}|
\nonumber \\
&&\hskip0.6in -\ \langle \psi_{W,k} |\psi_\Sigma \rangle
|\psi_{W,k} \rangle \langle \psi_\Sigma|]
\nonumber \\
&&+\ O(t^2).
\label{4.16}
\end{eqnarray}
which instantiates (\ref{4.5}).

It is critical here that the inner products appearing in the $it$ term in (\ref{4.16}) take the standard form (\ref{3.4}), not the digital form (\ref{3.5}). Matrix exponentiation algorithms applied within standard quantum simulation and quantum phase estimation algorithms are specifically designed to compute such inner products indirectly from the digital forms of the states. The difference now is that (\ref{4.15}) and (\ref{4.16}) avoid explicit implementation of the matrix form of $\hat G$, implementing it implicitly only through measurement results. Thus, a chosen measurement operator $\hat M$ acting only on the last $p$ qubits, denoted now by the working subspace ${\cal H}_\Sigma$, yields the expectation value
\begin{eqnarray}
M(t) &=& \langle \psi_{W,\Sigma}(t)|
\hat I_W \otimes \hat M |\psi_{W,\Sigma}(t) \rangle
\nonumber \\
&=& \mathrm{tr}[\hat M \hat F_{S,\Sigma}(t)],
\label{4.17}
\end{eqnarray}
in which $\hat I_W$ is the identity operator acting on all other registers. The rules of quantum measurement dictate the form (\ref{3.4}) for the inner product which here directly operates on the qubit state. In principle, one could replace $\hat I_W \otimes \hat M$ by a more complex operator, acting on the full space and implementing the qD/A conversion after the fact, but there does not appear to be any advantage for this. Moreover, generalization to $N$ data vector states, implementing the Suzuki--Trotter evolution (\ref{4.8}), is certainly unfeasible, as will now be discussed.

Now let
\begin{equation}
|\Psi_W^{(N)} \rangle = |\psi_W \rangle_N \ldots |\psi_W \rangle_2 |\psi_W \rangle_1
\label{4.18}
\end{equation}
correspond to $N$ copies of the state (\ref{3.8}), and define
\begin{equation}
|\Psi^{(N)}(t) \rangle = \prod_{l=1}^N e^{i \frac{t}{N} \hat S_l}
|\Psi_W^{(N)} \rangle |\psi_\Sigma \rangle,
\label{4.19}
\end{equation}
in which $\hat S_l$ is the swap operator acting on $|\psi_\Sigma \rangle$ and the second register of $|\psi_W \rangle_l$, and the product is understood to order larger $l$ to the left. Here and below, upper case $\Psi$ (and later $\Phi$) will be used to distinguish such higher dimensional product states. The reduced density matrix
\begin{equation}
\hat F_\Sigma(t) = \lim_{N \to \infty}
\mathrm{tr}_{1,2}^N\left[|\Psi^{(N)}(t) \rangle
\langle \Psi^{(N)}(t)|\right],
\label{4.20}
\end{equation}
obtained by averaging over all of the extra $|\psi_W \rangle$ state degrees of freedom, reproduces (\ref{4.8}), and by construction coincides in the limit with
\begin{equation}
\hat F_\Sigma(t) = |\psi_{\Sigma,G}(t) \rangle \langle \psi_{\Sigma,G}(t)|.
\label{4.21}
\end{equation}
derived directly from (\ref{3.12}). Note that even though $\hat F_\Sigma(t)$ corresponds to a pure state, it is clear, e.g., from (\ref{4.16}) that it is nontrivially produced by the trace operation: the state $|\Psi^{(N)}(t) \rangle$ does \emph{not} approximate some direct product form $|\chi^{(N)}(t) \rangle |\psi_{\Sigma,G}(t) \rangle$, with the trace operation corresponding simply to dropping the $N(p + \kappa)$ qubit prefactor state $|\chi^{(N)}(t) \rangle$. Rather, (\ref{4.21}) emerges from a nontrivial average over the state of these qubits, with its pure state form being a carefully designed consequence of the alternative Suzuki--Trotter decomposition (\ref{4.8}).

The generalization of the measurement (\ref{4.17}), still operating only on the working subspace qubits, is
\begin{eqnarray}
M(t) &=& \mathrm{tr}[\hat M \hat F_\Sigma(t)]
\nonumber \\
&=& \lim_{N \to \infty} \langle \Psi^{(N)}(t)|
\hat I_W^{(N)} \otimes \hat M |\Psi^{(N)}(t) \rangle \ \ \ \ \ \
\label{4.22}
\end{eqnarray}
in which $\hat I_W^{(N)}$ is the identity operator on the $|\Psi_W^{(N)} \rangle$ subspace. It is emphasized again that the measurement (\ref{4.22}) exhibiting the density matrix (\ref{4.20}) requires the quantum analog form of the states.

\subsection{Qubit recycling and quantum dissipation}
\label{sec:qdiss}

At first sight, the formulation (\ref{4.20}) and (\ref{4.22}) appears untenable, requiring careful control of a diverging number $N(p + \kappa)+p$ of qubits. In fact, since the measurement is applied only on the $p$ dimensional working subspace ${\cal H}_\Sigma$, only an additional $p+\kappa$ qubits, acting as a fixed data subspace, to be denoted ${\cal H}_W$, are required that are then recycled $N$ times through the data loading and qD/A conversion steps (\ref{3.6}).

The key observation is that following each $e^{i\frac{t}{N}\hat S_l}$ operation in (\ref{4.19}), serving to entangle $|\psi_W \rangle_l$ with the last $p$ qubits of the state $|\Psi^{(N)}(t) \rangle$, none of the previous $(l-1)(p+\kappa)$ qubits are ever touched by the simulation algorithm again. In particular, although the state of the latter remains entangled with the other qubits, they may be viewed as physically isolated. Moreover, once isolated, no subsequent operation performed on them can have any effect on the state of the last $p$ qubits---in the sense that there can be no impact on the result of any measurement acting within ${\cal H}_\Sigma$.

This observation follows formally from the general unitary dynamics property. Let the Hilbert space be written as a direct product ${\cal H} = {\cal H}_\mathrm{ad} \otimes {\cal H}_\Sigma$ of the working space and all additional degrees of freedom, and let $|\Psi_0 \rangle$ be any state in ${\cal H}$. A unitary operator $\hat U_\mathrm{ad}$ acting only on ${\cal H}_\mathrm{ad}$ generates the state
\begin{equation}
|\Psi \rangle = \hat U_\mathrm{ad} \otimes \hat I_\Sigma |\Psi_0 \rangle,
\label{4.23}
\end{equation}
and a measurement operation acting only on ${\cal H}_\Sigma$ generates the result
\begin{eqnarray}
M &=& \langle \Psi|\hat I_\mathrm{ad} \otimes \hat M|\Psi \rangle
\nonumber \\
&=& \langle \Psi_0|(\hat U_\mathrm{ad}^\dagger \hat U_\mathrm{ad}) \otimes \hat M|\Psi_0 \rangle
\nonumber \\
&=& \langle \Psi_0|\hat I_\mathrm{ad} \otimes \hat M|\Psi_0 \rangle = M_0.
\label{4.24}
\end{eqnarray}
The measurement result is therefore preserved as claimed, and is in particular independent of the degree of entanglement between ${\cal H}_\mathrm{ad}$ and ${\cal H}_\Sigma$ present in $|\Psi_0 \rangle$.

For the present qubit recycling application, ${\cal H}_\mathrm{ad} = {\cal H}_E \otimes {\cal H}_W$ is the product of the $p + \kappa$ data qubit state space and that of all other `environmental' degrees of freedom (to be defined below). An alternative to the construction (\ref{4.19}), producing the identical density matrix (\ref{4.16}), is the following iterative procedure. The first step is identical to the first product in (\ref{4.16}), generating the state
\begin{equation}
|\Phi^{(1)} \rangle = e^{i\frac{t}{N}\hat S}
|\psi_W \rangle |\psi_\Sigma \rangle.
\label{4.25}
\end{equation}
One next prepares a new data state $|\psi_W \rangle_1 \in {\cal H}_{W,1}$, and applies a swap operation to load it into the data qubit space, generating the state
\begin{eqnarray}
|\Phi^{(1)\prime} \rangle &=&
\hat S_{W,1}|\psi_W \rangle_1 |\Phi^{(1)} \rangle
\nonumber \\
&\equiv& |\psi_W \rangle |\Phi^{(1)}_S \rangle,
\label{4.26}
\end{eqnarray}
in which $\hat S_{W,1}$ is the full swap operator acting on ${\cal H}_{W,1} \otimes {\cal H}_W$. The state $|\Phi^{(1)}_S \rangle \in {\cal H}_{W,1} \otimes {\cal H}_\Sigma$ is identical in structure to $|\Phi^{(1)} \rangle$, but now entangles ${\cal H}_\Sigma$ with ${\cal H}_{W,1}$ in place of ${\cal H}_W$. Finally, one applies the evolution operation to obtain
\begin{equation}
|\Phi^{(2)} \rangle = e^{i\frac{t}{N}\hat S} |\Phi^{(1)\prime} \rangle
\label{4.27}
\end{equation}
with $\hat S$ continuing to act on the last two registers in ${\cal H}_W \otimes {\cal H}_\Sigma$. Iterating this procedure, one obtains the sequence of states
\begin{equation}
|\Phi^{(l+1)} \rangle
= e^{i\frac{t}{N}\hat S} \hat S_{W,l}
|\psi_W \rangle_l |\Phi^{(l)} \rangle,\ \ l=1,2,\ldots,N-1,
\label{4.28}
\end{equation}
in which $|\Phi^{(l)} \rangle \in \otimes_{k=1}^{l-1} {\cal H}_{W,k} \otimes {\cal H}_W \otimes {\cal H}_\Sigma$ and $\hat S_{W,l}$ acts on ${\cal H}_{W,l} \otimes {\cal H}_W$. In each iteration there is an intermediate state $|\Phi^{(l)\prime} \rangle = |\psi_W \rangle |\Phi_S^{(l)} \rangle$ in which $|\Phi_S^{(l)} \rangle \in \otimes_{k=1}^l {\cal H}_{W,k} \otimes {\cal H}_\Sigma$ is identical in structure to $|\Phi^{(l)} \rangle$ but now entangling ${\cal H}_{W,l}$ in place of ${\cal H}_W$.

Noting the commutation identity $e^{i\frac{t}{N} \hat S} \hat S_{W,l} = \hat S_{W,l} e^{i\frac{t}{N} \hat S_l}$, one obtains the final state
\begin{eqnarray}
|\Phi^{(N)}(t) \rangle &=& \hat S_W^{(N)}
\left(\prod_{l=1}^N e^{i\frac{t}{N}\hat S_l} \right) |\Psi_W^{(N)} \rangle
|\psi_\Sigma \rangle
\nonumber \\
&=& \hat S_W^{(N)} |\Psi^{(N)}(t) \rangle,
\label{4.29}
\end{eqnarray}
in which one may identify the environment subspace ${\cal H}_E = \otimes_{l=1}^{N-1} {\cal H}_{W,l}$ and
\begin{equation}
\hat S_W^{(N)} = \hat S_{W,N-1} \hat S_{W,N-2} \ldots \hat S_{W,1}
\label{4.30}
\end{equation}
is a unitary operator that performs the full sequence of swaps with the data qubit space ${\cal H}_W$. Since $\hat S_W^{(N)}$ acts as the identity on ${\cal H}_\Sigma$, preservation of the reduced density matrix follows immediately:
\begin{eqnarray}
&&\lim_{N \to \infty} \mathrm{tr}_{1,2}^N
\left[|\Phi^{(N)}(t) \rangle \langle \Phi^{(N)}(t)| \right]
\nonumber \\
&&\ \ \ \ =\ \lim_{N \to \infty} \mathrm{tr}_{1,2}^N
\left[\hat S_W^{(N) \dagger} \hat S_W^{(N)}
|\Psi^{(N)}(t) \rangle \langle \Psi^{(N)}(t)| \right]
\nonumber \\
&&\ \ \ \ =\ \hat F_\Sigma(t),
\label{4.31}
\end{eqnarray}
in which the cyclic property of the trace has been used for operators restricted to ${\cal H}_\mathrm{ad}$.

The key difference with the state (\ref{4.19}) is that $|\psi_W \rangle$ is imprinted on the same set of working qubits on each iteration. At first sight nothing appears to be gained, since a new set of data qubits continues to be introduced at each iteration. However, the technical simplification is that the expanding environmental state does not actually have to be maintained once the $\hat S_{W,l}$ operation is performed---the information may be permitted to dissipate away into the broader environment. Only the original data and working qubits need to be carefully controlled.

One may confirm this formally as follows. Let $|\Psi_{E,0} \rangle |\Psi_W^{(N)} \rangle \in {\cal H}_E$ be an initial environmental state, expanded from (\ref{4.18}) to include the state $|\Psi_{E,0} \rangle$ of all other degrees of freedom in the apparatus. In place of (\ref{4.29}), let
\begin{equation}
|\Phi_D^{(N)}(t) \rangle = \prod_{l=1}^N \hat U_{D,l}
\hat S_{W,l} e^{i\frac{t}{N}\hat S_l}
|\Psi_{E,0} \rangle |\Psi_W^{(N)} \rangle
|\psi_\Sigma \rangle
\label{4.32}
\end{equation}
in which a sequence of unitary dissipation operators ${\bf U}_{D,l}$ have been introduced whose only constraint is that they act as the identity on $\otimes_{k=l+1}^{N-1} {\cal H}_{W,k} \otimes {\cal H}_W \otimes {\cal H}_\Sigma$ (i.e., they do not touch any later-processed data vectors). In particular, $\hat U_{D,l}$ commutes with all $\hat S_k$ with $k \geq l+1$, and one obtains
\begin{equation}
|\Phi_D^{(N)}(t) \rangle = \hat U_D^{(N)}
|\Psi_{E,0} \rangle |\Psi^{(N)}(t) \rangle
\label{4.33}
\end{equation}
in which $\hat S_W^{(N)}$ in (\ref{4.29}) is replaced by the more general unitary operator
\begin{equation}
\hat U_D^{(N)} = \prod_{l=1}^N \hat U_{D,l} \hat S_{W,l}
\label{4.34}
\end{equation}
still acting as the identity on ${\cal H}_\Sigma$. The density matrix, now including a trace operation over all of the environmental states
\begin{widetext}
\begin{eqnarray}
\lim_{N \to \infty} \mathrm{tr}_{E;1,2}
\left[|\Phi_D^{(N)}(t) \rangle \langle \Phi_D^{(N)}(t)| \right]
&=& \lim_{N \to \infty} \mathrm{tr}_{E;1,2}
\left[\hat U_D^{(N)\dagger} \hat U_D^{(N)}
|\Psi_{E,0} \rangle \langle \Psi_{E,0}|
\otimes |\Psi^{(N)}(t) \rangle \langle \Psi^{(N)}(t)| \right]
\nonumber \\
&=& \langle \Psi_{E,0}|\Psi_{E,0} \rangle
\lim_{N \to \infty}
\mathrm{tr}_{1,2}^N\left[|\Psi^{(N)}(t) \rangle
\langle \Psi^{(N)}(t)| \right]
\nonumber \\
&=& \hat F_\Sigma(t),
\label{4.35}
\end{eqnarray}
\end{widetext}
continues to be preserved. Physically, this means that, once the new data vector is swapped in, one need only maintain error correction on the space ${\cal H}_W \otimes {\cal H}_\Sigma$ during the course of the computation. The transfer of information to the environment, a form of quantum dissipation, though unitary as required by many body quantum dynamics, is effectively unrecoverable. However, this has no impact on the desired result of the measurement operation (\ref{4.24}). This includes, for example, cases where classical measurement results are derived from the environmental qubits---such peripheral measurement outcomes have no impact on working space measurement outcomes \cite{W2019}.

\section{Generalized quantum phase estimation}
\label{sec:qphase}

The previous section showed how to construct a working $p$ qubit state, strongly entangled with a very high dimensional environment, measurements of which allow one to probe the desired state (\ref{3.12}) via its density matrix---see (\ref{4.17}). We now turn to adaptation of this construction to the HHL algorithm which is used to diagonalize and then invert the STAP covariance matrix. The matrix diagonalization step is accomplished using the quantum phase estimation algorithm whose generalization to the density matrix implementation is now described. It will be shown that the evolution operations (\ref{4.8}) and (\ref{4.12}) need to be organized in a very specific hierarchical fashion in order to maintain its computational efficiency.

\subsection{Conventional phase estimation}
\label{sec:qphaseconv}

Conventional phase estimation refers to a quantum algorithm that effectively diagonalizes a given unitary operator $\hat U$ \cite{NC2000}. Thus, an eigenstate $|\psi\rangle$ of $\hat U$ obeys
\begin{equation}
\hat U |\psi \rangle = e^{2\pi i \varphi} |\psi\rangle
\label{5.1}
\end{equation}
defining a (normalized) eigenphase $0 \leq \varphi < 1$. An algorithm outputting (an estimate of) $\varphi$ is constructed as follows. Each major step is called out in order to highlight the corresponding step required in a generalized algorithm for the present environment-entangled application.

\subsubsection{Lowest level oracle producing powers of $\hat U$}
\label{subsec:lloracleUpowers}

The algorithm relies on an oracle operator $\hat O_U$ which accesses $\hat U$ to produce the operation
\begin{equation}
\hat O_U (c_0|0 \rangle + c_1 |1\rangle)|J\rangle |\psi \rangle
= c_0 |0 \rangle |J \rangle |\psi \rangle
+ c_1 |1 \rangle |J \rangle \hat U^J |\psi \rangle,
\label{5.2}
\end{equation}
controlled by the first qubit state. Here $c_0,c_1$ are arbitrary amplitudes, and $|J \rangle$, $0 \leq J \leq M-1$ is a $m$ qubit binary register with $M = 2^m$. In particular
\begin{equation}
\hat O_U \hat H |0 \rangle |2^{j-1} \rangle |\psi \rangle
= \frac{|0 \rangle + e^{2\pi i 2^{j-1} \varphi} |1 \rangle}{\sqrt{2}}
|2^{j-1} \rangle |\psi \rangle.
\label{5.3}
\end{equation}

\subsubsection{Higher level oracle and binary power product state}
\label{subsec:hloraclebinary}

Defining, respectively, the $m$ and $m^2$ qubit registers
\begin{equation}
|0^{\otimes m} \rangle = \otimes_{j=1}^m |0\rangle_j,\ \
|2^{\otimes m} \rangle = \otimes_{j=1}^m |2^{j-1} \rangle_j
\label{5.4}
\end{equation}
and the product oracle operator
\begin{equation}
\hat {\bf O}_U = \left(\otimes_{j=1}^m \hat O_U^{(j)} \right)
\hat H^{\otimes m},
\label{5.5}
\end{equation}
whose factors act on the corresponding factors in (\ref{5.4}) (but on the same state $|\psi \rangle$), one obtains
\begin{equation}
\hat {\bf O}_U |0^{\otimes m} \rangle |2^{\otimes m} \rangle |\psi \rangle
= |\varphi_F \rangle |2^{\otimes m} \rangle |\psi \rangle
\label{5.6}
\end{equation}
in which
\begin{equation}
|\varphi_F \rangle = \otimes_{j=1}^m
\frac{|0\rangle_j + e^{2\pi i 2^{j-1} \varphi} |1\rangle_j}{\sqrt{2}}
= \frac{1}{\sqrt{M}} \sum_{J=0}^{M-1} e^{2\pi i J \varphi} |J \rangle.
\label{5.7}
\end{equation}
is a Fourier series.

\subsubsection{Fourier transform to eigenphase basis}
\label{subsec:fteigenbasis}

If $\varphi = M_\varphi/M = 0.\varphi_1 \varphi_2 \ldots \varphi_m$ is an exact $m$ digit binary fraction (with $M_\varphi = \sum_{j=1}^m \varphi_j 2^{m-j}$), then $|\varphi_F \rangle$ is precisely the quantum Fourier transform of the state $|\varphi \rangle = \otimes_{j=1}^m |\varphi_j \rangle$. More generally, the inverse transform
\begin{eqnarray}
|\tilde \varphi \rangle = \hat U_F^\dagger |\varphi_F \rangle
&=& \sum_{Q=0}^{M-1} \Delta_M(\varphi - Q/M) |Q \rangle
\nonumber \\
\Delta_M(s) &\equiv& \frac{1}{M} \frac{e^{2\pi i Ms} - 1}{e^{2\pi i s} - 1}
\label{5.8}
\end{eqnarray}
produces $|\tilde \varphi \rangle = |\varphi \rangle$ if $\varphi$ is a precise binary fraction, but is otherwise a superposition of states that is strongly peaked about the nearest $m$-digit binary approximation \cite{foot:Deltafn}. Analysis of the measurement statistics on this state produces precise error estimates \cite{NC2000}.

For general eigenstate superposition input state
\begin{equation}
|\psi \rangle = \sum_u A_u |\psi_u \rangle,
\label{5.9}
\end{equation}
with associated eigenvalues $\varphi_u$, one obtains
\begin{eqnarray}
\hat U_F^\dagger \hat {\bf O}_U |0^{\otimes m} \rangle
|2^{\otimes m} \rangle |\psi \rangle
= |2^{\otimes m} \rangle |{\bm \psi}_\varphi \rangle &&
\nonumber \\
|{\bm \psi}_\varphi \rangle = \sum_u A_u
|\tilde \varphi_u \rangle |\psi_u \rangle &&
\label{5.10}
\end{eqnarray}
in which each $|\tilde \varphi_u \rangle$ is strongly peaked about integer register values closest to $M \varphi_u$, and in a slight abuse of notation the states have been reordered so that $|2^{\otimes m} \rangle$ can be factored out.

\subsubsection{Full phase estimation operator}
\label{subsec:phestop}

Since the state $|2^{\otimes m} \rangle$ is unchanged it may be dropped from both sides to simplify the notation. With this understanding, the phase estimation algorithm, represented now by a unitary operator $\hat {\bf O}_\varphi$, produces the action
\begin{equation}
|\psi_\varphi \rangle \equiv \hat {\bf O}_\varphi |0^{\otimes m} \rangle |\psi \rangle
= \sum_u \langle \psi_u|\psi \rangle |\tilde \varphi_u \rangle |\psi_u \rangle,
\label{5.11}
\end{equation}
with state $|2^{\otimes m} \rangle$ now understood as an internal set of preset control qubits \cite{foot:phaseest}.

\subsection{Generalized phase estimation}
\label{sec:qphasegen}

The phase estimation algorithm would proceed entirely conventionally if one had access to $\hat U_\mathrm{est}(t) = e^{it \hat G_\mathrm{est}}$ [see (\ref{3.12})]. In particular, if one chooses $\hat U = \hat U_\mathrm{est}(\epsilon)$ for some sufficiently small time $\epsilon$, one obtains
\begin{eqnarray}
|\psi_{G,\Sigma}^\varphi \rangle
&\equiv& \hat {\bf O}_\varphi |0^{\otimes m} \rangle |\psi_\Sigma \rangle
\nonumber \\
&=& \sum_u \langle \psi^G_u|\psi_\Sigma \rangle
|\tilde \lambda_u \rangle |\psi^G_u \rangle
\label{5.12}
\end{eqnarray}
in which $2\pi \lambda_u,|\psi^G_u \rangle$ are the eigenvalues and eigenstates of $\hat G_\mathrm{est}$, and to simplify the notation $|\tilde \lambda_u \rangle$ is the state formed in estimating the phase $\varphi_u = \epsilon \lambda_u$. The eigenvalues are now approximated via the narrowly peaked superposition states $|\tilde \lambda_u \rangle$. Of course only $K$ of these eigenvalues should be nonzero.

Lacking an efficient quantum algorithm for constructing $\hat U_\mathrm{est}(t)$, we now describe the requirements for an alternative construction using the reduced density matrix formulation described in Sec.\ \ref{sec:qsim}. It will be seen that there are several important generalizations required that were not anticipated in the literature \cite{QPCA2013,QSVM2014,QML2017}.

\subsubsection{Generalized lowest level oracle}
\label{subsec:lloracle}

The objective is to construct the analogue of (\ref{5.11}) using the reduced density matrix construction. To this end, the basic evolution operator is defined by
\begin{equation}
|\Psi(t) \rangle = \hat U(t) |\Psi(0) \rangle
\label{5.13}
\end{equation}
in which the `pure' state $|\Psi(t) \rangle$ is given by (\ref{4.19}), with the superscript dropped for notational simplicity---some sufficiently large value of $N$ is now implicit in the notation. Later we will include qubit recycling and dissipation [with associated states (\ref{4.29}) and (\ref{4.33}), respectively]. Recall here that $|\Psi(0) \rangle = |\Psi_W \rangle |\psi_\Sigma \rangle$ is an initial product state with $|\Psi_W \rangle$ combining all $N$ required copies of the target-free data. Below we will see that phase estimation requires imposition of additional structure on this state.

We arrive now at the first critical difference with the conventional algorithm: it is clear that $\hat U(\epsilon J) \neq \hat U(\epsilon)^J$ because the two act on entirely different data subspaces. It follows that the eigenvectors of $\hat U(t)$ vary with time, and its eigenvalues $e^{2\pi i\varphi(t)}$ are not in general linear in $t$: $e^{2\pi i\varphi(\epsilon J)} \neq e^{2\pi iJ \varphi(\epsilon)}$. However this is not necessarily a significant concern since our interest is only in phase estimation at the level of the reduced density matrix. We proceed therefore by defining, in place of (\ref{5.2}),
\begin{eqnarray}
&&\hat O_U(\epsilon) (c_0|0\rangle + c_1|1 \rangle)
|J \rangle |\Psi(0) \rangle
\nonumber \\
&&\ \ \ \ \ \ =\ [c_0 |0 \rangle
+ c_1 |1 \rangle \hat U(J \epsilon)]
|J \rangle |\Psi(0) \rangle
\nonumber \\
&&\ \ \ \ \ \ =\ c_0 |0 \rangle |J \rangle |\Psi(0) \rangle
+ c_1 |1 \rangle |J \rangle |\Psi(\epsilon J) \rangle,\ \ \ \ \ \ \ \ \ \
\label{5.14}
\end{eqnarray}
which implements the time evolution, conditioned on the first qubit, without any further assumptions on the product structure (or lack thereof) of $\hat U(t)$.

\begin{figure*}

\includegraphics[width=5.5in,viewport=40 70 920 480,clip]{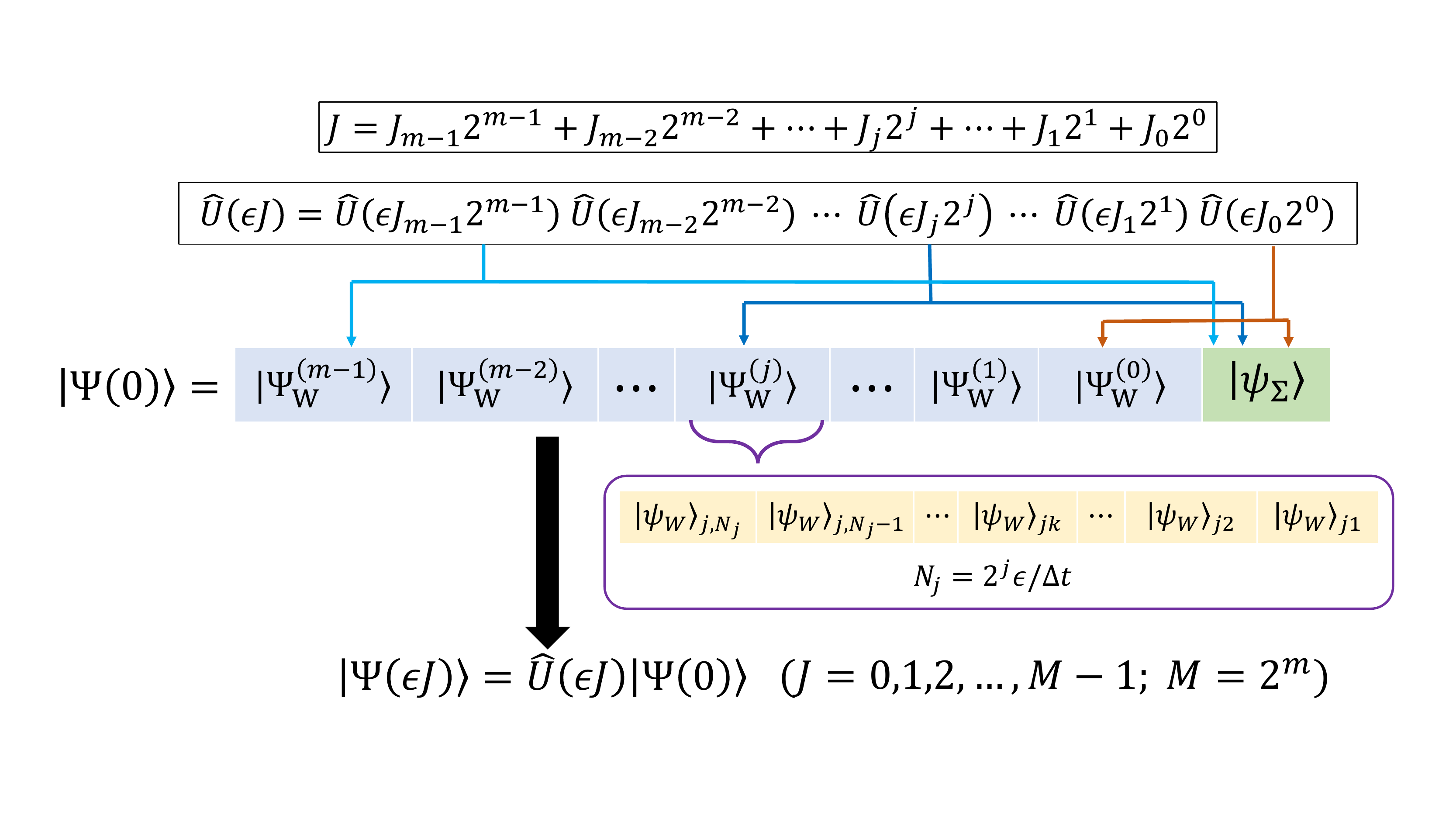}

\caption{Generalized quantum phase estimation quantum data structure implementing (\ref{5.16})--(\ref{5.18}), maintaining the quantum advantage based on the binary decomposition (top line) of the sequence of simulation times $t_J = \epsilon J$. For each $J$ the evolution operator $\hat U(t_J)$ is decomposed into a product of binary time interval operators (second line; here $\hat U(0)$ is the identity operator whenever $J_j = 0$). As indicated by the arrows, each $\hat U(\epsilon J_j 2^j)$ entangles a particular input environmental data state $|\Psi_W{(j)} \rangle$ (third line) with the evolving working subspace ${\cal H}_\Sigma$. The latter is initialized with the imaging data state $|\psi_\Sigma \rangle$ [defined by (\ref{3.2})]. As indicated by the magenta box, each $|\Psi_W{(j)} \rangle$ is in turn a product of $N_j = 2^j \epsilon/\Delta t$ copies of the underlying environmental data states $|\psi_W \rangle$ [defined by (\ref{3.8})] required to implement the generalized Suzuki--Trotter evolution operation (\ref{4.19}) with time step $\Delta t$. As discussed in the text, this hierarchical data structure is required to ensure proper parallel construction of the entangled state (\ref{5.15})---a uniform superposition of every evolved state $|\Psi(t_J) \rangle$ (bottom line)---with the same data state $|\Psi_W{(j)} \rangle$ applied consistently for every $J$ for which $J_j = 1$. The evolution operators may also include qubit recycling and dissipation (see Sec.\ \ref{sec:qdiss}) without affecting this underlying structure.}

\label{fig:datastruct}
\end{figure*}

\subsubsection{Generalized higher level oracle}
\label{subsec:hloracle}

We next address the generalization of the product operator (\ref{5.5}). We will see that in order to properly implement the exponential speedup encoded in the $|2^{\otimes m} \rangle$ state there is further hierarchical structure underlying the high-dimensional state (\ref{5.13}) that must be imposed, or the desired algorithm will fail. Specifically, the composite oracle (\ref{5.5}) has action
\begin{eqnarray}
|\Psi_F \rangle &\equiv&
\hat {\bf O}_U(\epsilon) |0^{\otimes m} \rangle |\Psi(0) \rangle
\nonumber \\
&=& \otimes_{j=1}^m \frac{1}{\sqrt{2}}
\left[|0 \rangle_j + |1 \rangle_j \hat U(\epsilon 2^{j-1}) \right]
|\Psi(0) \rangle
\nonumber \\
&=& \frac{1}{\sqrt{M}} \sum_{J=0}^{M-1}
|J \rangle \hat U(\epsilon J) |\Psi(0) \rangle
\nonumber \\
&=& \frac{1}{\sqrt{M}} \sum_{J=0}^{M-1}
|J \rangle |\Psi(J\epsilon) \rangle
\label{5.15}
\end{eqnarray}
in which, for simplicity, the $|2^{\otimes m} \rangle$ register has again been dropped from the notation. The subscript $F$ is intended to highlight the parallel to the conventional state (\ref{5.7}). In order to obtain the last two lines one must identify
\begin{equation}
\hat U(\epsilon J) = {\prod_{j=1}^m}' \hat U(\epsilon 2^{j-1}),
\label{5.16}
\end{equation}
in which the prime on the product (ordered, by convention, with larger $j$ to the left) indicates that the only $j$ appearing are those for which $J_{j-1} = 1$ in the binary expansion $J = J_{m-1} J_{m-2} \ldots J_0$.

The product decomposition (\ref{5.15}) at first sight appears inconsistent with the discussion above (\ref{5.14}). However, it is in fact valid if one organizes the data qubits in a particular way: for the given choice of register length $m$ let the generalized phase estimation data qubit Hilbert space be decomposed in the form
\begin{equation}
{\cal H}^{(\mathrm{GPE})}_W = {\cal H}_W^{(m-1)} \otimes {\cal H}_W^{(m-2)}
\otimes \ldots \otimes {\cal H}_W^{(0)},
\label{5.17}
\end{equation}
with correspondingly ordered quantum data structure
\begin{eqnarray}
|\Psi_W^{(\mathrm{GPE})} \rangle 
&=& |\Psi_W^{(m-1)} \rangle |\Psi_W^{(m-2)} \rangle
\ldots |\Psi_W^{(0)} \rangle
\nonumber \\
|\Psi_W^{(j)} \rangle &=& \otimes_{k=1}^{2^j N_\epsilon} |\psi_W \rangle_{jk},
\label{5.18}
\end{eqnarray}
with each $|\Psi_W^{(j)} \rangle \in {\cal H}_W^{(j)}$. Here $N_\epsilon = \epsilon/\Delta t$ where $\Delta t$ is the underlying Suzuki--Trotter time step, and ${\cal H}^{(j)}$ is therefore in turn a direct product of $2^j N_\epsilon$ individual data subspaces [containing each individual data state $|\psi_W \rangle_{jk}$---see (\ref{3.8})]. With this construction, the factor $\hat U(\epsilon 2^{j-1})$ in (\ref{5.15}), whenever present, acts on the subspace ${\cal H}_W^{(j-1)} \otimes {\cal H}_\Sigma$ to evolve the state according to (\ref{4.19}). In this way, the states $|\Psi(J\epsilon) \rangle$ are constructed in a consistent fashion, critically avoiding different operators in the product (\ref{5.15}) mistakenly reprocessing the same (now entangled) data vector.

It is emphasized again that this construction, summarized in Fig.\ \ref{fig:datastruct}, clearly has deep hardware implications in terms of organization of data loading, is critical to maintaining the binary product structure, and below we will see that it is critical as well to implementation of phase estimation for the reduced density matrix.

\subsubsection{Reduced density matrix factorization property}
\label{subsec:factor}

We next establish the following key reduced subspace factorization property:
\begin{eqnarray}
\hat F_\Sigma(J,J') &\equiv& \mathrm{tr}_W\left[|\Psi(J\epsilon) \rangle
\langle \Psi(J'\epsilon)| \right]
\nonumber \\
&=& |\psi_{\Sigma,G}(J\epsilon) \rangle \langle \psi_{\Sigma,G}(J'\epsilon)|,
\label{5.19}
\end{eqnarray}
with reduced space evolved states (\ref{3.12}) and the trace acting on the full ${\cal H}_W$ data subspace, generalizing the obvious equality when $J = J'$. The proof is obtained by comparing the binary expansions of $J$, $J'$ and using the definitions (\ref{4.5}) and (\ref{4.11}) of the two-sided and one-sided evolution operators, respectively. Defining again ${\cal L}_{t,G}^{(0)} = {\cal L}_{t,G}$, one obtains [compare (\ref{4.13})]
\begin{eqnarray}
\hat F_\Sigma(J,J') &=& \prod_{j=1}^m
({\cal L}_{\Delta t,G}^{(\nu_j)})^{2^j N_\epsilon}
[|\psi_\Sigma \rangle \langle \psi_\Sigma|]
\nonumber \\
&=& |\psi_{\Sigma,G}(J\epsilon) \rangle \langle \psi_{\Sigma,G}(J'\epsilon)|
\label{5.20}
\end{eqnarray}
in which larger $j$ are again to the left, and we define
\begin{equation}
\nu_j = \left\{\begin{array}{ll}
0, & J_j = J'_j = 1 \\
1, & J_j = 1,\ J'_j = 0 \\
2, & J_j = 0,\ J'_j = 1,
\end{array} \right.
\label{5.21}
\end{equation}
and, of course, for $J_j = J'_j = 0$ there is no operation. Thus, the hierarchical organization of the data space allows one to advance time sequentially in increasing binary steps. When the binary digits match one applies (\ref{4.5}), and when they fail to match one applies one or the other of (\ref{4.11}). In both cases the factorization property is preserved, and (\ref{5.20}) yields the second line of (\ref{5.19}).

\subsubsection{Qubit recycling and dissipation}
\label{subsec:qrecycdiss}

We next observe that the factorization property is preserved by global qubit recycling and dissipation operations. Thus, comparing (\ref{4.29}) and (\ref{4.30}), let
\begin{equation}
|\Phi(t) \rangle = \hat S_W |\Psi(t) \rangle
\label{5.22}
\end{equation}
correspond to the state in which the necessary number of swap operations (operating only within the data space ${\cal H}_W$) has been applied to implement the qubit recycling operation described in Sec.\ \ref{sec:qdiss}. It then follows that
\begin{eqnarray}
&&\mathrm{tr}_W[|\Phi(J\epsilon) \rangle
\langle \Phi(J'\epsilon)|]
\nonumber \\
&&\ \ \ \ \ \ =\ \mathrm{tr}_W\left[\hat S_W^\dagger \hat S_W
|\Phi(J\epsilon) \rangle \langle \Phi(J'\epsilon)|\right]
\nonumber \\
&&\ \ \ \ \ \ =\ \hat F_\Sigma(J,J'),
\label{5.23}
\end{eqnarray}
in which the cyclic property of the trace (valid, in this case, for operators restricted to ${\cal H}_W$) and the unitary property of the swap operators have been used. Note that it is critical here that the \emph{same} swap operator be applied on both left and right---hence that $\hat S_W$ in (\ref{5.22}) be independent of $t$, applied to all data subspaces in ${\cal H}_W$, not just to those that have been processed up until any particular time.

Similarly, comparing (\ref{4.33}) and (\ref{4.34}), let
\begin{equation}
|\Phi_D(t) \rangle = \hat U_D
|\Psi_{E,0} \rangle |\Psi(t) \rangle
\label{5.24}
\end{equation}
correspond to the state combining the necessary number of swap and dissipation operations, acting on the space ${\cal H}_W \otimes {\cal H}_E$, that now includes the broader environment. The cyclic property of the trace again eliminates $\hat U_D$, and one obtains
\begin{eqnarray}
&&\mathrm{tr}_{E,W}[|\Phi_D(J\epsilon) \rangle
\langle \Phi_D(J'\epsilon)|]
\nonumber \\
&&\ \ \ \ \ =\ \mathrm{tr}_W[|\Psi(J\epsilon) \rangle
\langle \Psi(J'\epsilon)|]
\mathrm{tr}_E[|\Psi_{E,0} \rangle \langle \Psi_{E,0}|]
\nonumber \\
&&\ \ \ \ \ =\ \hat F_\Sigma(J,J')
\label{5.25}
\end{eqnarray}
It is again critical that the same swap--dissipation operator be applied on both left and right, hence covering the entire time evolution range, not limited by particular values of $J,J'$.

\subsubsection{Fourier transform basis}
\label{subsec:ftbasis}

The Fourier transform operation on the $|J \rangle$ register in (\ref{5.15}) now proceeds exactly as in Sec.\ \ref{subsec:fteigenbasis}, generating the state
\begin{eqnarray}
|\Psi_\varphi \rangle &=& \hat U_F^\dagger |\Psi_F \rangle
\nonumber \\
&=& \frac{1}{\sqrt{M}} \sum_{Q=0}^{M-1} |Q \rangle |\hat \Psi(Q) \rangle,
\label{5.26}
\end{eqnarray}
where the subscript $\varphi$ is intended to highlight the parallel with the conventional state (\ref{5.11}), and the inverse Fourier transform states are defined by
\begin{equation}
|\hat \Psi(Q) \rangle
= \frac{1}{\sqrt{M}} \sum_{J=0}^{M-1}
e^{-2\pi i J Q/M} |\Psi(J \epsilon) \rangle.
\label{5.27}
\end{equation}
These states inherit the factorization property in the form
\begin{eqnarray}
\tilde F_\Sigma(Q,Q') &\equiv& \mathrm{tr}_W\left[|\hat \Psi(Q) \rangle
\langle \hat \Psi(Q')| \right]
\nonumber \\
&=& |\hat \psi_{\Sigma,G}(Q/\epsilon) \rangle
\langle \hat \psi_{\Sigma,G}(Q'/\epsilon)|
\label{5.28}
\end{eqnarray}
with reduced state Fourier transform
\begin{equation}
|\hat \psi_{\Sigma,G}(Q/\epsilon) \rangle
= \frac{1}{\sqrt{M}} \sum_{J=0}^{M-1}
e^{-2\pi i J Q/M} |\psi_{\Sigma,G}(J \epsilon) \rangle.
\label{5.29}
\end{equation}

The final result (\ref{5.28}) continues to hold if one uses $|\Phi \rangle$ or $|\Phi_D \rangle$ in (\ref{5.26}), generating states $|\hat \Phi(Q) \rangle$, $|\hat \Phi_D(Q) \rangle$ that produce the identical form for $\tilde F_\Sigma(Q,Q')$.

\subsubsection{Reduced density matrix phase estimation}
\label{subsec:rdmphest}

The factorization property (\ref{5.28}) is the key enabler of the desired phase estimation associated with the reduced subspace evolution (\ref{3.12}). It follows from (\ref{5.15}) that the reduced density matrix
\begin{eqnarray}
\hat F_{\Sigma,G}^F &\equiv& \mathrm{tr}_W
[|\Psi_F \rangle \langle \Psi_F|]
\nonumber \\
&=& |\psi^F_{\Sigma,G} \rangle \langle \psi^F_{\Sigma,G}|,
\label{5.30}
\end{eqnarray}
is the pure state generated by
\begin{equation}
|\psi^F_{\Sigma,G} \rangle = \frac{1}{\sqrt{M}} \sum_{J=0}^{M-1}
|J \rangle |\psi_{G,\Sigma}(\epsilon J) \rangle
\label{5.31}
\end{equation}
Similarly, from (\ref{5.26}) and (\ref{5.28}), in the Fourier basis one obtains the pure state
\begin{eqnarray}
\hat F^\varphi_{\Sigma,G} &\equiv& \mathrm{tr}_W
\left[|\Psi_\varphi \rangle \langle \Psi_\varphi| \right]
\nonumber \\
&=& |\psi^\varphi_{\Sigma,G} \rangle \langle \psi^\varphi_{\Sigma,G}|,
\label{5.32}
\end{eqnarray}
generated by
\begin{equation}
|\psi^\varphi_{\Sigma,G} \rangle = \frac{1}{\sqrt{M}} \sum_{Q=0}^{M-1}
|Q \rangle |\hat \psi_{\Sigma,G}(Q/\epsilon) \rangle.
\label{5.33}
\end{equation}

If one substitutes the eigenfunction expansion
\begin{equation}
|\psi_{G,\Sigma}(t) \rangle = \sum_u \langle \psi^G_u|\psi_\Sigma \rangle
e^{2\pi i \lambda_u t}|\psi^G_u \rangle,
\label{5.34}
\end{equation}
of $\hat G_\mathrm{est}$, then
\begin{eqnarray}
|\psi^F_{\Sigma,G} \rangle
&=& \sum_u \langle \psi^G_u|\psi_\Sigma \rangle
|\lambda_{u,F} \rangle |\psi^G_u \rangle
\nonumber \\
|\lambda_{u,F} \rangle &\equiv& \frac{1}{\sqrt{M}}
\sum_{J=0}^{M-1} e^{2\pi i \epsilon \lambda_u J} |J \rangle
\label{5.35}
\end{eqnarray}
and
\begin{eqnarray}
|\psi^\varphi_{\Sigma,G} \rangle
&=& \sum_u \langle \Psi^G_u|\psi_\Sigma \rangle
|\tilde \lambda_u \rangle |\psi^G_u \rangle
\nonumber \\
|\tilde \lambda_u \rangle
&\equiv& \sum_{Q=0}^{M-1} \Delta_M(\epsilon \lambda_u - Q/M) |Q \rangle
\label{5.36}
\end{eqnarray}
in which $\Delta_M(s)$ is defined in (\ref{5.8}) \cite{foot:Deltafn}. The latter reproduces the conventional phase estimator output state (\ref{5.11}).

To summarize, the factorization property allows one to transmit the phase estimation of $\hat U(t) = e^{i t \hat G_\mathrm{est}}$ to the reduced density matrix $\hat F^\varphi_{\Sigma,G}$, and hence to measurements on the reduced space ${\cal H}_\varphi \otimes {\cal H}_\Sigma$ where ${\cal H}_\varphi$ represents the $|\tilde \lambda_u \rangle$ (or $|Q \rangle$) register. Thus, the unitary operations performed on the full state $|\Psi(t)\rangle$ to obtain the state $|\Psi_\varphi \rangle$ allow one to access the eigenfunction expansion of $\hat G_\mathrm{est}$ via measurements of the form
\begin{equation}
\langle \Psi_\varphi| \hat I_\mathrm{ad} \otimes \hat M |\Psi_\varphi \rangle
= \mathrm{tr}[\hat M \hat F_{\Sigma,G}^\varphi]
= \langle \psi_{\Sigma,G}^\varphi| \hat M |\psi_{\Sigma,G}^\varphi \rangle
\label{5.37}
\end{equation}
in which $\hat M$ is now any measurement operator on ${\cal H}_\varphi \otimes {\cal H}_\Sigma$, and $\hat I_\mathrm{ad}$ is the identity operator on all additional ($|\psi_W \rangle$ state) degrees of freedom. The identical conclusion holds if one instead uses the qubit recycling or dissipation states.

\section{Quantum linear algebra}
\label{sec:qlinalg}

Given the generalized phase estimation algorithm, we seek now to generalize as well the HHL algorithm implementing $\hat G_\mathrm{est}^{-1}$ \cite{HHL2009}. The construction continues to rely on the factorization property (\ref{5.19}) but, as alluded to in Sec.\ \ref{subsec:treverse}, there are additional `time-reversal' considerations as well.

\subsection{Conventional HHL algorithm}
\label{sec:hhlconv}

Similar to the approach taken in Sec.\ \ref{sec:qphase} we first summarize the conventional HHL algorithm and then show how to generalize each step. One begins with the phase estimation output state (\ref{5.11}) [which becomes (\ref{5.12}) in our application], written in the form
\begin{eqnarray}
|\psi_\varphi \rangle &=& \sum_u A_u |\tilde \lambda_u \rangle |\psi_u \rangle
\nonumber \\
&=& \frac{1}{\sqrt{M}}\sum_{Q=0}^{M-1} {\cal A}_Q |Q \rangle |\psi_Q \rangle.
\label{6.1}
\end{eqnarray}
with $A_u = \langle \psi_u |\psi \rangle$. The $Q$-states are defined by
\begin{equation}
|\psi_Q \rangle = \frac{1}{{\cal A}_Q}
\sum_u A_u \Delta_M(\epsilon \lambda_u - Q/M) |\psi_u \rangle,
\label{6.2}
\end{equation}
in which $\Delta_M(s)$, defined in (\ref{5.8}), is strongly peaked around the origin \cite{foot:Deltafn}) and the normalization defined by
\begin{equation}
{\cal A}_Q^2 = \sum_u |A_u|^2 |\Delta_M(\epsilon \lambda_u - Q/M)|^2
\label{6.3}
\end{equation}
which is in turn strongly peaked at values $Q/\epsilon M \simeq \lambda_u$ approximating an eigenvalue.

\subsubsection{Ancillary qubit addition and rotation}
\label{subsec:qancillary}

We seek to derive from the form (\ref{6.1}) (an approximation to) the restricted inverse state
\begin{equation}
\hat G^{-1} |\psi \rangle = \sum_u \frac{A_u}{\lambda_u}
(1 - \delta_{\lambda_u 0}) |\psi_u \rangle
\label{6.4}
\end{equation}
which involves extraction of the multiplier via controlled operations on the state $|\tilde \lambda_u \rangle$, and then somehow dropping this register. This is accomplished as follows \cite{HHL2009}. Adjoin an extra qubit to the $|Q \rangle$ register in (\ref{6.1}), and perform the operation
\begin{equation}
\hat R |0 \rangle |Q \rangle
= [\cos(\theta_Q/2) |0 \rangle + \sin(\theta_Q/2) |1 \rangle] |Q \rangle,
\label{6.5}
\end{equation}
corresponding to the rotation $e^{i \frac{1}{2} \theta_Q \hat Y}$ controlled by the register $|Q \rangle$, in which the rotation angle is here defined by
\begin{equation}
\sin(\theta_Q/2) = \frac{C \epsilon}{\sin(2\pi Q/M)} (1 - \delta_{Q 0}),
\label{6.6}
\end{equation}
but could also take any other efficiently computable form (depending on the application). This form is designed to maintain periodicity in $Q$ while also approximating $C/\lambda_u$ when $Q/M = \epsilon \lambda_u$, under the additional condition that $\epsilon$ is chosen small enough to ensure $\epsilon \lambda_u \ll 1$. The constant $C$ depends on the matrix condition number, being chosen so that $0 < \sin(\theta_Q) < 1$ spans a reasonable range as the $|Q \rangle$ register ranges over the corresponding nonzero $\lambda_u$.

Applying the transformation (\ref{6.5}) to $|0 \rangle |\psi_\varphi \rangle$, one obtains the state
\begin{eqnarray}
|\psi_{R,\varphi} \rangle &\equiv& |0 \rangle \frac{1}{\sqrt{M}}
\sum_{Q=0}^{M-1} \cos(\theta_Q/2) |Q \rangle |\psi_Q \rangle
\nonumber \\
&&+\ |1 \rangle \frac{1}{\sqrt{M}}
\sum_{Q=0}^{M-1} \sin(\theta_Q/2) |Q \rangle |\psi_Q \rangle.\ \ \ \ \ \
\label{6.7}
\end{eqnarray}
in which the second line contains the desired state approximating (\ref{6.4}).

\subsubsection{Inverse phase estimation}
\label{subsec:phestinverse}

The next step is to reverse the phase estimation, effectively restoring all $|Q \rangle \to |0 \rangle$, but leaving the new amplitude factors in place. This involves first Fourier transforming the $Q$ register [undoing (\ref{5.8})] then applying time reversed evolution [undoing (\ref{5.7})], and finally applying the $m$-fold Hadamard gate. For general coefficient $f(Q)$, the result is the transformation
\begin{eqnarray}
&&\hat {\bf O}_\varphi^\dagger \sum_{Q=0}^{M-1}
f(Q) |Q \rangle |\psi_Q \rangle
\nonumber \\
&&\ \ \ \ \ \ =\ \hat H^{\otimes m} \sum_u A_u f(\epsilon \lambda_u)
\frac{1}{\sqrt{M}} \sum_{J=0}^{M-1} |J \rangle |\psi_u \rangle
\nonumber \\
&&\ \ \ \ \ \ =\ |0^{\otimes m} \rangle
\sum_u A_u f(\epsilon \lambda_u M) |\psi_u \rangle
\label{6.8}
\end{eqnarray}
in which
\begin{eqnarray}
&&f(\epsilon \lambda_u M)\ =\ \sum_{Q=0}^{M-1} f(Q)
\Delta_M(\epsilon \lambda_u - \textstyle{\frac{Q}{M}})
e^{2\pi i J(Q/M - \epsilon \lambda_u)}
\nonumber \\
&&\ \ \ \ =\ \frac{1}{M} \sum_{J'=0}^{M-1}
e^{2\pi i (J'-J) \epsilon \lambda_u}
\sum_{Q=0}^{M-1} f(Q) e^{-2\pi i (J'-J) Q/M}
\nonumber \\
&&\ \ \ \ =\ \frac{1}{\sqrt{M}} \sum_{P=0}^{M-1} \hat f(P)
e^{2\pi i P \epsilon \lambda_u}
\label{6.9}
\end{eqnarray}
is indeed independent of $J$, which allows the final Hadamard operation in (\ref{6.8}) to produce the desired $|0^{\otimes m} \rangle$ state. Here (\ref{5.8}) has been substituted to obtain the second line, and $\hat f(P)$ is the inverse Fourier transform of $f(Q)$ (and is also periodic with period $M$, which enables the substitution $P = J'-J$ in the final sum). The last sum also serves to define $f(\epsilon \lambda_u M)$ as the natural analytic continuation of $f(Q)$ [exhibited here as the Fourier transform of $\hat f(P)$] to noninteger values of its argument.

\subsubsection{Final matrix-inverse state}
\label{subsec:matinvstate}

Applying the identity (\ref{6.8}) to (\ref{6.7}), one obtains the superposition
\begin{eqnarray}
|\psi_{R,F} \rangle &\equiv& \hat {\bf O}_\varphi^\dagger |\Psi_{R,\varphi} \rangle
\nonumber \\
&=& \sqrt{p_0} |0 \rangle |\psi_c \rangle + \sqrt{p_1} |1 \rangle |\psi_s \rangle
\label{6.10}
\end{eqnarray}
in which the subscript $F$ is motivated by the relation between the phase estimation states (\ref{5.7}) and (\ref{5.8}), and the component states are
\begin{eqnarray}
|\psi_c \rangle &=& \frac{1}{\sqrt{p_0}}
\sum_u A_u \cos(\theta_{\epsilon \lambda_u M}/2) |\psi_u \rangle
\nonumber \\
|\psi_s \rangle &=& \frac{1}{\sqrt{p_1}}
\sum_u A_u \sin(\theta_{\epsilon \lambda_u M}/2) |\psi_u \rangle
\nonumber \\
p_1 &=& 1 - p_0 = \sum_u |A_u|^2
\sin^2(\theta_{\epsilon \lambda_u M}/2).\ \ \ \ \ \
\label{6.11}
\end{eqnarray}
Here $\theta_{\epsilon \lambda_u M}$ is the analytic continuation of $\theta_Q$ as defined by applying the Fourier--inverse Fourier transform combination (\ref{6.9}) to the function $f(Q) = e^{i \theta_Q/2}$, and the now redundant overall $|0^{\otimes m} \rangle$ factor has been dropped. Using (\ref{6.6}), the multiplier in the $|\psi_s \rangle$ term will be very close to $C/\lambda_u$.

\subsubsection{Final state measurements}
\label{subsec:finalstatemeas}

The state (\ref{6.10}) remains a superposition of desired and extraneous states. The former is accessed through final measurements by simultaneously projecting onto the $|1 \rangle$ state. Thus, given a measurement operator $\hat M$ acting on the subspace ${\cal H}_\Sigma$ containing $|\psi_c \rangle$ and $|\psi_s \rangle$, one obtains
\begin{eqnarray}
\langle \Psi_{R,F}||1\rangle \langle 1| \otimes \hat M |\Psi_{R,F} \rangle
&=& p_1 \langle \psi_s| \hat M |\psi_s \rangle
\label{6.12} \\
&=& |C|^2 \langle \psi| \hat G^{-1} \hat M \hat G^{-1} |\psi \rangle.\ \ \ \ \ \
\nonumber
\end{eqnarray}
For example, using the form $\hat M = |\phi \rangle \langle \phi|$, the result is matrix element $|C|^2 |\langle \phi| \hat G^{-1} |\psi \rangle|^2$.

Note that the result (\ref{6.12}) is often phrased as two separate measurements \cite{HHL2009}, with the ancillary qubit measurement first, with probability $p_1$ of successfully observing `1', followed by a measurement on the $\hat G^{-1} |\psi \rangle$ state. This, however, is incorrect because it violates unitarity \cite{W2019}. A measurement result on an isolated ancillary qubit is uncorrelated with separate measurements on the remaining qubits. Formally, the first step corresponds to a unitary operation
\begin{equation}
|\Psi_{F,E} \rangle \equiv \hat U_{E,A} \otimes \hat I_\Sigma
|\psi_{R,F} \rangle |\Psi_E \rangle
\label{6.13}
\end{equation}
which entangles the ancillary qubit (only) with the macroscale environment state $|\Psi_E \rangle$ in an effectively irreversible fashion, and includes the ancillary qubit measurement readout with Born probabilities $p_0$ and $p_1$ for outcomes 0 and 1, respectively.

A subsequent measurement on the (so far untouched) ${\cal H}_\Sigma$ subspace yields the result
\begin{eqnarray}
&&\langle \Psi_{F,E}| \hat I_{1,E} \otimes \hat M |\Psi_{F,E} \rangle
\nonumber \\
&&\ \ \ \ =\ \langle \Psi_E| \langle \phi_{R,F}|
\hat U_{E,A}^\dagger \hat U_{E,A}
\otimes \hat M |\psi_{R,F} \rangle |\Psi_E \rangle
\nonumber \\
&&\ \ \ \ =\ \langle \psi_{R,F}| \hat I_A \otimes \hat M |\psi_{R,F} \rangle
\nonumber \\
&&\ \ \ \ =\ p_0 \langle \psi_c| \hat M |\psi_c \rangle
+ p_1  \langle \psi_s| \hat M |\psi_s \rangle
\label{6.14}
\end{eqnarray}
independent of the result of the first measurement, and still including a contribution from the extraneous state $|\psi_c \rangle$. Only the simultaneous measurement operation (\ref{6.11}) produces the desired result, probing the state $|\psi_s \rangle = C p_1^{-1/2} \hat G^{-1} |\psi \rangle$ alone.

\subsection{Generalized HHL algorithm}
\label{sec:genhhl}

Paralleling (\ref{6.1}) and (\ref{6.7}), we adjoin an extra qubit to the generalized phase estimation output state (\ref{5.26}), with (\ref{5.27}), and apply the identical conditioned rotation (\ref{6.5}) to obtain
\begin{eqnarray}
|\Psi_{R,\varphi} \rangle &=&
\hat R |0 \rangle |\Psi_\varphi \rangle
= |0 \rangle |\Psi_{\varphi,c} \rangle
+ |1 \rangle |\Psi_{\varphi,s} \rangle
\nonumber \\
|\Psi_{\varphi,c} \rangle &\equiv& \frac{1}{\sqrt{M}}
\sum_{Q=0}^{M-1} \cos(\theta_Q/2)
|Q \rangle |\hat \Psi(Q) \rangle
\nonumber \\
|\Psi_{\varphi,s} \rangle &\equiv& \frac{1}{\sqrt{M}}
\sum_{Q=0}^{M-1} \sin(\theta_Q/2)
|Q \rangle |\hat \Psi(Q) \rangle.
\label{6.15}
\end{eqnarray}

The Fourier transform step in the reverse phase estimation acts only on the $|Q \rangle$ register and hence proceeds as before. However, the time reversal operation requires adjustment. Taken, literally, the reverse operation effectively requires a time-reversal of the state evolution, even to negative values of time. The former is impossible if any form of dissipation is present, and even in the absence of dissipation requires perfect maintenance of a huge number of data qubits---defeating the purpose of the generalized Suzuki--Trotter approach. Negative times are generally impossible because it requires reverse-computing $|\psi_W \rangle$ states that do not exist.

To circumvent both of these issues, in a manner consistent with time reversal of the reduced density matrix, we instead apply the swap operator evolution (\ref{4.5}) with $t < 0$, thus continuing to adjoin new data copies as before while reversing the evolution of $\hat F(t)$. For the present application, we begin by adjoining an additional quantum data structure, of the identical form (\ref{5.17}) and (\ref{5.18}), defining the extended states
\begin{eqnarray}
|\Psi_\mathrm{ext}(Q) \rangle
= |\Psi_W \rangle |\Psi(Q) \rangle
\nonumber \\
|\Psi_\mathrm{ext}(J\epsilon) \rangle
= |\Psi_W \rangle |\Psi(J\epsilon) \rangle.
\label{6.16}
\end{eqnarray}
Working again with a general coefficient $f(Q)$, we define the generalized reverse-phase estimation operation
\begin{eqnarray}
|\Psi_{F,f} \rangle
&\equiv& \hat {\bf O}_{-\varphi} \sum_{Q=0}^{M-1}
f(Q) |Q \rangle |\Psi_\mathrm{ext}(Q) \rangle
\label{6.17} \\
&=& \hat H^{\otimes m} \frac{1}{M}
\sum_{J,J'=0}^{M-1} \hat f(J-J') |J' \rangle
|\Psi(J \epsilon,J' \epsilon) \rangle
\nonumber
\end{eqnarray}
in which $\hat f(P)$ is again the inverse Fourier transform of $f(Q)$, and
\begin{equation}
|\Psi(J \epsilon,J' \epsilon) \rangle
= \hat U(-\epsilon J') |\Psi_\mathrm{ext}(J \epsilon) \rangle
\label{6.18}
\end{equation}
is the generalized time reversed state. Analogous to (\ref{5.15}), one defines the binary decomposition
\begin{equation}
\hat U(-\epsilon J') = {\prod_{j=1}^m}' \hat U(-\epsilon 2^{j-1})
\label{6.19}
\end{equation}
where the prime again indicates that only terms with binary coefficients $J'_{j-1} = 1$ appear. The component operators entangle the working qubit subspace ${\cal H}_\Sigma$ with the appropriate components of the new copy of $|\Psi_W \rangle$ as described by (\ref{5.20}) and (\ref{5.21}). This structure ensures proper generalization of the factorization property (\ref{5.19}).

We will now show that the reduced density matrix constructed from the state (\ref{6.16}) allows one to access $\hat G_\mathrm{est}^{-1} |\psi_\Sigma \rangle$. It follows that measurements on this state, restricted to the subspace ${\cal H}_\Sigma$, reproduce the conventional HHL result (\ref{6.12}).

The reduced density matrices corresponding to the underlying time-reverse states (\ref{6.18}) are obtained within the Suzuki--Trotter iteration in the form
\begin{eqnarray}
\hat F_\Sigma(J_1,J_2;J'_1,J_2')
&=& \mathrm{tr}_E[|\Psi(J_1\epsilon, J_2\epsilon) \rangle
\langle \Psi(J_1'\epsilon, J_2'\epsilon)|]
\nonumber \\
&=& \prod_{j=1}^m (\hat {\cal L}^{(\nu_j)}_{-\Delta t,G})^{2^j N_\epsilon}
[\hat F_\Sigma(J_1,J'_1)]
\nonumber \\
\label{6.20}
\end{eqnarray}
in which the initial forward-time matrix $\hat F_\Sigma(J_1,J'_1)$ is defined by (\ref{5.20}), and $\nu_j$ here are defined by (\ref{5.21}), but with $J \to J_2$, $J' \to J_2'$. The structure is identical to (\ref{5.20}), except for the appearance here of the negative time step $-\Delta t$. This same structure guarantees that the factorization property holds in the form
\begin{eqnarray}
&&\hat F_\Sigma(J_1,J_2;J'_1,J_2')\
=\ e^{-iJ_2 \epsilon \hat G_\mathrm{est}}
\hat F_\Sigma(J_1,J'_1) e^{iJ_2' \epsilon \hat G_\mathrm{est}}
\nonumber \\
&&\ \ \ \ \ \ =\ |\psi_{\Sigma,G}[(J_1-J_2)\epsilon]\rangle
\langle \psi_{\Sigma,G}[(J_1'-J_2')\epsilon]|
\nonumber \\
&&\ \ \ \ \ \ =\ \hat F(J_1-J_2,J'_1-J'_2).
\label{6.21}
\end{eqnarray}
Using this result one obtains from (\ref{6.17}) the factored form
\begin{eqnarray}
\mathrm{tr}_E[|\Psi_{F,f} \rangle \langle \Psi_{F,g} |]
= |\psi_{\Sigma,f} \rangle \langle \psi_{\Sigma,g} |
\label{6.22}
\end{eqnarray}
in which $f(Q)$ and $g(Q)$ are arbitrary coefficients and
\begin{eqnarray}
|\psi_{\Sigma,f} \rangle &=& \hat H^{\otimes m} \frac{1}{M} \sum_{J,J'=0}^{M-1}
\hat f(J-J') |J' \rangle |\psi_{\Sigma,G}[\epsilon(J-J')] \rangle
\nonumber \\
&=& \hat H^{\otimes m} \frac{1}{\sqrt{M}} \sum_{J'=0}^{M-1} |J' \rangle
f(\epsilon M \hat G_{\mathrm{est}}) |\psi_\Sigma \rangle
\nonumber \\
&=& |0^{\otimes m} \rangle
f(\epsilon M \hat G_{\mathrm{est}}) |\psi_\Sigma \rangle
\nonumber \\
&=& |0^{\otimes m} \rangle \sum_u A_u
f(\epsilon \lambda_u M) |\psi_u^G \rangle,
\label{6.23}
\end{eqnarray}
and similarly for $|\psi_{\Sigma,g} \rangle$.

\subsubsection{Generalized HHL algorithm final state}
\label{subsec:genhhlfinal}

Of interest here is state
\begin{eqnarray}
|\Psi_{R,F} \rangle
&=& \hat {\bf O}_{-\varphi} |\Psi_{R,\varphi} \rangle
= |0\rangle |\Psi_{F,c} \rangle
+ |1\rangle |\Psi_{F,s} \rangle
\nonumber \\
|\Psi_{F,c} \rangle
&\equiv& \hat {\bf O}_{-\varphi} |\Psi_{\varphi,c} \rangle
\nonumber \\
|\Psi_{F,s} \rangle
&\equiv& \hat {\bf O}_{-\varphi} |\Psi_{\varphi,s} \rangle,
\label{6.24}
\end{eqnarray}
derived from (\ref{6.15}), which leads to
\begin{equation}
\mathrm{tr}_E[|\Psi_{R,F} \rangle \langle \Psi_{R,F}|]
= |0^{\otimes m} \rangle \langle 0^{\otimes m}|
\otimes |\psi^R_{\Sigma,F} \rangle \langle \psi^R_{\Sigma,F}|,
\label{6.25}
\end{equation}
in which, identical in form to (\ref{6.10}) and (\ref{6.11}),
\begin{eqnarray}
|\psi_{\Sigma,F}^R \rangle &=& |0 \rangle
\cos(\theta_{\epsilon M \hat G_\mathrm{est}}/2) |\psi_\Sigma \rangle
\nonumber \\
&&+\ |1 \rangle \sin(\theta_{\epsilon M \hat G_\mathrm{est}}/2) |\psi_\Sigma \rangle
\nonumber \\
&=& \sqrt{p_{\Sigma,0}} |0 \rangle |\psi_{\Sigma,c} \rangle
+ \sqrt{p_{\Sigma,1}} |1 \rangle |\psi_{\Sigma,s} \rangle,\ \ \ \ \ \
\label{6.26}
\end{eqnarray}
with working space states
\begin{eqnarray}
|\psi_{\Sigma,c} \rangle
&=& \frac{1}{\sqrt{p_{\Sigma,0}}} \sum_u A_u
\cos(\theta_{\epsilon \lambda_u M}) |\psi_u^G \rangle
\nonumber \\
|\psi_{\Sigma,s} \rangle
&=& \frac{1}{\sqrt{p_{\Sigma,1}}} \sum_u A_u
\sin(\theta_{\epsilon \lambda_u M}) |\psi_u^G \rangle
\nonumber \\
p_{\Sigma,1} &=& 1 - p_{\Sigma,0}
= \sum_u |A_u|^2 \sin^2(\theta_{\epsilon \lambda_u M}).\ \ \ \ \ \
\label{6.27}
\end{eqnarray}
Here $\lambda_u$, $|\psi_u^G \rangle$ are again the eigenvalues and eigenstates of $\hat G_\mathrm{est}$ and $A_u = \langle \Psi_u^G|\psi_\Sigma \rangle$ are the corresponding expansion coefficients. It is important to emphasize that the states (\ref{6.24}) \emph{do not} themselves contain factors $|0^{\otimes m} \rangle$. These emerge only as a consequence of the trace operation.

Equations (\ref{6.26}) and (\ref{6.27}) are the main results of this section, confirming that the generalized reverse-time Suzuki--Trotter dynamics encoded in (\ref{6.17}) via (\ref{6.20}) indeed generates a state with properties entirely equivalent to the conventional state (\ref{6.10}) at the reduced density matrix (hence measurement) level. Explicitly, projecting again onto the $|1 \rangle$ subspace one obtains
\begin{eqnarray}
&&\langle \Psi_{R,F}|
|1\rangle \langle 1| \otimes \hat I_\mathrm{ad} \otimes \hat M
|\Psi_{R,F} \rangle
=\ p_{\Sigma,1} \langle \psi_{\Sigma,s}|\hat M|\psi_{\Sigma,s} \rangle \
\nonumber \\
&&\hskip1.0in =\ |C|^2 \langle \psi_\Sigma |\hat G_\mathrm{est}^{-1}
\hat M \hat G_\mathrm{est}^{-1} |\psi_\Sigma \rangle
\label{6.28}
\end{eqnarray}
which coincides with (\ref{6.12}).

\subsubsection{Reverse-phase qubit recycling and dissipation}
\label{subsec:rphqrecycdiss}

Finally, for simplicity, the reverse-phase estimation as implemented in (\ref{6.17}) neglects qubit recycling and dissipation---see Sec. \ref{subsec:qrecycdiss}, especially (\ref{5.22}) and (\ref{5.24}). These (likely physically inescapable) operations, again acting only on the environmental and data qubit subspaces, may be included here as well and produce the identical measurement state outcomes (\ref{6.26})--(\ref{6.28}).

\section{Quantum implementation of target detection}
\label{sec:qtgtdetect}

We turn now to the final step, namely implementation of the data processing algorithm identifying likely target locations (indexed here by the $p$-qubit radar image pixel register $|x \rangle \equiv |{\bf x} \rangle |{\bf v} \rangle$) via the detection statistic (\ref{2.21}). The input to the numerator (\ref{2.20}), and to the denominator (\ref{2.22}), are provided by the generalized HHL algorithm developed in Sec.\ \ref{sec:genhhl}. As summarized in Sec.\ \ref{sec:qsearchqcount}, the classic Grover quantum search algorithm \cite{NC2000} may be naturally adapted to this problem.

As will be seen, construction of the detection statistic requires input states in digital form for convenient formulation of the oracle $\hat U_\gamma$ [equation (\ref{3.15})] implementing the logical function (\ref{3.14}) that encodes the detection criterion (\ref{3.13}). Depending on the implementation of the HHL algorithm, an initial qA/D conversion (App.\ \ref{app:qad}) may therefore be required.

The remainder of this section is devoted to the conventional construction of $\hat U_\gamma$. This is presented in some detail since it is quite intricate. However, the generalization to account for the reduced density matrix approach, detailed in Sec.\ \ref{sec:groveroraclegen}, is then relatively straightforward, with essentially one-to-one correspondence between the steps. Some additional details of its implementation for the target identification problem are discussed in Secs.\ \ref{sec:groversearchimplement} and \ref{sec:grsearchqcountgen}.

\subsection{Detection statistic state: conventional construction}
\label{sec:detectstateconv}

The first step is to construct the state encoding the difference $|\Sigma_G(x)|^2 - h_0 \Sigma_{G,0}(x)$. To construct the state corresponding to $|\Sigma_G(x)|^2$, it is assumed that one has access to two independent copies of the output state:
\begin{equation}
|\psi_{\Sigma,G} \rangle
= \hat G_\mathrm{est}^{-1} |\psi_\Sigma \rangle
= \frac{1}{\sqrt{N_D}} \sum_x |\Sigma_G(x) \rangle |x \rangle.
\label{7.1}
\end{equation}
In the conventional construction these states are obtained directly from the digital format initial state (\ref{3.1}) via application of the conventional HHL algorithm (see Secs.\ \ref{sec:qphaseconv}, \ref{sec:hhlconv}). We write $|\Sigma_G(x) \rangle = |\mathrm{Re}[\Sigma_G(x)] \rangle |\mathrm{Im}[\Sigma_G(x)] \rangle$, and the sign of each component is determined by a leading qubit. By flipping the sign of the imaginary part of the second copy (an $\hat X$ gate), one obtains the mapping
\begin{equation}
|\psi_{\Sigma,G} \rangle |\psi_{\Sigma,G} \rangle
\to |\psi_{\Sigma,G} \rangle |\psi_{\Sigma,G}^* \rangle.
\label{7.2}
\end{equation}
By performing controlled bitwise multiplication one obtains
\begin{eqnarray}
|\psi_{\Sigma,G} \rangle |\psi_{\Sigma,G}^* \rangle
&=& \frac{1}{N_D} \sum_{x,x'} |\Sigma_G(x) \rangle
|\Sigma_G(x') \rangle |x \rangle |x' \rangle
\nonumber \\
&\to& \frac{1}{N_D} \sum_{x,x'} |\Sigma_G(x) \rangle
|\Sigma_G(x) \Sigma_G(x')^* \rangle
\nonumber \\
&&\ \ \ \ \ \ \otimes\ |x \rangle |x \oplus x' \rangle
\nonumber \\
&\equiv& |\psi_{\Sigma,G}^\oplus \rangle
\label{7.3}
\end{eqnarray}
in which, in the second line, an additional controlled Boolean sum has been performed on the last two registers. Here and in several places below there is a slight abuse of notation due to convenient reordering of the registers. The digital representation of the $|\Sigma(x)|^2$ state may be read off the second register under the condition that the last register is $|0 \rangle$.

In a similar fashion, the digital format state
\begin{eqnarray}
|\psi_G \rangle &=& \frac{1}{N_D} \sum_{x,x'}
|\hat G_\mathrm{est}^{-1}(x,x') \rangle |x \rangle |x' \rangle
\nonumber \\
&\to& \frac{1}{N_D} \sum_{x,x'}
|\hat G_\mathrm{est}^{-1}(x,x') \rangle |x \rangle |x \oplus x' \rangle
\nonumber \\
&\equiv& |\psi_G^\oplus \rangle
\label{7.4}
\end{eqnarray}
produces the digital representation of $G_\mathrm{est}^{-1}(x,x)$ through conditioning on the last register being $|0 \rangle$. The state $|\psi_G \rangle$ may be constructed as follows. First note that, for the special case $\Sigma(x) = \delta_{xx'}$, the HHL algorithm performs the transformation
\begin{equation}
|1 \rangle |x' \rangle \to |\psi_{x',G} \rangle \equiv
\sum_x |\hat G_\mathrm{est}^{-1}(x,x') \rangle |x \rangle,
\label{7.5}
\end{equation}
in which $|1 \rangle$ is an $m$ qubit register. One may therefore construct the desired state in the form
\begin{eqnarray}
|\psi_G \rangle &=& \hat G_\mathrm{est}^{-1} |\psi_\mathrm{diag} \rangle
\nonumber \\
|\psi_\mathrm{diag} \rangle &\equiv& |1 \rangle
\prod_{j=1}^p e^{-i \frac{\pi}{2} \hat X_{1,j} \hat Y_{2,j}}
|0 \rangle |0 \rangle
\nonumber \\
&=& |1 \rangle \otimes_{j=1}^p
\frac{|0\rangle_{1,j} |0 \rangle_{2,j} + |1\rangle_{1,j} |1 \rangle_{2,j}}{\sqrt{2}}
\nonumber \\
&=& \frac{1}{\sqrt{N_D}} \sum_{x'} |1 \rangle |x' \rangle |x' \rangle,
\label{7.6}
\end{eqnarray}
which consists of applying the HHL algorithm (in parallel) to the first two registers of the state $|\psi_\mathrm{diag} \rangle$.

The final detection statistic state emerges from the following sequence of controlled bitwise multiplication, addition, and Boolean sum operations:
\begin{widetext}
\begin{eqnarray}
|h_0 \rangle |\psi_{\Sigma,G}^\oplus \rangle |\psi_G^\oplus \rangle
&=& |h_0 \rangle \frac{1}{N_D^2} \sum_{x,y,x',y'} |\Sigma_G(x) \rangle
\left||\Sigma_G(x \oplus y)|^2 \right\rangle
\left|\hat G_\mathrm{est}^{-1}(x',x' \oplus y') \right\rangle
|x \rangle |y \rangle |x'\rangle |y'\rangle
\nonumber \\
&\to& |h_0 \rangle \frac{1}{N_D^2} \sum_{x,y,x',y'} |\Sigma_G(x) \rangle
\left||\Sigma_G(x \oplus y)|^2 \right\rangle
\left|h_0 \hat G_\mathrm{est}^{-1}(x',x' \oplus y') \right\rangle
|x \rangle |y \rangle |x' \rangle |y' \rangle
\nonumber \\
&\to& |h_0 \rangle \frac{1}{N_D^2} \sum_{x,y,x',y'} |\Sigma_G(x) \rangle
|\psi_\mathrm{DS}(x,y,x',y') \rangle
\left|h_0 \hat G_\mathrm{est}^{-1}(x',x' \oplus y') \right\rangle
|x \rangle |y \rangle |x \oplus x' \rangle |y' \rangle
\nonumber \\
&\equiv& |h_0 \rangle |\psi_\mathrm{DS}^\oplus \rangle
\label{7.7}
\end{eqnarray}
\end{widetext}
in which $|h_0 \rangle$ is an additional threshold parameter register and the extended detection statistic register in the third line is defined by
\begin{equation}
|\psi_\mathrm{DS}(x,y,x',y') \rangle =
\left||\Sigma_G(x \oplus y)|^2 - h_0 \hat G_\mathrm{est}^{-1}(x',x' \oplus y') \right\rangle.
\label{7.8}
\end{equation}
Controlling on $|0 \rangle$ for the last three registers in $|\psi_\mathrm{DS}^\oplus \rangle$, the register state
\begin{equation}
|\psi_\mathrm{DS}(x) \rangle \equiv |\psi_\mathrm{DS}(x,0,x,0) \rangle
\label{7.9}
\end{equation}
encodes precisely the digital representation of the input to the logical function (\ref{3.14}).

\subsection{Grover search oracle construction}
\label{sec:oracleconstruct}

To simplify the notation, write the detection statistic state in the form
\begin{eqnarray}
|\psi_\mathrm{DS}^\oplus \rangle &=& \frac{1}{N_D^2} \sum_{x,y,x',y'}
|\gamma(x,y,x',y') \rangle |A(x,y,x',y') \rangle
\nonumber \\
&&\otimes\ |x \rangle |y \rangle |x \oplus x' \rangle |y' \rangle
\label{7.10}
\end{eqnarray}
in which $|\gamma = 0,1 \rangle$ is the overall sign qubit of the register (\ref{7.8}), and $|A \rangle$ combines all remaining registers not explicitly displayed. It is the need for direct access to the sign qubit that requires the digital format. The objective is to design a unitary operation
\begin{equation}
|x_0 \rangle |q \rangle |\psi_\mathrm{DS}^\oplus \rangle
\to |x_0 \rangle |q \oplus \gamma(x_0) \rangle
|\psi_\mathrm{DS}^\oplus(x_0) \rangle,
\label{7.11}
\end{equation}
controlled by the state $|\psi_\mathrm{DS}^\oplus \rangle$, in which $|\psi_\mathrm{DS}^\oplus(x_0) \rangle$ is a certain projected state, depending on $x_0$, defined in (\ref{7.23}) below. To accomplish this, define the projection operators
\begin{eqnarray}
\hat P_a &=& \frac{1}{N_D} \prod_{j=1}^p (1 + \hat Z^a_j)
\nonumber \\
\hat P_{ab} &=& \frac{1}{N_D} \prod_{j=1}^p (1 + \hat Z^a_j \hat Z^b_j),
\label{7.12}
\end{eqnarray}
with actions
\begin{eqnarray}
\hat P_a |x \rangle_a &=& \delta_{x0} |0 \rangle_a
\nonumber \\
\hat P_{ab} |x \rangle_a |x' \rangle_b &=& \delta_{x x'} |x \rangle_a |x \rangle_b,
\label{7.13}
\end{eqnarray}
where $\hat Z^a_j, \hat Z^b_j$ act on qubit $j$ in the registers $|x \rangle_a, |x' \rangle_b$, respectively. Finally, let
\begin{equation}
\hat P = \hat P_{01} \hat P_2 \hat P_3 \hat P_4,
\label{7.14}
\end{equation}
acting on $|x_0 \rangle |q \rangle |\psi_\mathrm{DS}^\oplus \rangle$, be the simultaneous projection onto the subspace defining the support of (\ref{7.9}), with subscript 0 referring to the $|x_0 \rangle$ register and the remaining values to the last four registers in (\ref{7.10}). Let $\hat X_q$ be the bit flip operator acting on $|q \rangle$, and define the controlled bit flip unitary operator
\begin{equation}
\hat Q_\gamma = \left(\frac{\hat I_\gamma + \hat Z_\gamma}{2} \hat X_q
+ \frac{\hat I_\gamma - \hat Z_\gamma}{2} \hat I_q \right) \hat P
+ \hat I_\gamma \hat I_q  (\hat I - \hat P),
\label{7.15}
\end{equation}
in which the subscripts indicate action on the $|q \rangle$ or $|\gamma \rangle$ registers. One obtains
\begin{eqnarray}
\hat Q_\gamma |x_0 \rangle |q \rangle |\psi_\mathrm{DS}^\oplus \rangle
&=& |x_0 \rangle |q \oplus \gamma(x_0) \rangle
\hat P(x_0) |\psi_\mathrm{DS}^\oplus\rangle
\nonumber \\
&&+\ |x_0 \rangle |q \rangle [\hat I-\hat P(x_0)]
|\psi_\mathrm{DS}^\oplus \rangle
\ \ \ \ \ \
\label{7.16}
\end{eqnarray}
in which
\begin{equation}
\hat P(x_0) |\psi_\mathrm{DS}^\oplus \rangle
= |\gamma(x_0) \rangle |A(x_0) \rangle
|x_0 \rangle |0 \rangle |0 \rangle |0 \rangle
\label{7.17}
\end{equation}
is the projection onto the $x_0$ subspace, with shorthand $\gamma(x_0) = \gamma(x_0,0,x_0,0)$ and $A(x_0) = A(x_0,0,x_0,0)$. Thus, the desired oracle output (\ref{7.11}) resides only in this projected subspace.

The final step is to rotate the $|x_0 \rangle$ state in (\ref{7.17}) to $|0 \rangle$ so that all choices for $x_0$ lead to a common output channel. The unitary operator
\begin{equation}
\hat Q = \prod_{j=1}^p \left(\frac{\hat I_j + \hat Z_j}{2} \hat I_j'
+ \frac{\hat I_j - \hat Z_j}{2} \hat X_j'  \right)
\label{7.18}
\end{equation}
performs the Boolean sum
\begin{equation}
\hat Q |x \rangle |x' \rangle
= |x \rangle |x \oplus x' \rangle,
\label{7.19}
\end{equation}
which uniquely maps $|x_0 \rangle |x_0 \rangle \to |x_0 \rangle |0 \rangle$. It follows that the unitary operator
\begin{equation}
\hat U_\gamma = \hat Q \hat Q_\gamma
\label{7.20}
\end{equation}
produces the transformation
\begin{eqnarray}
|\psi_\mathrm{DS}^\gamma \rangle &\equiv& \hat U_\gamma
|x_0 \rangle |q \rangle |\psi_\mathrm{DS}^\oplus \rangle
\label{7.21}\\
&=& |x_0 \rangle |q \oplus \gamma(x_0) \rangle
|\psi_\mathrm{DS}^\oplus(x_0) \rangle
+ |x_0 \rangle |q \rangle
|\psi_\mathrm{DS}^{\oplus \perp} \rangle,
\nonumber
\end{eqnarray}
in which we define the projected states
\begin{eqnarray}
|\psi_\mathrm{DS}^\oplus(x_0) \rangle
&=& \hat Q(x_0) \hat P(x_0) |\psi_\mathrm{DS}^\oplus \rangle
\nonumber \\
&=& \frac{1}{N_D^2} |\gamma(x_0) \rangle |A(x_0) \rangle |0^{\otimes 4} \rangle
\nonumber \\
|\psi_\mathrm{DS}^{\oplus \perp}(x_0) \rangle
&=& \hat Q(x_0) [\hat I - \hat P(x_0)] |\psi_\mathrm{DS}^\oplus \rangle,
\label{7.22}
\end{eqnarray}
where, to condense the notation, we have defined
\begin{equation}
\hat Q(x_0) |x' \rangle = |x_0 \oplus x' \rangle,
\label{7.23}
\end{equation}
By construction, $|\psi_\mathrm{DS}^{\oplus \perp}(x_0) \rangle$ contains no $|0^{\otimes 4} \rangle$ component.

A general input state of the form
\begin{equation}
|\psi_b \rangle = \sum_{x_0} b(x_0) |x_0 \rangle |q(x_0) \rangle
\label{7.24}
\end{equation}
then produces output
\begin{eqnarray}
\hat U_\gamma |\psi_b \rangle |\psi_\mathrm{DS}^\oplus \rangle
&=& \sum_{x_0} b(x_0) |x_0 \rangle |q(x_0) \oplus \gamma(x_0) \rangle
|\psi_\mathrm{DS}^\oplus(x_0) \rangle
\nonumber \\
&&+\ \sum_{x_0} b(x_0) |x_0 \rangle |q(x_0) \rangle
|\psi_\mathrm{DS}^{\oplus \perp}(x_0) \rangle,
\nonumber \\
\label{7.25}
\end{eqnarray}
in which the first line contains the desired oracle output as the first two registers, and the second line is again orthogonal to $|0^{\otimes 4} \rangle$.

If one uses input qubit $|q \rangle = |q_- \rangle = \hat H |1 \rangle$ (which is subsequently factored out and dropped) one recovers the sign flip operator (\ref{3.16}) in the form
\begin{eqnarray}
\hat U_\gamma |\psi_b \rangle
&=& \sum_{x_0} b(x_0) |x_0 \rangle
\label{7.26} \\
&&\otimes\ \left[(-1)^{\gamma(x_0)}|\psi_\mathrm{DS}^\oplus(x_0) \rangle
+ |\psi_\mathrm{DS}^{\oplus \perp}(x_0) \rangle \right],
\nonumber
\end{eqnarray}
also leaving the $\perp$ subspace invariant.

\subsection{Grover search implementation}
\label{sec:groversearchimplement}

The Grover operator $\hat U_\mathrm{gr}$ continues to be defined by (\ref{3.19}). It is easy to check that, for any initial angle $\phi$, its action on the uniform superposition states (\ref{3.18}) takes the form
\begin{eqnarray}
|\psi_\phi(\theta_\gamma) \rangle
&\equiv& \hat U_\mathrm{gr} [\cos(\phi/2) |\gamma_0 \rangle
+ \sin(\phi/2) |\gamma_1 \rangle] |\psi_\mathrm{DS}^\oplus \rangle
\nonumber \\
&=& \cos(\theta_\gamma + \phi/2) |\gamma_0 \rangle_\mathrm{DS}
\nonumber \\
&&+\ \sin(\theta_\gamma + \phi/2) |\gamma_1 \rangle_\mathrm{DS}
\nonumber \\
&&+\ \cos(\theta_\gamma - \phi/2) |\gamma_0 \rangle_\mathrm{DS}^\perp
\nonumber \\
&&+\ \sin(\theta_\gamma - \phi/2) |\gamma_1 \rangle_\mathrm{DS}^\perp
\label{7.27}
\end{eqnarray}
in which the angle $\theta_\gamma(N_\gamma)$ is defined by (\ref{3.22}), and the uniform input state $|\xi_0 \rangle$ corresponds to $\phi = \theta_\gamma$. To further simplify the notation, for any state $|\phi_b \rangle = \sum_{x_0} b(x_0) |x_0 \rangle$ we have defined the shorthand
\begin{eqnarray}
|\phi_b \rangle_\mathrm{DS} &=& \sum_{x_0} b(x_0) |x_0 \rangle
|\psi_\mathrm{DS}^\oplus(x_0) \rangle
\nonumber \\
|\phi_b \rangle_\mathrm{DS}^\perp &=& \sum_{x_0} b(x_0) |x_0 \rangle
|\psi_\mathrm{DS}^{\oplus \perp}(x_0) \rangle.
\label{7.28}
\end{eqnarray}
The focus, of course, will be on the $|\psi_\mathrm{DS}^\oplus \rangle$ terms. By iteration one obtains
\begin{eqnarray}
|\psi^{(k)}_{\gamma,\phi} \rangle
&\equiv& \hat U_\mathrm{gr}^k [\cos(\phi/2) |\gamma_0 \rangle
+ \sin(\phi/2) |\gamma_1 \rangle] |\psi_\mathrm{DS}^\oplus \rangle
\nonumber \\
&=& |\psi_\phi(k \theta_\gamma) \rangle.
\label{7.29}
\end{eqnarray}

The key idea is to choose the iteration number $k$ so that $\cos(k \theta_\gamma + \phi/2)$ is close to zero \cite{NC2000}, yielding a projection onto the solution vector $|\gamma_1 \rangle_\mathrm{DS}$. For this value of $k$, a measurement in the computational basis $|x \rangle$ will with high probability produce one of the solution states. If $N_\gamma/N_D \ll 1$ is small, as expected in our application with threshold $h_0$ chosen sufficiently large, then using input state $|\xi_0 \rangle$ we consider the choice
\begin{equation}
k_G(N_\gamma) = \frac{\pi}{2\theta_\gamma} - \frac{1}{2}
= \frac{\cos^{-1}(\sqrt{N_\gamma/N_D})}{2 \sin^{-1}(\sqrt{N_\gamma/N_D})}
\label{7.30}
\end{equation}
rounded to the nearest integer (rounding down if $k$ is exactly half-integer, so as to reduce the number of iterations). This yields $(k_G + \frac{1}{2}) \theta_\gamma \simeq \frac{\pi}{2}$, hence $\cos(k_G \theta_\gamma + \phi/2) = \cos[(k_G + \frac{1}{2}) \theta_\gamma] \simeq 0$ as desired. The angular error in the final state is at most $\theta_\gamma/2 \simeq \sqrt{N_\gamma/N_D}$, yielding error probability at most $N_\gamma/N_D$. The quadratic speed-up follows from $k_G \simeq \frac{\pi}{4} \sqrt{N_D/N_\gamma}$---the number of iterations of $\hat U_\mathrm{gr}$ that must be applied.

We may now consider a measurement operator of the form
\begin{equation}
\hat {\cal M} = \hat M \otimes \hat I_\gamma \otimes \hat I_A
\otimes |0^{\otimes 4} \rangle \langle 0^{\otimes 4}|
\label{7.31}
\end{equation}
in which $\hat M$ acts on the $|x_0 \rangle$ register qubits. With the choice $k = k_G$, the result will lie close to
\begin{equation}
\langle \psi^{(k_G)}_{\gamma,\phi}|
\hat {\cal M} |\psi^{(k_G)}_{\gamma,\phi} \rangle
= \frac{1}{N_D^2} \langle \gamma_1|\hat M|\gamma_1 \rangle,
\label{7.32}
\end{equation}
thereby providing information about the above-threshold target pixels. Validity of the output, e.g., the pixel index of one of these targets, may be checked either via a classical computation or using the oracle, and failure simply requires that the algorithm be rerun some $O(1)$ number of times until a correct solution is found.

\subsection{Quantum counting}
\label{sec:qcount}

The choice (\ref{7.30}) for the Grover algorithm iteration number $k_G$ requires \emph{a priori} knowledge of the number of solutions $N_\gamma$. For the target detection application, this value (the number of bright targets) is certainly not generally known in advance (and could be zero).

The value of $N_\gamma$ may in fact be determined by applying the (conventional) phase estimation algorithm to the action of the Grover iteration operator $\hat U_\mathrm{gr}$ on the projected subspace:
\begin{equation}
\hat U_\mathrm{gr} |\gamma_\pm \rangle |\psi_\mathrm{DS}^\oplus \rangle
= e^{\pm i\theta_\gamma} |\gamma_\pm \rangle_\mathrm{DS}
+ e^{\mp i\theta_\gamma} |\gamma_\mp \rangle_\mathrm{DS}^\perp
\label{7.33}
\end{equation}
leading to
\begin{eqnarray}
\hat U_\mathrm{gr}^{2k} |\gamma_\pm \rangle |\psi_\mathrm{DS}^\oplus \rangle
&=& e^{\pm i 2k \theta_\gamma} |\gamma_\pm \rangle_\mathrm{DS}
+ |\gamma_\pm \rangle_\mathrm{DS}^\perp
\label{7.34} \\
\hat U_\mathrm{gr}^{2k+1} |\gamma_\pm \rangle |\psi_\mathrm{DS}^\oplus \rangle
&=& e^{\pm i (2k+1) \theta_\gamma} |\gamma_\pm \rangle_\mathrm{DS}
+ e^{\mp i \theta_\gamma} |\gamma_\mp \rangle_\mathrm{DS}^\perp.\ \ \
\nonumber
\end{eqnarray}
The eigenvalue equation (\ref{3.20}) is now exhibited in a generalized sense, with eigenvalues $e^{\pm i \theta_\gamma}$ applying in the projected $|0^{\otimes 4} \rangle$ subspace. It will now be shown that this is sufficient for constructing a phase estimation algorithm for $\theta_\gamma$, and hence for $N_\gamma$ via the relations (\ref{3.22}). The key observation is that
\begin{equation}
|\xi_0 \rangle = \frac{1}{\sqrt{2}}
\left(e^{i\theta_\gamma/2} |\gamma_+\rangle
+ e^{-i\theta_\gamma/2} |\gamma_-\rangle \right).
\label{7.35}
\end{equation}
is a superposition of the two eigenvectors. The algorithm is based only on the state $|\xi_0 \rangle$, hence avoids direct estimation of the number of states for which $\gamma(x) = 0$---as would be required by a classical algorithm.

Following the general procedure described in Sec.\ \ref{sec:qphase}, adjoining the extra control qubit registers, the controlled Grover oracle $\hat O_\mathrm{gr}$ is defined by (\ref{5.2}), with $\hat U_\mathrm{gr}$ substituted for $\hat U$. The corresponding product oracle operator $\hat {\bf O}_\mathrm{gr}$ is defined by (\ref{5.5}) and produces the states [compare (\ref{5.6}) and (\ref{5.7})]
\begin{widetext}
\begin{equation}
|\psi^F_{\gamma,\pm} \rangle \equiv
\hat {\bf O}_\mathrm{gr} |0^{\otimes t} \rangle
|\gamma_\pm \rangle |\psi_\mathrm{DS}^\oplus \rangle
\ =\ |\theta^F_{\gamma,\pm} \rangle
|\gamma_\pm \rangle_\mathrm{DS}
+ H^{\otimes (t-1)}|0^{\otimes (t-1)} \rangle
\left(|0 \rangle_1 |\gamma_\pm \rangle_\mathrm{DS}^\perp
+ e^{\mp i \theta_\gamma} |1 \rangle_1 |\gamma_\mp \rangle_\mathrm{DS}^\perp \right),
\label{7.36}
\end{equation}
where
\begin{equation}
|\theta^F_{\gamma,\pm} \rangle = \frac{1}{\sqrt{T}}
\sum_{J=0}^{T-1} e^{\pm i J \theta_\gamma} |J \rangle,
\label{7.37}
\end{equation}
in which a $t$-qubit register, with $T = 2^t$, is being used to estimate $\theta_\gamma$, but the state $|2^{\otimes t} \rangle$ common to all terms in (\ref{7.36}) has been dropped to condense the notation. Here $H^{\otimes (t-1)}|0^{\otimes (t-1)} \rangle = (2/T)^{1/2} \otimes_{j=2}^t (|0\rangle_j + |1\rangle_j)$ contains all the qubits corresponding to even powers of $\hat U_\mathrm{gr}$ which, via (\ref{7.34}), act as the identity on the $\perp$ subspace. Only the $j=1$ qubit, associated with odd powers, produces nontrivial mixing of the $|\gamma_\pm \rangle_\mathrm{DS}^\perp$ states.

Applying the inverse Fourier transform to the first register one obtains
\begin{equation}
|\tilde \psi_{\gamma,\pm} \rangle \equiv
\hat U_F^\dagger |\psi^F_{\gamma,\pm} \rangle
\ =\ |\tilde \theta_{\gamma,\pm} \rangle |\gamma_\pm \rangle_\mathrm{DS}
+ \frac{1+e^{\mp i \theta_\gamma}}{2\sqrt{2}}
|0 \rangle |\gamma_\pm \rangle^\perp_\mathrm{DS}
+ \frac{1-e^{\mp i \theta_\gamma}}{2\sqrt{2}}
|T/2 \rangle |2^{\otimes t} \rangle |\gamma_\mp \rangle^\perp_\mathrm{DS},
\label{7.38}
\end{equation}
\end{widetext}
in which
\begin{eqnarray}
|\tilde \theta_{\gamma,+} \rangle
&=& \hat U_F^\dagger |\theta^F_{\gamma,+} \rangle
= |\tilde \theta_\gamma/2\pi \rangle
\nonumber \\
|\tilde \theta_{\gamma,-} \rangle
&=& \hat U_F^\dagger |\theta^F_{\gamma,-} \rangle
= |1 - \tilde \theta_\gamma/2\pi \rangle
\label{7.39}
\end{eqnarray}
are the desired eigenvalue approximations defined by (\ref{5.8}). The pair of Fourier states $|0 \rangle$, $|T/2 = 2^{t-1} \rangle$ emerge from the odd--even structure in the $\perp$ subspace.

Clearly either of the states (\ref{7.39}) can be used to estimate $\theta_\gamma$. In particular, using (\ref{7.35}), the phase estimation output is the superposition
\begin{eqnarray}
|\xi_0 \rangle &\to& \frac{1}{\sqrt{2}}
\left(e^{i\theta_\gamma/2} |\tilde \theta_\gamma/2\pi \rangle
|\gamma_+ \rangle_\mathrm{DS} \right.
\nonumber \\
&&+\ \left. e^{-i\theta_\gamma/2} |1 - \tilde \theta_\gamma/2\pi \rangle
|\gamma_- \rangle_\mathrm{DS} \right)
\label{7.40}
\end{eqnarray}
in the $|0^{\otimes 4} \rangle$ subspace. It follows that measurement operators of the form (\ref{7.31}), but with $\hat M$ now operating on the first register, directly probe the states (\ref{7.39}), and the measurement will project onto one or the other, each with probability $\frac{1}{2}$. Since $0 \leq \theta_\gamma/2\pi < \frac{1}{2}$, there is no ambiguity between the two. Detailed error analysis \cite{NC2000} shows that (by judicious choice of $t \sim p/2$) one may estimate $\theta_\gamma$, and hence $N_\gamma$, with sufficient accuracy to ensure high probability success of the counting algorithm using $O(\sqrt{N_D})$ Grover oracle calls. Using this $N_\gamma$ value to construct the superposition states (\ref{3.18}), the target search algorithm constructed in Sec.\ \ref{sec:groversearchimplement} will similarly succeed with high probability, and the overall algorithm efficiency continues to scale as $\sqrt{N_D}$.

Obviously, the error and algorithm efficiency analysis for the present problem must be additionally informed by the new subspace structure, which will certainly affect the probability of various measurement outcomes. This is an important topic for future investigation, but lies beyond the scope of the present work.

\section{Grover search target detection oracle: generalized construction}
\label{sec:groveroraclegen}

We now adapt the `conventional' target search algorithm construction of the previous section to that based on the output of the generalized HHL algorithm, in which access to the (analog form) state $|\psi_{\Sigma,G}^{\cal A} \rangle$ is only through the reduced density matrix, and is equivalent to the state $|\psi_{\Sigma,s} \rangle$ [see (\ref{6.27})]. Since $|\psi_{\Sigma,s} \rangle$  is linearly related to the full state $|\Psi_{F,s} \rangle$ [see (\ref{6.24}) and (\ref{6.15})], linear operations on the latter will be reflected on the former. It follows that adapting the series of steps in the previous subsection leads to the appropriate generalized algorithm.

\subsection{Generalized detection statistic state construction}
\label{sec:gendetstatstate}

The first step in the construction of the detection statistic is to apply the qA/D conversion to the working space qubits, generating the transformation
\begin{equation}|
\Psi_{F,s} \rangle \to |\Psi_{\Sigma,G} \rangle,
\label{8.1}
\end{equation}
in which the trace operation
\begin{equation}
\mathrm{tr}_E[|\Psi_{\Sigma,G} \rangle \langle \Psi_{\Sigma,G}|]
= |\psi_{\Sigma,G} \rangle \langle \psi_{\Sigma,G}|
\label{8.2}
\end{equation}
now generates the digital form (\ref{7.1}) of the output state (see App.\ \ref{app:dmqad}).

Lacking the full digital form of the environmental state, the complex conjugate state must be computed separately, starting from the complex conjugation of the digital format data states $|\psi_\Sigma^* \rangle$, $|\psi_W^* \rangle$, followed by qD/A conversion, and then using time reversed evolution $t \to -t$ in (\ref{3.12}) and in the alternative Suzuki--Trotter evolution (\ref{4.8}) and (\ref{4.13}). From this construction, it is assumed that the two states $|\Psi_{F,s} \rangle$, $|\Psi_{F,s}^* \rangle$ are separately available, and rely on independent data and environmental subspaces.

As will now be shown, bitwise multiplication on the working space register produces the states generalizing (\ref{7.2}) and (\ref{7.3}). In the much larger environment plus working space, one may decompose
\begin{equation}
|\Psi_{\Sigma,G} \rangle = \frac{1}{\sqrt{N_D}}
\sum_x |\Psi_{\Sigma,G}(x) \rangle |x \rangle
\label{8.3}
\end{equation}
in which
\begin{equation}
|\Psi_{\Sigma,G}(x) \rangle = \frac{1}{\sqrt{M}}
\sum_A |\Psi_{\Sigma,G}^A(x) \rangle |A \rangle,
\label{8.4}
\end{equation}
is an entangled superposition of working subspace $m$ qubit digital register states $|A \rangle$ (with value of $m$ determined by the desired resolution of the qA/D conversion---see Sec.\ \ref{app:qa2d}), with very high dimensional analog states $|\Psi_{\Sigma,G}^A(x) \rangle$ lying entirely in the environmental subspace ${\cal H}_E$. The generalization of the product state transformation (\ref{7.3}) is
\begin{widetext}
\begin{eqnarray}
|\Psi_{\Sigma,G} \rangle |\Psi_{\Sigma,G}^* \rangle
&=& \frac{1}{M N_D} \sum_{x,x'} \sum_{A,A'}
|\Psi_{\Sigma,G}^A(x) \rangle |\Psi_{\Sigma,G}^{A' *}(x') \rangle
|A \rangle |A' \rangle |x \rangle |x' \rangle
\nonumber \\
&\to& \frac{1}{M N_D} \sum_{x,x'} \sum_{A,A'}
|\Psi_{\Sigma,G}^A(x) \rangle |\Psi_{\Sigma,G}^{A' *}(x') \rangle
|A \rangle |A A' \rangle |x \rangle |x \oplus x' \rangle
\nonumber \\
&\equiv& |\Psi^\oplus_{\Sigma,G} \rangle,
\label{8.5}
\end{eqnarray}
\end{widetext}
obeying, via (\ref{8.2}), the trace identities
\begin{eqnarray}
\mathrm{tr}_E[|\Psi^A_{\Sigma,G}(x) \rangle
\langle \Psi^{A'}_{\Sigma,G}(x')|]
&=& M \delta_{A,\Sigma_G(x)} \delta_{A',\Sigma_G(x')}
\nonumber \\
\mathrm{tr}_E[|\Psi_{\Sigma,G}(x) \rangle \langle \Psi_{\Sigma,G}(x')|]
&=& |\Sigma_G(x) \rangle \langle \Sigma_G(x')|
\nonumber \\
\mathrm{tr}_E[|\Psi^\oplus_{\Sigma,G} \rangle \langle \Psi^\oplus_{\Sigma,G}|]
&=& |\psi^\oplus_{\Sigma,G} \rangle \langle \psi^\oplus_{\Sigma,G}|,
\label{8.6}
\end{eqnarray}
and similarly for $|\Psi_{\Sigma,G}^* \rangle$. The traces here are performed independently over the first two registers in (\ref{8.5}).

Similarly, defining
\begin{equation}
|\Psi_{x',G} \rangle = \frac{1}{\sqrt{N_D}}
\sum_x |\Psi_{x',G}(x,x') \rangle |x \rangle
\label{8.7}
\end{equation}
to be the generalized HHL output state generated with (working space) input $\Sigma(x) = \delta_{xx'}$, equation (\ref{7.4}) is generalized in the form
\begin{eqnarray}
|\Psi_G^\oplus \rangle &=& \frac{1}{N_D}
\sum_{x,x'} |\Psi_{x',G}(x) \rangle |x \rangle |x \oplus x' \rangle
\nonumber \\
|\Psi_{x',G}(x) \rangle &=& \frac{1}{\sqrt{M}}
\sum_A |\Psi_{x',G}^A(x) \rangle |A \rangle
\label{8.8}
\end{eqnarray}
obtained from the identical working space input state (\ref{7.6}). The trace identities follow in the form
\begin{eqnarray}
\mathrm{tr}_E[|\Psi_{x',G}^A(x) \rangle \langle \Psi_{y',G}^{A'}(y)|]
&=& M \delta_{A,\hat G_\mathrm{est}^{-1}(x,x')}
\delta_{A',\hat G_\mathrm{est}^{-1}(y,y')}
\nonumber \\
\mathrm{tr}_E[|\Psi_{x',G}(x) \rangle \langle \Psi_{y',G}(y)|]
&=& |\hat G_\mathrm{est}^{-1}(x,x') \rangle \langle \hat G_\mathrm{est}^{-1}(y,y')|
\nonumber \\
\mathrm{tr}_E[|\Psi_{x',G} \rangle \langle \Psi_{y',G}|]
&=& |\psi_{x',G} \rangle \langle \psi_{y',G}|
\nonumber \\
\mathrm{tr}_E[|\Psi_G^\oplus \rangle \langle \Psi_G^\oplus|]
&=& |\psi_G^\oplus \rangle \langle \psi_G^\oplus|.
\label{8.9}
\end{eqnarray}

Finally, the generalization of the detection statistic state (\ref{7.7}) is obtained from the corresponding sequence of transformations
\begin{widetext}
\begin{eqnarray}
&&|h_0 \rangle |\Psi_{\Sigma,G}^\oplus \rangle |\Psi_G^\oplus \rangle
\ =\  |h_0 \rangle \frac{1}{M^{3/2} N_D^2} \sum_{x,y,x',y'} \sum_{A,A',B}
|\Psi^A_{\Sigma,G}(x) \rangle |\Psi^{A' *}_{\Sigma,G}(x \oplus y) \rangle
|\Psi_{x' \oplus y',G}^B(x') \rangle
|A \rangle |A A' \rangle |B \rangle
|x \rangle |y \rangle |x' \rangle |y' \rangle
\nonumber \\
&&\hskip1.025in \to\ |h_0 \rangle |\Psi^\oplus_\mathrm{DS} \rangle
\label{8.10} \\
&&|\Psi^\oplus_\mathrm{DS} \rangle \ \equiv \
\frac{1}{M^{3/2} N_D^2} \sum_{x,y,x',y'} \sum_{A,A',B}
|\Psi^A_{\Sigma,G}(x) \rangle |\Psi^{A' *}_{\Sigma,G}(x \oplus y) \rangle
|\Psi_{x' \oplus y',G}^B(x') \rangle
|A \rangle |A A' - h_0 B \rangle |h_0 B \rangle
|x \rangle |y \rangle |x \oplus x' \rangle |y' \rangle,
\nonumber
\end{eqnarray}
\end{widetext}
in which controlled bitwise multiplication and addition are all restricted to the $A,A',B$ registers. The fundamental trace identity
\begin{equation}
\mathrm{tr}_E[|\Psi^\oplus_\mathrm{DS} \rangle \langle|\Psi^\oplus_\mathrm{DS}|]
= |\psi^\oplus_\mathrm{DS} \rangle \langle \psi^\oplus_\mathrm{DS}|.
\label{8.11}
\end{equation}
follows from (\ref{8.6}) and (\ref{8.9}), with the various delta-functions acting, in particular, to map $|AA'-h_0B \rangle$ onto the state (\ref{7.8}). It follows that the conventional state (\ref{7.7}) is reproduced in reduced density matrix form.

The environmental average is now over three independent subspaces corresponding to the first three registers in $|\Psi^\oplus_\mathrm{DS} \rangle$. Independence here is defined in the following important restricted sense. Independent data loading subspace structures (\ref{5.17}) and (\ref{5.18}) must be maintained for each (which may have important implications for the computational hardware structure), but dissipation dynamics may subsequently mix them arbitrarily with each other and with the broader environment. Thus, arbitrary unitary transformations $\hat U_\mathrm{diss}$ may be applied to the first three registers (the non-working space qubits), in particular mixing the three subspaces arbitrarily. The cyclic property of the trace, for operators restricted to the environmental subspace, guarantees that only $\hat U_\mathrm{diss}^\dagger \hat U_\mathrm{diss} = \hat I_E$ enters, leaving the working subspace reduced density matrix invariant.

\subsection{Generalized oracle construction}
\label{sec:genoracleconstruct}

We next show how the state $|\Psi^\oplus_\mathrm{DS} \rangle$ is used to construct a generalized Grover oracle. Paralleling (\ref{7.10}), we write this state in the form
\begin{eqnarray}
|\Psi^\oplus_\mathrm{DS} \rangle
&=& \frac{1}{M^{3/2} N_D^2} \sum_{x,y,x',y'} \sum_{A,A',B}
|\gamma(AA' - h_0 B) \rangle
\nonumber \\
&&\otimes\ |{\cal A}(x,y,x',y';A,A',B) \rangle
|x \rangle |y \rangle |x \oplus x' \rangle |y' \rangle
\nonumber \\
\label{8.12}
\end{eqnarray}
in which $\gamma(A,A',B) = \frac{1}{2}[1+\mathrm{sgn}(AA' - h_0 B)]$ is the signature qubit for the $|AA' - h_0 B \rangle$ register and all remaining registers are combined into $|{\cal A} \rangle$. With the projection operators (\ref{7.12})--(\ref{7.14}), and the unitary operators (\ref{7.15}), (\ref{7.18}), and (\ref{7.20}) acting on the working space qubits exactly as before, the explicit expression for
\begin{eqnarray}
|\Psi_\mathrm{DS}^\gamma \rangle \equiv \hat U_\gamma
|x_0 \rangle |q \rangle |\Psi^\oplus_\mathrm{DS} \rangle,
\label{8.13}
\end{eqnarray}
generalizing (\ref{7.21}), is quite complicated, with the qubit $|q \rangle \to |q \oplus \gamma(AA'-h_0B) \rangle$ moving inside the sum in the $\hat P_0 |\Psi^\oplus_\mathrm{DS} \rangle$ term. However, since these operators act only on the working space qubits, they factor out of the trace operation,
\begin{eqnarray}
\mathrm{tr}_E[|\Psi^\gamma_\mathrm{DS} \rangle \langle \Psi^\gamma_\mathrm{DS}|]
&=& \hat Q \hat U_\gamma |x_0 \rangle |q \rangle
\mathrm{tr}_E[|\Psi^\oplus_\mathrm{DS} \rangle \langle \Psi^\oplus_\mathrm{DS}|]
\nonumber \\
&&\ \ \ \ \ \ \otimes\ \langle q| \langle x_0| \hat U_\gamma^\dagger \hat Q^\dagger
\nonumber \\
&=& |\psi_\mathrm{DS}^\gamma \rangle \langle \psi_\mathrm{DS}^\gamma|,
\label{8.14}
\end{eqnarray}
precisely reproducing (\ref{7.21}) in reduced density matrix form. The sign flip implementation (\ref{7.26}), using $|q \rangle = |q_- \rangle$, then immediately follows as well.

Most importantly, one sees that a measurement operator of the form
\begin{equation}
\hat {\cal M} = \hat I_E \otimes \hat I_\gamma \otimes \hat I_{\cal A}
\otimes |0^{(4)} \rangle \langle 0^{(4)}| \otimes \hat M
\label{8.15}
\end{equation}
acting (through $\hat M$) only on the $|x_0 \rangle |q \rangle$ oracle qubit subspace, and projecting onto the $|0^{(4)} \rangle$ pixel qubit subspace, yields
\begin{equation}
\langle \Psi^\gamma_\mathrm{DS}| \hat {\cal M} |\Psi^\gamma_\mathrm{DS} \rangle
= \frac{1}{M^3 N_D^4} \langle q \oplus \gamma(x_0) |\langle x_0|
\hat M |x_0 \rangle |q \oplus \gamma(x_0) \rangle,
\label{8.16}
\end{equation}
directly accessing the oracle output.

\subsection{Generalized Grover search and quantum counting implementation}
\label{sec:grsearchqcountgen}

We finally generalize the implementation (\ref{7.27}) of the Grover operator (\ref{3.19}), and the phase estimation procedure (\ref{7.36})--(\ref{7.39}) underlying the quantum counting relation (\ref{3.22}). The procedure is actually now straightforward, being a direct analogue of the steps in the previous section mapping operations on the larger space detection statistic state $|\Psi^\oplus_\mathrm{DS} \rangle$, defined in (\ref{8.12}), to the equivalent operation on the output state $|\psi^\oplus_\mathrm{DS} \rangle$ of the environmental trace operation, defined in (\ref{7.7}).

To begin, since the operator $\hat U_\mathrm{gr}$, defined by (\ref{3.19}) and implemented in (\ref{7.27}), acts only on the control and working space qubits, it also factors out of the environmental trace. Thus, the states
\begin{equation}
|\Psi^{(k)}_{\gamma,\phi} \rangle
= \hat U_\mathrm{gr}^k[\cos(\phi/2) |\gamma_0 \rangle
+ \sin(\phi/2) |\gamma_1 \rangle] |\Psi^\oplus_\mathrm{DS} \rangle,
\label{8.17}
\end{equation}
generalizing (\ref{7.27}) and (\ref{7.29}), obey
\begin{equation}
\mathrm{tr}_E\left[|\Psi^{(k)}_{\gamma,\phi} \rangle
\langle \Psi^{(k)}_{\gamma,\phi}| \right]
= |\psi^{(k)}_{\gamma,\phi} \rangle \langle \psi^{(k)}_{\gamma,\phi}|.
\label{8.18}
\end{equation}
In particular, for known target number $N_\gamma$, the choice $k = k_G(N_\gamma)$ defined by  (\ref{7.30}) produces the desired state dominated by the target-present pixel superposition $|\gamma_1 \rangle$. Measurement--projection operators with structure (\ref{7.31}), but now including an environmental identity factor $\hat I_E$, allow one to access the matrix element on the right hand side of (\ref{7.32}).

The quantum counting algorithm proceeds in an identical fashion. The product oracle operator $\hat {\bf O}_\mathrm{gr}$ acts only on the control qubits (now including $|0^{\otimes t} \rangle |2^{\otimes t} \rangle$) and on the working space qubits. Therefore, the phase estimation operation (\ref{7.36}) followed by inverse Fourier transform (\ref{7.38}), produce, respectively, states $|\Psi^F_{\gamma,\pm} \rangle$ and $|\tilde \Psi_{\gamma,\pm} \rangle = \hat U_F^\dagger |\Psi^F_{\gamma,\pm} \rangle$, with properties
\begin{eqnarray}
\mathrm{tr}_E\left[|\Psi^F_{\gamma,\pm} \rangle
\langle \Psi^F_{\gamma,\pm}| \right]
&=& |\psi^F_{\gamma,\pm} \rangle \langle \psi^F_{\gamma,\pm}|
\nonumber \\
\mathrm{tr}_E\left[|\tilde \Psi_{\gamma,\pm} \rangle
\langle \tilde \Psi_{\gamma,\pm}| \right]
&=& |\tilde \psi_{\gamma,\pm} \rangle \langle \tilde \psi_{\gamma,\pm}|
\nonumber \\
&=& \hat U_F^\dagger |\psi^F_{\gamma,\pm} \rangle \langle \psi^F_{\gamma,\pm}| \hat U_F.
\ \ \ \ \ \
\label{8.19}
\end{eqnarray}
Using $|\xi_0 \rangle$ as the input state, measurement--projection operators of the form (\ref{7.31}), with an additional $\hat I_E$ operator, and $\hat M$ now acting only on the image pixel register $|x_0 \rangle$, again allow one to access properties of the angle eigenstates (\ref{7.39}). The algorithm accuracy discussion below (\ref{7.40}) remains valid. Of course the algorithm efficiency analysis will have additional complications associated with the reduced density matrix representation.

\section{Concluding remarks}
\label{sec:conclude}

Motivated by potentially numerically intensive image processing and target identification applications, we have explored here the possibility of quantum enhancements of the STAP algorithm through adaptation of quantum machine learning algorithms \cite{QML2017}. A key insight is that linear algebra based on rather large matrices lies at the core of many advanced machine learning algorithms, and the quantum implementation then often relies on adaptation of the HHL algorithm \cite{HHL2009}.

A major barrier to the STAP application is that the underlying covariance matrix is not sparse, precluding direct application of the quantum phase estimation algorithm lying at the heart of the HHL algorithm. To circumvent this, an alternative Trotter--Suzuki simulation method (termed density matrix exponentiation) has been proposed, at the expense of strongly entangling the `working qubits' with the environment \cite{QPCA2013}---in this case the very large number qubits involved in importing and processing new copies of the imaging data as the simulation steps forward in time. As a result, the desired quantum computational output state exists only at the level of the reduced density matrix obtained by averaging over the environmental degrees of freedom. This average is automatically accounted for in any measurement performed on the working qubits, however there are a number of new features underlying this procedure, critical to any eventual hardware implementation, that have not been previously explored:

\paragraph{Quantum Digital--Analog conversion:} In order to be consistent with the Hilbert space inner product defined by the Born rule, though typically loaded in `digital' qubit register format (\ref{3.1}), the data must be converted to standard wavefunction `analog' format (\ref{3.2}). The latter is required for correct formulation of the Trotter--Suzuki simulation described in Sec.\ \ref{sec:altST}, producing `analog' inner products such as those in (\ref{4.16}). On the other hand, the Grover search algorithm at the heart of target detection is formulated using the digital format, therefore requiring the reverse qA/D conversion of the HHL algorithm output. Although the qD/A conversion is quite efficient (App.\ \ref{app:qd2a}), the reverse has much higher overhead (App.\ \ref{app:qa2d}), requiring a large number of HHL algorithm parallel output copies. The additional computational burden will need to be evaluated in future work. It is also possible that the much higher density of data coded into an analog state (single complex number representing an entire qubit register) correspondingly increases the level of error correction needed.

\paragraph{Data qubit structure for phase estimation:} For direct computation of the reduced density matrix at a single time the organization of the sequential loading and processing of the data qubits is not important. However, as described in Sec.\ \ref{sec:qphase}, the generalization of the phase estimation algorithm requires a factorization property for state correlations at different times. As illustrated in Fig.\ \ref{fig:datastruct}, this imposes a specific hierarchical subspace structure (\ref{5.17})--(\ref{5.18}) on the data qubits---even if their states are permitted to dissipate to the broader environment after they have been processed. This will place strong additional constraints on the hardware implementation.

\paragraph{Detection statistic oracle for Grover search:} The detection statistic (\ref{2.21}), used to evaluate presence of absence of a target, must now be derived from the density matrix, i.e., from a working space qubit measurement. This precludes direct unitary implementation of the logical function $\gamma$ defined in (\ref{3.14}). Similar to generalized phase estimation, an indirect approach, described in Secs.\ \ref{sec:qtgtdetect} and \ref{sec:groveroraclegen}, is required.

\paragraph{Projected subspace measurements:} At many points in the algorithm, desired qubit states emerge not as independent direct products, but rather require projection of a strongly entangled state onto a particular subspace---for example the outputs of the HHL algorithm [equation (\ref{6.10}) or (\ref{6.26})] and quantum digital--analog conversion algorithm [equations (\ref{A3}), and (\ref{A16}) or (\ref{A27})]. Since projections are non-unitary operations they can be performed only as part of a classical measurement. The projection must therefore be performed simultaneously with measurements on the projected state of interest, not in some sequential fashion, or the desired information will be lost (see, e.g., the discussion in Sec.\ \ref{subsec:finalstatemeas}). This is a perhaps subtle point that has not been properly appreciated in the literature. It does not change any of the published fundamental quantum probabilistic conclusions, but does strongly impact the design of the quantum computation output measurement.

\subsection{Future work}

The aim of this paper has been to define in detail the major quantum algorithm components for a particular machine learning application, filling in a number significant gaps in the literature. General algorithm component efficiencies have been verified, but detailed quantum supremacy estimates, based, e.g., on size, structure, and formatting of the data vectors $|\psi_W \rangle$, and on the new measurement subspace structures (e.g., described in Secs.\ \ref{sec:groversearchimplement} and \ref{sec:grsearchqcountgen}), are beyond the scope of this paper and would be an interesting topic for future investigation.

There are other machine learning applications, such as principal component analysis \cite{QPCA2013} and support vector machines \cite{QSVM2014}, which also rely on density matrix exponentiation to implement some form of the HHL algorithm. It would be interesting to investigate these as well.

Finally, relevant to the current noisy intermediate scale quantum (NISQ) era, variational quantum--classical hybrid versions of the HHL algorithm have recently been reported \cite{VQLS2019}. It would be interesting to explore further adaptation of such algorithms to machine learning problems.

\appendix

\section{Quantum analog--digital conversion}
\label{app:qad}

It was pointed out in Sec.\ \ref{sec:qimpsummary} that the alternative Suzuki--Trotter evolution (\ref{4.8}) requires that the evolving state be written in the analog representation (\ref{3.2}), and the HHL algorithm output state derived in Sec.\ \ref{sec:qlinalg} appears in the same representation. Given that the radar data is most conveniently loaded in the digital form (\ref{3.1}), a quantum digital-to-analog conversion step is required. Conversely, as described in Secs.\ \ref{sec:qtgtdetect} and \ref{sec:groveroraclegen}, the detection statistic that forms the basis for the target identification algorithm requires the digital representation (\ref{7.8}) and (\ref{7.9}), or (\ref{8.10}), for the corresponding quantum state. A quantum analog-to-digital conversion of the HHL algorithm output is therefore required. The two transformations clearly must be very different because the two states lie in entirely different Hilbert spaces.

In this Appendix we summarize possible implementations of the two algorithms. There may well be more efficient versions, and this would be an interesting topic for future work.

\subsection{Quantum digital-to-analog conversion}
\label{app:qd2a}

Consider a general state with representations
\begin{eqnarray}
|\psi^{\cal A} \rangle &=& \sum_x \psi(x) |x \rangle
\nonumber \\
|\psi^{\cal D} \rangle &=& \frac{1}{\sqrt{N}}
\sum_x |\psi(x) \rangle |x \rangle.
\label{A1}
\end{eqnarray}
with normalization $\sum_x |\psi(x)|^2 = 1$. Here $x$ is an arbitrary index, represented here as an $n$-qubit register, and $\psi(x)$ is represented by an $m$-qubit register. The two representations have different numbers of qubits, so there can be no unitary transformation directly outputting one from the other even though the two have identical physical content, specifying the same wavefunction $\psi(x)$.

The ${\cal D}$ representation is highly redundant, reducing the full $M = 2^m$ degrees of freedom available in the first register Hilbert space to a single $m$-bit complex number. One way to accomplish this reduction is as follows. Similar to the HHL algorithm step (\ref{6.5}), we append an ancilla qubit, and perform the rotation
\begin{equation}
\hat R |0 \rangle_a |\psi(x) \rangle
= \left[|0 \rangle \sqrt{1-C^2|\psi(x)|^2}
+ |1 \rangle C \psi(x) \right] |\psi(x) \rangle.
\label{A2}
\end{equation}
controlled by the wavefunction register, in which $\hat R = e^{i\frac{\theta}{2} \hat {\bf n} \cdot {\bm \sigma}}$ with $\sin(\theta/2) = C|\psi|$, $\hat n_z = 0$, $\hat n_y - i \hat n_x = \psi/|\psi|$. The constant $C$ is chosen so that $0 \leq C|\psi| \leq 1$ substantially covers the permitted interval as $x$ is varied.

Next, apply a Hadamard gate product to the wavefunction register to obtain
\begin{eqnarray}
H^{\otimes m} \hat R |\psi^{\cal D} \rangle
&=& \frac{1}{\sqrt{M}} |0 \rangle_a \sum_{J=0}^{M-1} |J \rangle |\phi_J \rangle
\nonumber \\
&&+\ \frac{1}{\sqrt{M}} |1_a \rangle \sum_{J=0}^{M-1} |J \rangle |\psi_J \rangle
\nonumber \\
&=& \frac{C}{\sqrt{NM}} |1 \rangle_a |0^{\otimes m} \rangle |\Psi^{\cal A} \rangle
+ |\Psi^{\cal D}_\perp \rangle \ \ \ \ \ \
\label{A3}
\end{eqnarray}
in which
\begin{eqnarray}
|\phi_J \rangle &=& \frac{1}{\sqrt{N}}
\sum_x \sqrt{1-C^2 |\psi(x)|^2}
(-1)^{\sigma_J(x)} |x \rangle
\nonumber \\
|\psi_J \rangle &=& \frac{C}{\sqrt{N}} \sum_x \psi(x)
(-1)^{\sigma_J(x)} |x \rangle.
\label{A4}
\end{eqnarray}
with bitwise register inner product signature
\begin{equation}
\sigma_J(x) = \sum_{j=1}^n J_j \psi_j(x).
\label{A5}
\end{equation}
The second line of (\ref{A3}) exhibits the desired state $|\Psi^{\cal A} \rangle$ as the $J = 0$ term, which can therefore be extracted through projected measurements on the $|1 \rangle_a |0 \rangle$ subspace, with $|\Psi^{\cal D}_\perp \rangle$ representing the orthogonal subspace consisting of all other terms in the first line of (\ref{A3}).

There are many alternatives to the Hadamard product (which will be important for the inverse conversion). For example, the Fourier transform
\begin{equation}
\hat U_F |\psi(x) \rangle = \frac{1}{\sqrt{M}}
\sum_{J=0}^{M-1} e^{2\pi i J\psi(x)}
\label{A6}
\end{equation}
corresponding to integer $K(x) = M \psi(x)$ (guaranteed by the $m$-bit representation), also leads to the form (\ref{A3}) but now with
\begin{eqnarray}
|\phi_J \rangle &=& \frac{1}{\sqrt{N}}
\sum_x \sqrt{1-C^2 |\psi(x)|^2}
e^{2\pi i J\psi(x)} |x \rangle
\nonumber \\
|\psi_J \rangle &=& \frac{C}{\sqrt{N}} \sum_x \psi(x)
e^{2\pi i J \psi(x)} |x \rangle.
\label{A7}
\end{eqnarray}
More generally, one may use any orthogonal function representation
\begin{eqnarray}
\hat U_P |\psi(x) \rangle
&=& \frac{1}{\sqrt{M}} \sum_{J=0}^{M-1} p_J[\psi(x)] |J \rangle
\nonumber \\
|\psi_J \rangle &=& \frac{1}{\sqrt{N}}
\sum_x \psi(x) p_J[\psi(x)] |x \rangle,
\label{A8}
\end{eqnarray}
with basis functions $p_J$ constrained by the orthogonality condition
\begin{equation}
\frac{1}{M} \sum_{J=0}^{M-1} p_J(\psi) p_J(\phi) = \delta_{\psi \phi}
\label{A9}
\end{equation}
for integer $0 \leq M\psi,M\phi \leq M-1$. In all cases (with convention $p_0(\psi) = 1$), the second line of (\ref{A3}) exhibits the desired state $|\Psi^{\cal A} \rangle$ as the $J = 0$ term. Note that if $p_J(\psi) = \psi$ for some value of $J$ (e.g., orthogonal polynomial basis), this would obviate the need for the rotated ancilla qubit introduced in (\ref{A2}). Instead one simply projects along $|J \rangle$.

\subsection{Quantum analog-to-digital conversion}
\label{app:qa2d}

It is important to note the highly nonlinear structure of the $J \neq 0$ states defined above. It follows that if Hamiltonian evolution is applied to the $|x \rangle$ register [either directly, or via the alternative Trotter--Suzuki formulation (\ref{4.19})], then this dynamics operates quite differently on the different $|\Psi_J \rangle$ (which, for example, have substantially different eigenfunction content due to the additional nonlinear modulation). It follows that one cannot simply apply the inverse operation $\hat R^\dagger \hat U_P^\dagger$ to reconstruct the register state $|\psi(x,t) \rangle$. The dynamics fails to update the bit structure of $|\psi(x) \rangle$ encoded in the basis functions $p_J[\psi(x)]$ to that of $|\psi(x,t) \rangle$:  $\psi_J(x,t) \neq p_J[\psi(x,t)] \psi(x,t)$.

It is for this basic reason that a reverse conversion algorithm is required, based only on the state $|\psi^{\cal A} \rangle$. The key insight is that in order to expand the representation from $n$ to $n+m$ qubits many copies of the state $|\psi^{\cal A} \rangle$ are required. Thus the previous ${\cal D}$ to ${\cal A}$ conversion was accomplished through an effective reduction in the number of qubits: elimination of the $|\psi(x) \rangle$ register by projection on the $|0^{\otimes m} \rangle$ subspace. In order to reverse this procedure, one must reconstruct all of $J \neq 0$ terms in (\ref{A3}) from multiple copies of $|\psi(x) \rangle$ alone. We begin with the most natural polynomial basis, and then discuss the possibility of a more efficient implementation using quantum Fourier transform.

The power wavefunctions
\begin{equation}
|\psi^J \rangle = \frac{1}{{\cal N}_J}
\sum_x \psi(x)^J |x \rangle,
\label{A10}
\end{equation}
may be derived as a subspace of the tensor product state
\begin{eqnarray}
|\psi^{\otimes J} \rangle &\equiv&
|\psi \rangle_1 |\psi \rangle_2 \ldots |\psi \rangle_J
\nonumber \\
&=& \sum_{x_1,x_2,\ldots,x_M}
\psi(x_1) \psi(x_2) \ldots \psi(x_J)
\nonumber \\
&&\ \ \ \ \ \ \ \times\ |x_1 \rangle |x_2 \rangle \ldots |x_J \rangle
\nonumber \\
&\to& \sum_{x_1,x_2,\ldots,x_M}
\psi(x_1) \psi(x_2) \ldots \psi(x_M)
\nonumber \\
&&\ \ \ \ \ \ \times\ |x_1 \rangle |x_1 \oplus x_2 \rangle
\ldots |x_1 \oplus x_J \rangle
\nonumber \\
&=& {\cal N}_J |\psi^J \rangle |0^{\otimes (J-1)} \rangle
+ |\psi_\perp^{\otimes J} \rangle
\label{A11}
\end{eqnarray}
in which the unitary Boolean sum operation is applied to rotate the desired state along $|0^{\otimes (J-1)} \rangle$. One limits $0 \leq J \leq M-1$ to finite values by assuming, as previously, that $\psi(x)$ takes only $M$ binary fraction values.

Given the collection $|\psi^J \rangle |0^{\otimes (J-1)} \rangle$, one may apply a unitary transformation to produce the orthogonal polynomial states $|p_J(\psi) \rangle$ (each in a particular projected subspace, adjoining extra $|0 \rangle$ qubits if necessary). Orthogonality of polynomial basis functions (e.g., Legendre polynomials) is usually defined by continuous integration over the unit interval, so the replacement (\ref{A9}) by a sum over binary fractions may produce somewhat different polynomial coefficients (especially for large $J$).

Given this collection, one may construct the state (again, along some particular projected subspace)
\begin{eqnarray}
|\Psi_P \rangle &=& \frac{1}{\sqrt{M}} \sum_{J=0}^{M-1} |J \rangle |p_J(\psi) \rangle
\nonumber \\
&=& \frac{1}{\sqrt{N}} \sum_x |\psi_P(x) \rangle |x \rangle
\label{A12}
\end{eqnarray}
in which (for each $x$)
\begin{equation}
|\psi_P \rangle = \frac{1}{\sqrt{M}} \sum_{J=0}^{M-1} p_J(\psi) |J \rangle.
\label{A13}
\end{equation}
Finally, by applying the inverse polynomial basis operation
\begin{equation}
\hat U_P^\dagger |J \rangle = \frac{1}{\sqrt{M}}
\sum_\phi p_J(\phi)^* |\phi \rangle
\label{A14}
\end{equation}
the orthogonality relation (\ref{A9}) produces the pure register value
\begin{eqnarray}
\hat U_P^\dagger |\psi_J \rangle &=& \sum_\phi \frac{1}{M}
\sum_{J=0}^M p_J(\psi) p_J^*(\phi) |\phi \rangle
\nonumber \\
&=& |\psi \rangle,
\label{A15}
\end{eqnarray}
and hence the desired digital representation
\begin{eqnarray}
\hat U_P^\dagger |\Psi_P \rangle &=& \frac{1}{\sqrt{N}}
\sum_x |\psi(x) \rangle |x \rangle
\nonumber \\
&=& |\psi^{\cal D} \rangle.
\label{A16}
\end{eqnarray}

\subsubsection{Fourier basis construction}
\label{subsec:fbasisconstruct}

Absent some kind of hierarchical structure, all $M$ polynomial basis functions need to be constructed separately, which limits the algorithm efficiency. On the other hand, Fourier representations often allow an exponential speed-up (enabling, e.g., the phase estimation algorithm discussed in Sec.\ \ref{sec:qphaseconv}). This possibility is now explored in the present context, but unfortunately does not appear to work.

Approximating exponentials by polynomials relies on the usual identity
\begin{equation}
e^{2\pi i t \psi(x)} = \lim_{N_\psi \to \infty}
\left[1 + i \frac{2\pi t}{N_\psi} \psi(x) \right]^{N_\psi},
\label{A17}
\end{equation}
applied in parallel here simultaneously for all $x$, and using an appropriate finite value of $N_\psi$, controlled by the $O(t^2/N_\psi)$ error. The basis we consists of the states
\begin{equation}
|E_J(\psi) \rangle = \frac{1}{\sqrt{N}}
\sum_x e^{2\pi i J \psi(x)} |x \rangle
\label{A18}
\end{equation}
obtained by applying (\ref{A17}) for integer values of $t$. However, rather than construct these for all $J$, we seek a method that reduces it to the binary powers $J = 2^{j-1}$, $j=1,2,\ldots,m$. By adjoining an extra qubit to each, one seeks then an efficient construction of the state
\begin{eqnarray}
|\Psi_P \rangle &=& \sum_x \otimes_{j=1}^m
\frac{1}{\sqrt{2}} \left(|0 \rangle
+ e^{2\pi i 2^{j-1} \psi(x)} |1 \rangle \right)
|x \rangle
\nonumber \\
&=& \frac{1}{\sqrt{M}} \sum_{J=0}^{M-1} |J \rangle |E_J(\psi) \rangle,
\label{A19}
\end{eqnarray}
by bringing the product inside the $x$ sum, the result of which is precisely the Fourier basis version of (\ref{A12}). The inverse Fourier transform operation acting on the $|J \rangle$ register now produces the desired digital form (\ref{A16}), and corresponds precisely to the phase estimation step (\ref{5.8}) performed in parallel for each $x$:
\begin{equation}
|\psi(x) \rangle = \hat U_F^\dagger
\frac{1}{\sqrt{M}} \sum_{J=0}^{M-1} e^{i2\pi J \psi(x)} |J \rangle,
\label{A20}
\end{equation}
with, as usual, the notation $|\psi(x) \rangle \equiv |M_\psi(x) \rangle$ representing the integer $M_\psi(x) = M \psi(x)$.

Construction of (\ref{A19}) in the given product form would appear to require a controlled operation on the indices $j$, $x$ as well as the value $\psi(x)$. Thus, beginning with the state
\begin{equation}
|0^{\otimes m} \rangle H^{\otimes n} |{\bf 0} \rangle
= \frac{1}{\sqrt{N}} \sum_x \otimes_{j=1}^m |0 \rangle_j |x \rangle
\label{A21}
\end{equation}
one can create the internal product via the rotations $|0\rangle \to \frac{1}{\sqrt{2}} \left(|0 \rangle_j +  e^{2\pi i 2^{j-1} \psi(x)} |1 \rangle_j \right)$ in the first line of (\ref{A19}) only if one has available not only the given $|0 \rangle_j |x \rangle$ register values, but also a $|\psi(x) \rangle$ register. The latter clearly fails for the ${\cal A}$ representation.

If, as seems likely, one is not able to construct $|\Psi_P \rangle$ in such a hierarchical fashion but instead requires each $|J \rangle$ term to be constructed independently (in which case there is no advantage to the Fourier representation), then there will be a critical balance between the accuracy of $\psi(x)$ (number of bits) and the algorithm efficiency. For example, $m=10$ bits yields $M = 1024$ values of $J$. Parallelization in $x$ is always preserved, so if $M/N \ll 1$ quantum efficiency might still be maintained.

\subsubsection{Analog-to-digital Summary}
\label{subsec:adsummary}

In conclusion, in order to construct the ${\cal D}$ representation of a time-evolved ${\cal A}$ representation state $|\psi^{\cal A}(t) \rangle$, one must run the time evolution in parallel on a sufficiently large number of copies of the initial state $|\psi^{\cal A}(0) \rangle$ to be able to construct the polynomial basis function states $|p_J[\psi(t)] \rangle$, from there the superposition state $|\Psi_P(t) \rangle$ via (\ref{A12}), and then finally the ${\cal D}$ representation state $|\Psi^{\cal D}(t) \rangle$ via (\ref{A15}).

As a final comment, we note that the alternative Suzuki--Trotter dynamics described in Sec.\ \ref{sec:qsim} derives the desired state from a trace over a large orthogonal space of environmental degrees of freedom. Here the desired states (polynomials in $|\psi \rangle$) are also entangled in a much larger space, but are instead obtained by projection along a particular $|0 \rangle$ product state axis. In both cases, projected measurements are required to extract output from the final computational state.

\subsection{Reduced density matrix analog-to-digital conversion}
\label{app:dmqad}

We finally verify the generalized construction, including environmental trace operation for reduced density matrix. The key is that analog-to-digital conversion only needs to be performed on the working space qubits, not on the huge environmental space qubits.

We construct as many copies as necessary of the state $|\hat {\bm \Psi}_{R,F} \rangle$ (Sec.\ \ref{subsec:genhhlfinal}), each of which includes an independent environmental subspace, generally undergoing dissipation. Polynomials of such states are constructed in the same fashion as (\ref{A11}), except that the Boolean sum is performed only on the measurement space degrees of freedom.

In somewhat more detail, the final output of the generalized HHL algorithm is a state
\begin{equation}
|\Psi \rangle = \frac{1}{\sqrt{N}} \sum_x |\Psi(x) \rangle |x \rangle
\label{A22}
\end{equation}
with the (analog form) factorization property
\begin{equation}
\mathrm{tr}_E[|\Psi(x) \rangle \langle \Psi(x')|]
= \Sigma_G(x)  \Sigma_G(x')^*
\label{A23}
\end{equation}
where, recall, $\Sigma_G(x) = [{\bf G}^{-1} {\bm \Sigma}](x)$ is the desired filtered state originally defined in (\ref{2.2}). The product states, following the Boolean operation, take the form
\begin{eqnarray}
|\Psi \rangle_1 \ldots |\Psi \rangle_J
&\to& {\cal N}_J |\Psi^J \rangle |0^{\otimes (J-1)} \rangle
+ |\Psi^{\perp \otimes J} \rangle
\nonumber \\
|\Psi^J \rangle &\equiv& {\cal N}_J
\sum_x \otimes_{j=1}^J |\Psi(x) \rangle_j
|x \rangle |x \rangle \ \ \ \ \ \
\label{A24}
\end{eqnarray}
with factorization property (following from identical considerations applied in Sec.\ \ref{subsec:factor})
\begin{equation}
\mathrm{tr}_E[|\Psi(x) \rangle_j {}_k\langle \Psi(x')|]
= \Sigma_G(x) \Sigma_G(x')^*
\label{A25}
\end{equation}
leading to
\begin{eqnarray}
\mathrm{tr}_E[|\Psi^J \rangle \langle \Psi^J|]
&=& |\psi_\Sigma^J \rangle \langle \psi_\Sigma^J|
\nonumber \\
|\psi_\Sigma^J \rangle &\equiv& \sum_x
\Sigma_G(x)^J |x \rangle.
\label{A26}
\end{eqnarray}
Thus, the factorization property again guarantees the desired output product states.

The subsequent linear operations also factor out of the environmental trace operation. Thus, linear combinations, followed by the inverse polynomial basis operation generates the state
\begin{equation}
|\Psi^{\cal D} \rangle = U_P^\dagger \frac{1}{\sqrt{M}}
\sum_{J=0}^{M-1} |p_J(\Psi^J) \rangle  |J \rangle
\label{A27}
\end{equation}
with reduced density matrix
\begin{equation}
\mathrm{tr}_E[|\Psi^{\cal D} \rangle \langle \Psi^{\cal D}|]
= |\Psi_{G,\Sigma}^{\cal D} \rangle \langle \Psi_{G,\Sigma}^{\cal D}|
\label{A28}
\end{equation}
constructed from the desired ${\cal D}$ format state. It is critical here that this construction involves only the working qubit subspace. The underlying unitary structure guarantees that, if this transformation were applied as well to a maintained data qubit space, it would simply cancel out in the trace operation. Such an additional operation is therefore redundant and unnecessary. This property permits us to continue to allow qubit recycling and dissipation without affecting the reduced density matrix.


\begin{thebibliography}{}

\bibitem{qcsize} E. Conover, ``Google moves toward quantum supremacy with 72-qubit computer,'' \href{https://www.sciencenews.org/article/google-moves-toward-quantum-supremacy-72-qubit-computer}{Science News, March 5, 2018}. S. Shankland, ``IBM's biggest-yet 53-qubit quantum computer will come online in October,'' \href{https://www.cnet.com/news/ibm-new-53-qubit-quantum-computer-is-its-biggest-yet/}{CNET}.


\bibitem{qsupreme} For recent discussion on achievement of quantum supremacy, see: ``Quantum supremacy using a programmable superconducting processor,'' F. Arute, \emph{et al.}, F. Arute, K. Arya, R. Babbush, \emph{et al.}, ``Quantum supremacy using a programmable superconducting processor,'' \href{https://doi.org/10.1038/s41586-019-1666-5}{Nature \textbf{574}, 505–-510 (2019)}; E. Pednault, J. A. Gunnels, G. Nannicini, L. Horesh, R. Wisnieff, ``Leveraging Secondary Storage to Simulate Deep 54-qubit Sycamore Circuits,'' \href{https://arxiv.org/abs/1910.09534}{arXiv:1910.09534 [quant-ph]}.

\bibitem{Klemm2004} R. Klemm, \href{https://www.researchgate.net/publication/235759447_Applications_of_Space-Time_Adaptive_Processing}{\emph{Applications of Space-Time Adaptive Processing,} (IEE Publishing, 2004)}.

\bibitem{Melvin2004} W. L. Melvin, ``A STAP Overview,'' \href{http://ieee-aess.org/sites/ieee-aess.org/files/documents/2004 January Systems Magazine.pdf}{IEEE A\&E Systems Magazine \textbf{19}, 19 (2004)}.

\bibitem{NC2000} M. A. Nielsen and I. L. Chuang, \emph{Quantum computation and quantum information} (Cambridge University Press, 2000); ISBN 978-0-521-63503-5; \href{https://www.worldcat.org/title/quantum-computation-and-quantum-information/oclc/844974180}	{OCLC 844974180}.

\bibitem{HHL2009} A. W. Harrow, A. Hassidim, and S. Lloyd, ``Quantum algorithm for linear systems of equations,'' \href{https://doi.org/10.1103/10.1103/PhysRevLett.103.150502}{Phys.\ Rev.\ Lett.\ \textbf{103}, 150502 (2009)}; \href{https://arxiv.org/abs/0811.3171}{arXiv:quant-ph/0811.3171}.

\bibitem{QML2017} J. Biamonte, P. Wittek, N. Pancotti, P. Rebentrost, N. Wiebe, and S. Lloyd, ``Quantum machine learning,'' \href{https://doi.org/10.1038/nature23474}{Nature (London) \textbf{549}, 195--202 (2017)}. \href{https://arxiv.org/abs/1611.09347}{arXiv:quant-ph/1611.09347}.

\bibitem{QPCA2013} S. Lloyd, M. Mohseni, and P. Rebentrost, ``Quantum principal component analysis,'' \href{https://doi.org/10.1038/nphys3029}{Nature Physics \textbf{10}, 631--633 (2014)}. \href{https://arxiv.org/abs/1307.0401}{arXiv:quant-ph/1307.0401}.

\bibitem{QSVM2014} P. Rebentrost, M. Mohseni, and S. Lloyd, ``Quantum support vector machine for big data classification,'' \href{https://doi.org/10.1103/PhysRevLett.113.130503}{Phys.\ Rev.\ Lett.\ \textbf{113}, 130503 (2014)}. \href{https://arxiv.org/abs/1307.0471}{arXiv:quant-ph/1307.0471}.

\bibitem{Chirp} Chirp signals and, among other things, serve to vastly decrease transmitter instantaneous power requirements. See, e.g., \url{https://en.wikipedia.org/wiki/Chirp} and references and links therein.

\bibitem{foot:centerfreq} The signal carrier frequency (typically from a few to a few tens of GHz) is removed by the analog hardware, enabling significantly lower frequency digital time sampling.

\bibitem{QRAM2008} V. Giovannetti, S. Lloyd, L. Maccone, ``Quantum random access memory,'' \href{https://doi.org/10.1103/PhysRevLett.100.160501}{Phys.\ Rev.\ Lett.\ \textbf{100}, 160501 (2008)}. \href{https://arxiv.org/abs/0708.1879}{arXiv:quant-ph/0708.1879}.

\bibitem{foot:qcompress} It is assumed here that the compression operation (\ref{2.8}) is performed classically. There are extremely efficient algorithms for this, and it is not the rate limiting step. For example, although Fourier transforms may be involved, the record lengths are not long enough to make a quantum implementation desirable. In addition, the raw data would be need to be reloaded and reprocessed for each later quantum memory access, so it is likely more efficient to perform this step once and have the result classically available for reloading.

\bibitem{foot:motionmodel} It is assumed for simplicity here (and to maintain focus on the essentials of the quantum algorithm) that all data sets are collected effectively simultaneously on the scale of any target motion within the scene. For example, relevant pulse--scene interation times appearing in the signal model (\ref{2.2}) might be $t \sim 10^{-5}$ s. On the other hand, synthetic aperture radar (SAR) data collected along an aircraft trajectory may extend 1 s or more in time. An additional motion model ${\bf x}_j(T_m) \simeq {\bf x}_j + T_m {\bf v}_j$ (which might spread the target over several meters) would then need to be included if different signals $S_m(t)$ are collected at different transmission times $T_m$.

\bibitem{foot:cmplx} For complex data the signal register $|S \rangle = |S_\mathrm{Re} \rangle |S_\mathrm{Im} \rangle$ is also separated into real and imaginary parts.

\bibitem{foot:hadamard} The Hadamard gate $\hat H$ is defined by $\hat H|0 \rangle = (|0\rangle + |1 \rangle)/\sqrt{2}$, $\hat H|1 \rangle = (|0\rangle - |1 \rangle)/\sqrt{2}$ (equivalent to a 1-qubit Fourier transform), and the identity (\ref{3.7}), used also in (\ref{3.18}), which efficiently constructs the uniform register state superposition from an initial $|0 \rangle$ is extremely useful.

\bibitem{HS2005} See, e.g., N. Hatano and M. Suzuki, ``Finding Exponential Product Formulas of Higher Orders'' in \href{https://doi.org/10.1007/11526216_2}{\emph{Quantum Annealing and Other Optimization Methods}} Eds.\ A. Das and B. K. Chakrabarti (Springer, Berlin, 2005) pp.\ 37--68, and references therein. \href{https://arxiv.org/abs/math-ph/0506007}{arXiv:math-ph/0506007}.

\bibitem{A2004} G. Ahokas, ``Improved algorithms for approximate quantum Fourier transforms and sparse Hamiltonian simulations,'' M.Sc.\ Thesis, University of Calgary (2004).

\bibitem{BACS2006} D. W. Berry, G. Ahokas, R. Cleve, and B. C. Sanders,``Efficient quantum algorithms for simulating sparse Hamiltonians,'' \href{https://doi.org/10.1007/s00220-006-0150-x}{Commun.\ Math.\ Phys.\ \textbf{270}, 359--371 (2007)}; \href{https://arxiv.org/abs/quant-ph/0508139}{arXiv:quant-ph/0508139}.

\bibitem{W2019} P. B. Weichman, ``A quantum phase transition implementation of quantum measurement,'' \href{https://arxiv.org/abs/1912.08764}{arXiv:1912.08764 [cond-mat.stat-mech]}.

\bibitem{foot:checkoracle} One takes advantage here of the fact that it is easy to check that a given $x$ is a solution, even if finding the correct $x$ is very hard. Classically one might need to sequentially check each value of $x$, hence $O(N_D)$ steps. The quantum algorithm requires only $O(\sqrt{N_D})$ steps \cite{NC2000}---well short of exponential speedup, but significant nonetheless.

\bibitem{foot:Deltafn} As is well known, although vanishing at all nonzero diadic rationals $x = Q/M$, the function $\Delta_M(s)$ oscillates extremely rapidly about these points with slowly decreasing magnitude $\sim$\,$1/|s|$. There are quantum algorithms that include additional window functions in the inverse Fourier transform operation in (\ref{5.8}) that suppress the oscillation magnitude and produce better quantum error bounds on eigenvalue estimates derived from the states (\ref{5.11}) or (\ref{5.14}) \cite{NC2000,HHL2009}. Since the application of Fourier transforms in the present work is completely conventional, such windowing operations may be applied here as well [see, e.g., (\ref{5.36})], but this is ignored here for presentational simplicity.

\bibitem{foot:phaseest} It is worth noting that $\hat O_U \hat H^{\otimes m} |1\rangle |0^{\otimes m} \rangle |\psi_u \rangle = |1\rangle |\varphi_{F,u}\rangle |\psi_u \rangle$ \cite{foot:hadamard}, providing an apparently more compact alternative to construction of the Fourier series $|\varphi_{F,u} \rangle$ without the additional $|2^{\otimes m} \rangle$ register ($\hat O_U$ acts here instead directly on the $|J \rangle$ register, and the initial $|1 \rangle$ register is actually redundant). However this is much less efficient since the breakup into binary factors requires exponentially fewer calls for the $\hat U$ operation \cite{NC2000}.

\bibitem{VQLS2019} C. Bravo-Prieto, R. LaRose, M. Cerezo, Y. Subasi, L. Cincio, P. J. Coles, ``Variational quantum linear solver: A hybrid algorithm for linear systems,'' \href{https://arxiv.org/abs/1909.05820}{arXiv:1909.05820 [quant-ph]}.






\end{thebibliography}
\end{document}